\documentclass{article}
\usepackage{bm}% bold math
\usepackage{graphicx}
\usepackage{amsfonts}
\usepackage{amssymb}
\usepackage{amsmath}
\usepackage{apacite}\newcommand{\url}[1]{}

\newcommand{\bd}{\begin{document}}
\newcommand{\ed}{\end{document}}
\newcommand{\beq}{\begin{equation}}
\newcommand{\eeq}{\end{equation}}
\newcommand{\nid}{\noindent}
\newcommand{\ben}{\begin{enumerate}}
\newcommand{\een}{\end{enumerate}}
\newcommand{\bit}{\begin{itemize}}
\newcommand{\eit}{\end{itemize}}
\newcommand{\baR}{\begin{array}}
\newcommand{\eaR}{\end{array}}
\newcommand{\su}[1]{\section{#1}}
\newcommand{\ssu}[1]{\subsection{#1}}
\newcommand{\sssu}[1]{\subsubsection{#1}}
\newcommand{\deq}{\stackrel {\rm def}{=}}
\newcommand{\bea}{\begin{eqnarray}}
\newcommand{\eea}{\end{eqnarray}}
\newcommand{\SOS}[1]{~\\\centerline{\bf{\large [#1]}}\\~}
\newcommand{\st}[1]{{\nid\bf #1}}
\newcommand{\emp}[1]{{\it #1}}

\newcommand\D{\displaystyle}

\newcounter{CDef}[section]
\newtheorem{definition}{Definition}%[CDef]
\newcommand{\bdf}{\begin{definition}}
\newcommand{\edf}{\end{definition}}
\newcommand{\bpr} {\noindent {\bf Proof }}
\newcommand{\epr}{\rule{0.25cm}{0.25cm}\\}

\newtheorem{theorem}{Theorem}
\newcommand{\bth}{\begin{theorem}}
\newcommand{\enth}{\end{theorem}}
\newcounter{Cprop}[section]
\newtheorem{proposition}{Proposition}%[Cprop]
\newcommand{\bp}{\begin{proposition}}
\newcommand{\ep}{\end{proposition}}
\newcounter{Ccor}\newtheorem{corrollary}{Corrollary}%[CLem]
\newcommand{\bcor}{\begin{corrollary}}
\newcommand{\ecor}{\end{corrollary}}
\newcounter{CLem}\newtheorem{lemma}{Lemma}%[CLem]
\newcommand{\blem}{\begin{lemma}}
\newcommand{\elem}{\end{lemma}}
\newcounter{Cconj}\newtheorem{conjecture}{Conjecture}%[CLem]
\newcommand{\bconj}{\begin{conjecture}}
\newcommand{\econj}{\end{conjecture}}
\newcommand\rep{\rightharpoonup}
\newcommand\drep{\downharpoonright}

\newcommand\bbbr{{\sf I\!R}}
\newcommand\bbbrn{{\sf I\!R}^N}
\newcommand\bbbn{{\sf I\!N}}
\newcommand\Wb{\bar{W}}
\newcommand\Vm{V_{min}}
\newcommand\VM{V_{max}}
\newcommand\bt{\bar{\theta}}
\newcommand\sigt{\sigma_\theta}
\newcommand{\XI}{\mbox{\boldmath $\xi$}}

\newcommand\ei{{\bf e}_i}
\newcommand\x{{\bf x}}
\newcommand\X{{\bf X}}
\newcommand\xT{{\x^{(T)}}}
\newcommand\XTt{{\X^{(\tau)}(t)}}
\newcommand\XTtp{{\X^{(\tau)}(t+1)}}
\newcommand\VTt{{\V^{(\tau)}(t)}}
\newcommand\VTtp{{\V^{(\tau)}(t+1)}}
\newcommand\bolm{{\bf m}}
\newcommand\mT{\bolm^{(\tau)}}
\newcommand\miT{m_i^{(\tau)}}
\newcommand\mTp{\bolm^{(\tau+1)}}
\newcommand\Y{{\bf Y}}
\newcommand\F{{\bf F}}
\newcommand\bH{{\bf H}}
\newcommand\Fg{{\F_{\bg}}}
\newcommand\FgT{{\F_{\gT}}}
\newcommand\Fgt{{\F_{\bsg}^t}}
\newcommand\Fgta{{\F_{\bsg}^{t\ast}}}
\newcommand\Fgn{{\F_{\bsg}^n}}
\newcommand\Wij{W_{ij}}
\newcommand\WijT{W_{ij}^{(\tau)}}
\newcommand\WT{\cW^{(\tau)}}
\newcommand\GaT{\Gamma^{(\tau)}}
\newcommand\GijT{\Gamma_{ij}^{(T)}}
\newcommand\WTp{\cW^{(\tau+1)}}
\newcommand\dWT{\delta\cW^{(\tau)}}
\newcommand\WijTp{W_{ij}^{(\tau+1)}}
\newcommand\dWijT{\delta \WijT}
\newcommand\dWij{\delta \Wij}
\newcommand\poT{\pi_{\tom^{(\tau)}}}
\newcommand\poTp{\pi_{\tom^{(\tau+1)}}}
\newcommand\tin{\tau_i(\omega,n)}

\newcommand\ld{r_{d}}
\newcommand\lmT{\lambda_1^{(\tau)}}

\newcommand\tX{\tilde{\X}}

\newcommand\lb{\left\langle}
\newcommand\rb{\right\rangle}
\newcommand\rbT{\rb^{(\tau)}}
\newcommand\rbTp{\rb^{(\tau+1)}}

\newcommand\cA{{\cal A}}
\newcommand\cB{{\cal B}}
\newcommand\bC{{\bf C}}
\newcommand\cF{{\cal F}}
\newcommand\cL{{\cal L}}
\newcommand\cFg{{\cF_{\bg}(\phi)}}
\newcommand\cFgp{{\cF_{\bg'}(\phi)}}
\newcommand\cP{{\cal P}}
\newcommand\cR{{\cal R}}
\newcommand\cT{{\cal T}}

\newcommand\Po{{\cP_{\bom}}}
\newcommand\Pop{{\cP_{\bom'}}}
\newcommand\Pot{{\cP_{\bom(t)}}}
\newcommand\pTo{{\pi_{\tom}^{(T)}}}
\newcommand\pToT{{\pi_{\tom^{(\tau)}}^{(T)}}}
\newcommand\pTpoT{{\pi_{\tom^{(\tau+1)}}^{(T)}}}
\newcommand\po{{\pi_{\tom}}}
\newcommand\pop{{\pi_{\tom'}}}
\newcommand\piij{{\pi_{\tom;ij}}}
\newcommand\mula{{\mu^{Leb}_{\cA}}}
\newcommand\muo{{\mu_{\tom}}}
\newcommand\mg{{\mu_{\bg}}}
\newcommand\mpg{{\nu_{\bspsi}}}
\newcommand\mpgs{{\nu_{\bspsis}}}
\newcommand\Do{\cD(\bom)}
\newcommand\bDo{\bar{\cD}(\bom)}
\newcommand\Dop{\cD(\bom')}
\newcommand\bDop{\bar{\cD}(\bom')}

\newcommand\dT{{d_\Theta}}
\newcommand\dtl{\Delta^{(\tau)}[S']}

\newcommand\BA{\cB(\cA)}
\newcommand\ba{\boldsymbol\alpha}
\newcommand\Bg{\boldsymbol\Gamma}
\newcommand\bg{\boldsymbol\gamma}
\newcommand\bl{\boldsymbol\lambda}
\newcommand\bphi{\boldsymbol\phi}
\newcommand\Beta{\boldsymbol\eta}
\newcommand\bpsi{\boldsymbol\psi}
\newcommand\bPsi{\boldsymbol\Psi}
\newcommand\bom{\boldsymbol\omega}
\newcommand\bu{{\bf u}}
\newcommand\tbu{\tilde{\bu}}
\newcommand\tu{\tilde{u}}
\newcommand\sbg{\scriptstyle{\boldsymbol\gamma}}
\newcommand\gT{\bg^{(\tau)}}
\newcommand\sgT{\sbg^{(\tau)}}
\newcommand\gTp{\bg^{(\tau+1)}}
\newcommand\bsg{\sbg}
\newcommand\bsl{\scriptstyle{\boldsymbol\lambda}}
\newcommand\bspsi{\scriptstyle{\boldsymbol\psi}}
\newcommand\bspsis{\scriptstyle{\boldsymbol\psi^\ast}}
\newcommand\bls{\bsl^\ast}
\newcommand\ble{\bl^\ast}
\newcommand\bpsis{\bpsi^\ast}
\newcommand\bxi{\bm{\xi}}

\newcommand\tom{\omega}
\newcommand\sg{{\sigma_{\bg}}}
\newcommand\sTg{{\sigma_{\sgT}}}
\newcommand\sTpg{{\sigma_{\sgTp}}}
\newcommand\stg{{\sigma^t_{\bg}}}
\newcommand\sit{{\sigma^t}}
\newcommand\stgT{{\sigma^t_{\sgT}}}
\newcommand\stgo{{\sigma^t_{\bg}\tom}}
\newcommand\Gg{{G_{\bg}}}
\newcommand\Slp{\left\{0,1\right\}^\bbbn}
\newcommand\SG{\Sigma_{G}}
\newcommand\Spg{\Sigma_{\bg}}
\newcommand\Spgp{\Sigma_{\bg'}}
\newcommand\Spgm{\Sigma_{\bg_m}}
\newcommand\SpgT{\Sigma_{\bg}^{(T)}}
\newcommand\Spgn{\Sigma_{\bg}^{(n)}}
\newcommand\SpT{\Sigma^{(T)}}
\newcommand\Spn{\Sigma^{(n)}}
\newcommand\SgT{\Sigma_{\gT}}
\newcommand\Sgm{\Sigma_{\bg^{(m)}}}
\newcommand\SgTp{\Sigma_{\gTp}}
\newcommand\Minv{\cM^{inv}_{\bg}}
\newcommand\MinvT{\cM^{inv}_{\sgT}}
\newcommand\MinvTp{\cM^{inv}_{\sgTp}}
\newcommand\STpo{S^{(T)}\bpsi(\tom)}
\newcommand\Snpo{S^{(n)}\bpsi(\tom)}
\newcommand\STq{S^{(T)}_{\bg} \psi^{(4)}(.;\bl)}
\newcommand\pres{P\left[\bpsi\right]}
\newcommand\grad{\nabla_{\bg}}
\newcommand\gradPT{\nabla_{\bg}\PT}

\newcommand\phia{\phi_{i_1,j_1}}
\newcommand\phib{\phi_{i_2,j_2}}
\newcommand\psinl{\bpsi^{(l,n)}}
\newcommand\psiOl{\bpsi^{(l,0)}}
\newcommand\psilpO{\bpsi^{(l+1,0)}}
\newcommand\psilnl{\bpsi^{(l,n_l)}}
\newcommand\dpsirlpO{\delta\bpsi^{(l+1,0)}_{reg}}
\newcommand\dpsislpO{\delta\bpsi^{(l+1,0)}_{sing}}
\newcommand\psiT{\bpsi^{(\tau)}}
\newcommand\dpsiT{\delta\bpsi^{(\tau)}}
\newcommand\psiTp{\bpsi^{(\tau+1)}}
\newcommand\dPT{\delta P^{(\tau)}}
\newcommand\PT{P\left[\psiT\right]}
\newcommand\PTp{P\left[\psiTp\right]}
\newcommand\npT{\nu_{\psiT}}
\newcommand\npTp{\nu_{\psiTp}}

\newcommand\muu{\nu^{(1)}_{\bl;\bg}}
\newcommand\Pu{P^{(1)}_{\bg}(\bl)}
\newcommand\pu{\phi^{(1)}(\tom;\bl)}
\newcommand\mud{\nu^{(2)}_{\bl;\bg}}
\newcommand\Pd{P^{(2)}_{\bg(\bl)}}
\newcommand\pd{\phi^{(2)}(\tom;\bl)}
\newcommand\mut{\nu^{(3)}_{\bl;\bg}}
\newcommand\Pt{P^{(3)}_{\bg}(\bl)}
\newcommand\pt{\phi^{(3)}(\tom;\bl)}
\newcommand\muq{\nu^{(4)}_{\bl;\bg}}
\newcommand\Pq{P^{(4)}_{\bg}(\bl)}
\newcommand\pq{\phi^{(4)}(\tom;\bl)}
\newcommand\psigo{\psi(\tom;\small{\bg})}
\newcommand\phigo{\phi(\tom;\small{\bg})}
\newcommand\hchi{\hat\chi}

\newcommand\nua{\nu_{\cA}}

\newcommand\gijo{g_{ij}^{(0)}(W_{ij})}
\newcommand\gijou{g_{ij}^{(01)}(W_{ij},\omejt)}
\newcommand\gijuo{g_{ij}^{(10)}(W_{ij},\omeit)}
\newcommand\gijdu{g_{ij}^{(21)}(W_{ij},\omeit^2\times \omejt)}
\newcommand\gijud{g_{ij}^{(12)}(W_{ij},\omeit\times \omejt^2)}
\newcommand\gijuu{g_{ij}^{(11)}(W_{ij},\omeit\times \omejt)}
\newcommand\gijkl{g_{ij}^{(kl)}(W_{ij},\omeitk\times \omejtl)}
\newcommand\hijkl{h_{ij}^{(kl)}(W_{ij})}

\newcommand\aijo{a_{ij}^{(0)}}
\newcommand\aijou{a_{ij}^{(01)}(\omejt)}
\newcommand\aijuo{a_{ij}^{(10)}(\omeit)}
\newcommand\aijdu{a_{ij}^{(21)}(\omeit^2\times \omejt)}
\newcommand\aijud{a_{ij}^{(12)}(\omeit\times \omejt^2)}
\newcommand\aijuu{a_{ij}^{(11)}(\omeit\times \omejt)}
\newcommand\aijkl{a_{ij}^{(kl)}(\omeitk\times \omejtl)}

\newcommand\aijtou{a_{ij}^{(01)}(W_{ij},\pTo(\tomej))}
\newcommand\aijtuo{a_{ij}^{(10)}(W_{ij},\pTo(\tomei))}
\newcommand\aijtdu{a_{ij}^{(21)}(W_{ij},\pTo(\tomei^2\times \tomej))}
\newcommand\aijtud{a_{ij}^{(12)}(W_{ij},\pTo(\tomei\times \tomej^2))}
\newcommand\aijtuu{a_{ij}^{(11)}(W_{ij},\pTo(\tomei\times \tomej))}

\newcommand\paijou{a_{ij}^{(01)}(W_{ij},\po(\omej))}
\newcommand\paijuo{a_{ij}^{(10)}(W_{ij},\po(\omei))}
\newcommand\paijdu{a_{ij}^{(21)}(W_{ij},\po(\tomei^2\times \tomej))}
\newcommand\paijud{a_{ij}^{(12)}(W_{ij},\po(\tomei\times \tomej^2))}
\newcommand\paijuu{a_{ij}^{(11)}(W_{ij},\po(\tomei\times \tomej))}
\newcommand\paijkl{a_{ij}^{(kl)}(W_{ij},\po(\tomeik\times \tomejl))}

\newcommand\Catn{[\ba(t_1)\ba(t_2) \dots \ba(t_n)]}
\newcommand\omei{\omega_i}
\newcommand\omej{\omega_j}
\newcommand\omek{\omega_k}
\renewcommand\omit{\omega_i^{(\tau)}}
\newcommand\omjt{\omega_j^{(\tau)}}
\newcommand\tomei{\tom_i}
\newcommand\tomej{\tom_j}
\newcommand\tomt{\left[\tom\right]_0^T}
\newcommand\tomit{\left[\tom_i\right]_0^T}
\newcommand\tomjt{\left[\tom_j\right]_0^T}
\newcommand\tomeik{\tom_i^{(k)}}
\newcommand\tomejl{\tom_j^{(l)}}
\newcommand\omeit{\left[\omega_i\right]_{t-T_s,t}}
\newcommand\omejt{\left[\omega_j\right]_{t-T_s,t}}
\newcommand\omeitk{\left[\omega_i\right]^{k}_{t-T_s,t}}
\newcommand\omeitku{\left[\omega_i\right]^{k_1}_{t-T_s,t}}
\newcommand\omeitkd{\left[\omega_i\right]^{k_2}_{t-T_s,t}}
\newcommand\omejtl{\left[\omega_j\right]^{l}_{t-T_s,t}}
\newcommand\omejtlu{\left[\omega_j\right]^{l_1}_{t-T_s,t}}
\newcommand\omejtld{\left[\omega_j\right]^{l_2}_{t-T_s,t}}

\newcommand\ZTp{Z_{\bg}^{(T)}(\psi)}
\newcommand\ZT{Z^{(T)}(\psi)}
\newcommand\Zn{Z^{(n)}(\psi)}

\newcommand\Vr{V_{reset}}
\newcommand\V{{\bf V}}
\newcommand\hV{\hat{V}}

\newcommand\I{{\bf I}}
\newcommand\J{{\bf J}}
\newcommand\Iw{I^{(w)}}
\newcommand\ie{i^{(ext)}}
\newcommand\Ie{{\bf{I}}^{(ext)}}
\newcommand\Iei{I^{ext}_i}
\newcommand\Is{I^{(syn)}}
\newcommand\Isk{I^{(syn)}_k}
\newcommand\hIa{\hat{I}^{(ion)}}

\newcommand\FpT{\cF^{(\tau)}_{\small{\bphi}}}
\newcommand\FpTp{\cF^{(\tau+1)}_{\small{\bphi}}}

\newcommand\tk{\tau^{(k)}}
\newcommand\tik{t_i^{(k)}}
\newcommand\til{t_i^{(l)}}
\newcommand\tjn{t_j^{(n)}}
\newcommand\tpm{\tau^{\pm}}
\newcommand\bG{\bar{G}}
\newcommand\cC{{\cal C}}
\newcommand\cD{{\cal D}}
\newcommand\cE{{\cal E}}
\newcommand\cG{{\cal G}}
\newcommand\cH{{\cal H}}
\newcommand\cI{{\cal I}}
\newcommand\cN{{\cal N}}
\newcommand\cM{{\cal M}}
\newcommand\cQ{{\cal Q}}
\newcommand\cS{{\cal S}}
\newcommand\cU{{\cal U}}
\newcommand\cV{{\cal V}}
\newcommand\cW{{\cal W}}
\newcommand\dW{\delta \cW}
\newcommand\cWT{\cW^{(\tau)}}
\newcommand\cWTp{\cW^{(\tau+1)}}
\newcommand\cWs{\cW^{(\ast)}}
\newcommand\KT{\kappa^{(\tau)}}
\newcommand\KabT{\kappa_{i_1,j_1,i_2,j_2}^{(\tau)}}
\newcommand\GT{G_{\gT}}

\newcommand\Bev{\cB(\V,\epsilon)}
\newcommand\bmd{\cB_\cM(\delta)}

\newcommand\tost{\left[\tom\right]_{s,t}}
\newcommand\ton{\left[\tom\right]_{O,n}}
\newcommand\tot{\left[\tom\right]_{0,t}}
\newcommand\CoT{\left[\tom\right]_{0,T-1}}
\newcommand\Con{\left[\tom\right]_{0,n-1}}
\newcommand\totr{\left[\tom\right]_{t,t+r-1}}
\newcommand\totm{\left[\tom\right]_{0,t-1}}
\newcommand\torm{\left[\tom\right]_{0,r-1}}
\newcommand\tos{\left[\tom\right]_{0,s}}
\newcommand\toT{\left[\tom\right]_{0,T}}
\newcommand\bomt{\bom(t)}
\newcommand\bdom{\bar{\cD}(\bom)}
\newcommand\com{\cC(\bom)}
\newcommand\comt{\cC(\bom(t))}
\newcommand\dom{\cD(\bom)}
\newcommand\mjt{M_j(t,\tom)}

\newcommand\cMo{{\cM_{\bom}}}
\newcommand\cMot{\cM_{\bom(t)}}
\newcommand\Mot{\cM_{\tot}}
\newcommand\Motm{\cM_{\totm}}
\newcommand\MoT{\cM_{\toT}}
\newcommand\Mos{\cM_{\bom(s)}}
\newcommand\oM{\Omega}

\newcommand\be{{\bf e}}
\newcommand\FT{\F^T}
\newcommand\FTp{\F^{T+1}}
\newcommand\bcF{\bar{\cF}}
\newcommand\Ft{\F\left[.;.,\tom\right]}
\newcommand\Fto{\F\left[.;t,\tom\right]}
\newcommand\Ftot{\F^{t+1}_{\tom}}
\newcommand\Ftso{\F^{t,s}_{\tom}}
\newcommand\Fkto{F_k\left(t,\tom\right)}
\newcommand\Fkso{F_k\left(s,\tom\right)}
\newcommand\FkVto{F_k\left(t,\tom\right)\V}
\newcommand\FVTo{\F^{t+1}_{\tom}(\V)}
\newcommand\FTo{\F^{t}_{\tom}}
\newcommand\FkTo{F^{t}_{k;\tom}}
\newcommand\DFFTV{\left. D\FTF\right|_{\V}}
\newcommand\DFFTpV{\left. D\FTpF\right|_{\V}}

\newcommand\dAS{d(\oM,\cS)}
\newcommand\dOS{d(\oM,\cS)}
\newcommand\dASe{d^{(exp)}(\oM,\cS)}

\newcommand\GFTe{\Gamma_{\cF,T,\epsilon}}
\newcommand\GFTpe{\Gamma_{\cF,{T+1},\epsilon}}
\newcommand\GFTep{\Gamma_{\cF,T,\epsilon'}}

\newcommand\PF{\Pi_{\cF}}
\newcommand\PbF{\Pi_{\bcF}}
\newcommand\VF{{\V_{\cF}}}
\newcommand\VFb{{\V_{\bcF}}}
\newcommand\FF{{\F_{\cF}}}
\newcommand\FTF{{\FT_{\cF}}}
\newcommand\FTpF{{\FTp_{\cF}}}
\newcommand\FFb{{\F_{\bcF}}}
\newcommand\oGF{{\Omega(\GF)}}
\newcommand\oGFTe{{\Omega(\GFTe)}}
\newcommand\oFTe{{\Omega_{\cF}(\GFTe)}}
\newcommand\cPT{\cP^{(T)}}

\newcommand\Jkto{J_k(t,\tot)}

\newcommand\gkt{\gamma_k(t,\tot)}
\newcommand\gks{\gamma_k(s,\tos)}
\newcommand\skt{\sigma_k(t,\tot)}

\newcommand\heff{h^{(eff)}}
\newcommand\PrW{\cP_\cW}
\newcommand\PrI{\cP_{\Ie}}
\newcommand\PrWi{\cP_{\cW,\Ie}}

\newcommand\ltil{\left\{\til\right\}_t}
\newcommand\ltjn{\left\{\tjn\right\}_t}

\newcommand\wsg{\cW^s(\V)}
\newcommand\wsl{\cW^s_{loc}(\V)}
\newcommand\wslf{\cW^s_{loc}(\F(\V))}
\newcommand\wse{\cW^s_{\epsilon}(\V)}
\newcommand\wset{\cW^s_{\epsilon}(\V(t))}
\newcommand\psis{\psi^\ast}
\newcommand\nus{\nu^\ast}
\newcommand\bgs{\bg^\ast}

\newcommand{\Jb}{\bar{J}}
\newcommand{\Jbab}{\bar{J}_{\alpha\beta}}
\newcommand{\Wbab}{\bar{W}_{ab}}
\newcommand{\Jbcd}{\bar{J}_{\gamma\delta}}
\newcommand{\Jab}{\sigma_{ab}}
\newcommand{\Jdab}{\sigma^2_{ab}}
\newcommand{\ma}{m_\alpha}
\newcommand{\Ma}{M_\alpha}
\newcommand{\mua}{\mu_\alpha}
\newcommand{\muax}{\mua^X}
\newcommand{\muay}{\mua^Y}
\newcommand{\va}{v_\alpha}
\newcommand{\vax}{v_\alpha^X}
\newcommand{\Ca}{C_{\alpha\alpha}}
\newcommand{\Cax}{C_{\alpha\alpha}^X}
\newcommand{\Cbx}{C_{\beta\beta}^X}
\newcommand{\Cay}{C_{\alpha\alpha}^Y}
\newcommand{\Ia}{I_\alpha}
\newcommand{\Da}{\Delta_\alpha}
\newcommand{\Deab}{\Delta_{\alpha\beta}}
\newcommand{\Deabx}{\Delta_{\alpha\beta}^X}
\newcommand{\Deabv}{\Delta_{ab}}
\newcommand{\sa}{s_\alpha}
\newcommand{\ta}{\tau_\alpha}
\newcommand{\mub}{\mu_\beta}
\newcommand{\mb}{m_{\alpha\beta}}
\newcommand{\vb}{v_\beta}
\newcommand{\vbx}{v_\beta^X}
\newcommand{\Cb}{C_{\beta\beta}}
\newcommand{\Db}{\Delta_\beta}
\newcommand{\Mb}{M_\beta}
\newcommand{\qb}{q_\beta}
\newcommand{\gb}{\gamma_\beta(\mub,\vb;\Cb)}
\newcommand{\gbt}{\gamma_\beta(\mub,\vb;\Cb(t))}
\newcommand{\tb}{\tau_\beta}
\newcommand{\Sb}{S_\beta}
\newcommand{\Ha}{H_{\alpha}}
\newcommand{\Hax}{H_{\alpha}^X}
\newcommand{\Hab}{H_{\alpha\beta}}
\newcommand{\Habx}{H_{\alpha\beta}^X}
\newcommand{\Dab}{D_{\alpha\beta}}
\newcommand{\Dabx}{D_{\alpha\beta}^X}
\newcommand{\cJ}{{\cal J}}

\newcommand{\ind}[1]{\mathbbm{1}_{#1}}          %% Indicative function of an event

\newcommand{\derpart}[2]{ \frac{\partial #1}{\partial #2} } %% Partial derivative
\newcommand{\der}[2]{ \frac{\text{d} #1}{\text{d} #2} }  %% Partial derivative
\newcommand{\derparts}[2]{ \frac{\partial^2 #1}{\partial #2 ^2}}%% Second partial derivative w.r.t. the same variable
\newcommand{\derpartsxy}[3]{ \frac{\partial^2 #1}{\partial #2 \partial #3}} %% Second derivative w.r.t. 2 different variables
\newcommand{\argmin}{\mathop{\mathrm{Arg\,Min}}}
\newcommand{\eps}{\varepsilon}

\DeclareMathOperator{\eqlaw}{\stackrel{\mathcal{L}}{=}}         %%  equal in law

\newcommand{\Herm}[1]{\mathcal{H}_{#1}}            %% Hermite function with parameter
\newcommand{\Hermnu}{\mathcal{H}_{\nu}}            %% Hermite function with parameter
\newcommand{\erf}{\mathrm{erf}}              %% error function erf
\newcommand{\Cov}{\mathrm{Cov}}

\newcommand{\m}{\mathcal}

\newcommand {\image} {\includegraphics}

\newcommand{\mC}{\mathbf{C}}
\newcommand{\mD}{\mathbf{D}}
\newcommand{\mE}{\mathbf{E}}
\newcommand{\mX}{\mathbf{X}}
\newcommand{\mx}{\mathbf{x}}
\newcommand{\mY}{\mathbf{Y}}
\newcommand{\mL}{\mathbf{L}}
\newcommand{\mQ}{\mathbf{Q}}
\newcommand{\mI}{\mathbf{I}}
\newcommand{\mJ}{\mathbf{J}}
\newcommand{\mW}{\mathbf{W}}
\newcommand{\mU}{\mathbf{U}}
\newcommand{\mone}{\mathbf{1}}

\newcommand{\msig}{\boldsymbol{\sigma}}
\newcommand{\mtheta}{\boldsymbol{\theta}}
\newcommand{\mdelta}{\boldsymbol{\Delta}}
\newcommand{\mla}{\boldsymbol{\Lambda}}
\newcommand{\mga}{\boldsymbol{\Gamma}}
\newcommand{\mxi}{\boldsymbol{\xi}}

\newcommand{\proc}{\mathcal{M}_1^+(C([t_0,T], \R)}
\newcommand{\procP}[1]{\mathcal{M}_1^+(C([t_0,T], \R^{#1}))}
\newcommand{\proci}{\mathcal{M}_1^+(C((-\infty,T], \R)}
\newcommand{\prociP}[1]{\mathcal{M}_1^+(C((-\infty,T], \R^{#1})}
\newcommand{\procs}[1]{\mathcal{M}_1^+(C([0,#1], \R)}
\newcommand{\Exp}[1]{\mathbb{E}\left [ #1 \right]}
\newcommand{\ExpSimple}{\mathbb{E}}
\newcommand{\nvar}[1]{\left \| #1 \right\|_{\mathrm{var}}}
\newcommand{\nvarP}[2]{\left \| #1 \right\|_{\mathrm{var}, #2}}
\newcommand{\abs}[1]{\left \vert #1 \right \vert }
\newcommand{\pente}{\|S_{\beta}'\|_{\infty}}
\newcommand{\Fs}{F_{\text{stat}}}
\newcommand{\triplebar}[1]{\vert\vert\vert #1 \vert\vert\vert}
\newcommand{\procsP}[2]{\mathcal{M}_1^+(C([t_0,#1], \R^{#2})}
\newcommand{\dvar}[2]{d_{\rm{var}}\left(#1,\; #2\right)}
\newcommand{\dt}[2]{d_t\left(#1,\; #2\right)}
\newcommand{\norm}[1]{\left \| #1 \right \|}
\newcommand{\Si}{S}
\newcommand{\T}{\mathcal{T}}
\newcommand{\IT}{\mathcal{IT}}
\newcommand{\tV}{\widetilde{V}}

\newcommand {\VV} {\mathbf{V}}
\newcommand {\G} {\mathcal{G}}
\newcommand{\R}{\mathbbm{R}}
\newcommand{\N}{\mathbbm{N}}
\newcommand{\Lop}{\mathcal{L}}
\newcommand{\Pro}{\mathbbm{P}}
\newcommand{\f}{\varphi}
\newcommand{\Diag}{{\mathbf{\rm Diag}}}
\newcommand{\mA}{\mathbf{A}}

\begin{document}

\title{
How Gibbs distributions  may naturally arise from synaptic adaptation mechanisms.\\
A model-based argumentation.
}

\author{
B. Cessac
\thanks{Laboratoire J. A. Dieudonn\'e,
U.M.R. C.N.R.S. N° 6621,
Universit\'e de Nice Sophia-Antipolis,France},
\thanks{INRIA, 2004 Route des Lucioles, 06902 Sophia-Antipolis, France.}
H. Rostro $^\dag$,
J.C. Vasquez $^\dag$,
T. Vi\'eville $^\dag$
}

\date{\today}

\maketitle

\begin{abstract}
This paper addresses two questions in the context of neuronal networks dynamics,
using methods from dynamical systems theory and statistical physics: 
(i) How to characterize the statistical properties of sequences of action potentials
(``spike trains'') produced by neuronal networks ? and; (ii) what are the effects of synaptic
plasticity on these statistics ? We introduce a framework in which spike trains are associated
to a coding of membrane potential trajectories, and actually, constitute a symbolic coding 
in important explicit examples (the so-called gIF models). On this basis, we use the thermodynamic formalism from ergodic theory to
show how Gibbs distributions are natural probability measures to describe the statistics
of spike trains, given the empirical averages of prescribed quantities. As a second result, we show that Gibbs distributions
naturally arise when considering ``slow'' synaptic plasticity rules where the characteristic
time for synapse adaptation is quite longer than the characteristic time
for neurons dynamics.   
\end{abstract}


\su{Introduction.} 

\paragraph{Spike trains as a ``neural code''.}
Neurons activity results from complex and nonlinear mechanisms
\cite{hodgkin-huxley:52,cronin:87,dayan-abbott:01,gerstner-kistler:02b},
leading
to a wide variety of dynamical behaviours \cite{cronin:87,guckenheimer-labouriau:93}.
This activity is revealed by the  emission of action potentials or ``spikes''.
While the shape of an action potential is essentially always the same 
for a given neuron, 
the succession of spikes emitted by this neuron can have a wide variety
of patterns (isolated spikes, periodic spiking, bursting, tonic spiking, tonic bursting, etc \dots)
 \cite{izhikevich:04,brette-gerstner:05,touboul:08},
depending on physiological parameters, but also on excitations coming either from other
neurons or from external inputs. Thus, it seems natural to consider
spikes as ``information  quanta'' or ``bits'' and to seek the information exchanged by
neurons in the structure of spike trains. Doing this, one switches from the description
of neurons in terms of membrane potential dynamics, to a description in terms of
spike trains. This point of view is used, in experiments, by the analysis
of \textit{raster plots}, i.e. the activity of a neuron is represented by a mere vertical bar
each time this neuron emits a spike. Though this change of description raises many questions,
it is commonly admitted in the computational neuroscience community that spike trains
constitute a ``neural code''.

This raises however other questions. 
How is ``information'' encoded in a spike train:
 rate coding \cite{adrian-zotterman:1926}, temporal coding \cite{theunissen-miller:95},
rank coding \cite{perrinet-et-al:01,delorme-et-al:01}, correlation coding \cite{johnson:80} ?
How to measure the information content of a spike train ? 
There is a wide literature dealing with these questions 
\cite{nirenberg-latham:03,johnson:04,barbieri-et-al:04,nemenman-et-al:06,arabzadeh-et-al:06,sinanovic-johnson:06,gao-et-al:08,osbone-et-al:08},
which are inherently related to the notion of  \textit{statistical characterizations} of spike trains, 
see \cite{rieke-etal:96,dayan-abbott:01,gerstner-kistler:02b}
and references therein for a  review.
As a matter of fact, a prior to handle ``information'' in a spike train is
the definition of a suitable probability distribution that matches the empirical
averages obtained from measures. Thus, in some sense that we make precise in this paper,
the choice of a set of quantities to measure (observables) constrains the form
of the probability characterizing the statistics of spike trains. 

As a consequence, there is currently a wide debate on the canonical form of
these probabilities. While Poisson statistics, based on the mere knowledge of frequency rates,
are commonly used with some success in biological experiments \cite{georgeopoulos:82,georgeopoulos:07}, other
investigation evidenced
the role of spikes coincidence or correlations \cite{grammont-riehle:99,grammont-riehle:03}
 and some people have proposed non Poisson probabilities, such as
Ising-like Gibbs distributions, to interpret their data \cite{schneidman-etal:06,tkacik-etal:06}. It is important to note here that beyond the determination of
the ``right'' statistical model there is the aim of identifying which type of 
information is used by the brain to interpret the spike trains that it receives, coming from different sources. As an example 
choosing a model where only frequency rates matters
amounts to assuming that spikes coming from different
neurons are essentially treated independently.

 However, determining 
the form of the probability characterizing the statistics of spike trains 
is extremely difficult in real experiments. In the present paper, we focus
on neural networks \textit{models} considered as dynamical systems, with a good mathematical 
and numerical control on dynamics. In this context we argue that Gibbs distributions
are indeed natural candidates whenever a set of quantities to measure (observables)
has been prescribed. As a matter of fact Poisson distributions and 
Ising-like distributions are specific examples, but, certainly,  do not constitute
the general case.

\paragraph{Synaptic plasticity.}
The notion of neural code and information cannot be separated from
the capacity of neuronal networks to  evolve and adapt by \textit{plasticity}
mechanisms, and especially \textit{synaptic plasticity}.
 The latter occurs at many levels of organization and time scales in the nervous system 
\cite{bienenstock-etal:82}. 
It is of course
involved in memory and learning mechanisms, but it also alters excitability of brain area
and regulates behavioural states (e.g. transition between sleep and wakeful activity).
Therefore, understanding the effects of synaptic plasticity on neurons dynamics
is a crucial challenge. 
On experimental grounds, different synaptic plasticity mechanisms 
have been exhibited from the Hebbian's ones \cite{hebb:49} to Long Term
Potentiation (LTP) and Long Term Depression (LTD), and more recently to Spike Time Dependent
Plasticity (STDP) \cite{markram-etal:97,bi-poo:01} (see \cite{dayan-abbott:01,gerstner-kistler:02,cooper-etal:04}
for a review).  Modeling  these mechanisms  requires both a bottom-up and top-down approach.

This issue is
tackled, on theoretical grounds, by inferring ``synaptic updates rules'' or ``learning rules'' from
biological observations \cite{vondermalsburg:73,bienenstock-etal:82,miller-etal:89} and extrapolating, by theoretical or numerical investigations, the effects
of such synaptic rule on such neural network \textit{model}. 
This bottom-up  approach relies on the belief that
there are ``canonical neural models'' and ``canonical plasticity rules'' capturing the most essential
features of biology. Unfortunately, this results in a plethora of canonical ``candidates'' and
a huge number of papers and controversies.  In an attempt to clarify and unify the overall vision, some researchers have proposed to associate learning rules
and their dynamical effects to general principles, and especially to``variational'' or ``optimality'' principles, where some functional
has to be maximised or minimised \cite{dayan-hausser:04,rao-sejnowski:99,rao-sejnowski:01,bohte-mozer:07,chechik:03,toyoizumi-etal:05,toyoizumi-etal:07}. Therefore, in these ``top-down''
 approaches, plasticity rules ``emerge'' from first principles. 
Unfortunately, in most examples, the validations of these theories has been restricted to considering isolated neurons 
submitted to input spike trains with ad hoc statistics (typically, Poisson distributed with independent spikes  \cite{toyoizumi-etal:05,toyoizumi-etal:07}). 

However, addressing the effect of synaptic plasticity in neuronal networks 
 where dynamics is \textit{emerging} from collective effects and where spikes statistics are \textit{constrained} by this dynamics
seems  to be of central importance. This is the point of view raised in the present paper,
where, again, we focus on  models, which are
simplifications of real neurons. Even in this case,
 this issue is subject to two main difficulties.
On one hand, one must identify the generic collective dynamical regimes displayed by the 
model for different choices of parameters (including synaptic weights). On the other hand,
one must analyse the effects of varying synaptic weights when applying plasticity rules.
This requires to handle a complex interwoven evolution where neurons dynamics 
depends on synapses
and synapses evolution depends on neuron dynamics. The first 
aspect has been addressed by several 
authors using mean-field approaches (see e.g. \cite{samuelides-cessac:07} and references therein), ``Markovian approaches''
\cite{soula-chow:07}, or dynamical system theory (see \cite{cessac-samuelides:07} and references therein).
The second aspect has, up to our knowledge, been investigated theoretically in only a few
examples with Hebbian learning \cite{dauce-etal:98,siri-etal:07,siri-etal:08} or discrete
time Integrate and Fire models with an STDP like rule \cite{soula:05,soula-etal:06} and is further addressed here.

\paragraph{What the paper is about.}
To summarize the previous discussion the study of neuronal networks at the current stage is submitted
to two central questions:

\bit

\item How to characterize the statistics of spike trains in a network of neurons ?

\item How to characterize the effects of synaptic plasticity on this network dynamics and especially
on spike trains statistics ?
 
\eit

In this paper we suggest that these two questions are closely entangled and must both be addressed in the same
setting. In order to have a good control on the mathematics we mainly
consider the case of neural networks models where one has a full characterization
of the generic dynamics  \cite{cessac:08,cessac-vieville:08}. Thus, our aim is not to provide general statements
about  biological neural networks. We simply want to have a good mathematical control
of what is going on in specific models, with the hope that this analysis should shed some light on what happens
(or  \textit{does not} happen) in ``real world'' neural systems. However, though part of the results used here
are rigorous, this work relies also on ``working assumptions'' that we have not been able to check rigorously.
These assumptions, which are essentially used to apply the standard theorems in ergodic theory and thermodynamic
formalism, provide  a logical chain which drives us to important conclusions that could be checked in experimental data.

The paper is organised as follows. In section \ref{neurodyn}
we introduce a framework in which spike trains are associated
to a coding of membrane potential trajectories, and actually, constitute a symbolic coding in 
explicit examples. On this basis, we 
show how Gibbs distributions are natural probability measures to describe the statistics
of spike trains, given the data of known empirical averages. Several authors have discussed the relevance
of Gibbs distribution in the context of Hopfield model (see \cite{amit:89} and references
therein) and more recently to 
spike trains \cite{wood-et-al:06,kang-amari:08}. A breakthrough has been made in 
\cite{schneidman-etal:06,tkacik-etal:06}. Actually, our approach has been greatly influenced
by these two papers, though  our results  hold in a wider context.
In  section \ref{adapt}, we discuss the effect of synaptic adaptation
when considering ``slow'' synaptic plasticity rules where the characteristic
time for synapse adaptation is quite a bit longer that the characteristic time
for neurons dynamics. These rules are formulated in the context of thermodynamic
formalism, where we introduce  a functional, closely 
related to thermodynamic potentials like free energy in statistical physics, and
called ``topological pressure'' in ergodic theory.
In this setting we  show that the variations of synaptic weights leads
to variation of the topological pressure that can be smooth (``regular periods''), or singular (``phase
transitions''). Phase transitions are in particular associated to a change of ``grammar''
inducing modifications in the set of spike trains that the dynamics is able to produce.
We exhibit a functional, closely related to the topological pressure, that decreases
during regular periods. As a consequence, when the synaptic weights converge 
to a fixed value, this functional reaches a minimum. This minimum corresponds to a situation
where  spike trains statistics
are characterized by a Gibbs distribution, whose potential can be explicitly written.
An example with numerical computations is presented in section \ref{numex}.

\su{Neuron dynamics.}\label{neurodyn}

\ssu{Neural state.} \label{NeurStat}

We consider a set of $N$ neurons. Each neuron $i$ is characterized by 
its state, $X_i$, which  belongs to some compact set $\cI \in \bbbr^M$.
  $M$ is the number of variables characterizing
the state of one neuron (we assume that all neurons are described by the same number of variables).
A typical example is an integrate and fire model where $M=1$ and $X_i=V_i$ is the membrane potential
of neuron $i$ and $\cI=[\Vm,\VM]$ (see section \ref{Examples}). Other examples are provided by conductances based models of
 Hodgkin-Huxley type\footnote{Note 
that Hodgkin-Huxley equations are differential equations, while (\ref{DNN}) 
corresponds to a discrete time evolution. We  assume that we have discretized time with
a time scale that can be arbitrary small. A detailed discussion on these aspects
for conductance based IF models has been presented in \cite{cessac-vieville:08}.
Some further comments are given below.
} \cite{hodgkin-huxley:52}.
 Then
$X_i=(V_i,m_i,n_i,h_i)$ where $m_i,n_i$ are respectively the activation variable for Sodium
and Potassium channels and $h_i$ is the inactivation variable for the Sodium channel.

The evolution of these $N$ neurons is
given by a deterministic dynamical system
of form:

\beq\label{DNN}
\X(t+1)=\Fg\left[\X(t)\right]
\eeq  

\nid where  $\X=\left\{X_i\right\}_{i=1}^N$  represents the dynamical state of a network
of $N$ neurons at time $t$, while time is discrete (for a discussion on time discretisation in spiking neural
networks see \cite{cessac-vieville:08}).
Thus $\X \in \cM=\cI^N$ where $\cM$ is the phase space of (\ref{DNN}), and  $\Fg(\cM) \subset \cM$.
 The map $\Fg: \cM \to \cM$ depends on a set of parameters $\bg \in  \bbbr^P$.
The typical case considered here is $\bg = \left( \cW,\Ie \right)$ where  $\cW$ is the matrix of synaptic weights and $\Ie$ is some  external current,
assumed to be independent of time in the present paper
(see section \ref{Examples} for two explicit examples). Thus $\bg$ is a point in a $P=N^2+N$ dimensional space of control parameters.

\ssu{Natural partition.} %\label{NatPart}

Neurons are excitable systems. Namely, neuron $i$ ``fires'' (emits a spike or
action potential), whenever its state $X_i$ belongs to some connected
region $\cP_1$ of its phase space. Otherwise, it is quiescent ($X \in \cP_0=
\cI \setminus \cP_1$).
 In Integrate and Fire models
neuron $i$ fires whenever its membrane potential $V_i$
exceeds some threshold $\theta$. In this case, the corresponding
region is $\cP_1=[\theta,\VM]$. In Fitzhugh-Nagumo \cite{fitzhugh:55,fitzhugh:61,nagumo-etal:62}
or Hodgkin-Huxley model \cite{hodgkin-huxley:52}, the firing
corresponds to the crossing of a manifold called the threshold separatrix \cite{cronin:87} which separates the phase
space into two connected regions $\cP_0$ and $\cP_1$.
For $N$ identical neurons this leads to a ``natural partition''
$\cP$ of the product phase space $\cM$. Call $\Lambda=\left\{0,1\right\}^N$,
$\bom=\left(\omega_i\right)_{i=1}^N \in \Lambda$. Then, $\cP=\left\{\Po \right\}_{\omega \in \Lambda}$,
where $\Po = \cP_{\omega_1} \times \cP_{\omega_2} \times \dots \times \cP_{\omega_N}$.
Equivalently, if $\X \in \Po$, all neurons such that $\omega_i=1$ are firing
while neurons such that $\omega_k=0$ are quiescent. 
We call therefore $\bom$ a  ``spiking pattern''.

\ssu{Raster plots.} %\label{Raster}

To each initial condition $\X \in \cM$ 
 we associate a ``raster plot'' $\tom=\left\{\bom(t)\right\}_{t=0}^{+\infty}$
such that $\X(t) \in \Pot, \forall t \geq 0$. We write $\X \rep \tom$.
Thus, $\tom$ is the sequence of spiking patterns
displayed by the neural network when prepared with the initial condition $\X$.
On the other way round, we say that an infinite sequence $\tom=\left\{\bom(t)\right\}_{t=0}^{+\infty}$
is \textit{an admissible raster plot} if there exists $\X \in \cM$ such that   
$\X \rep \tom$. We call $\Spg$ the set of admissible raster plots for the set
of parameters $\bg$.

\begin{figure}[htb]
 \centerline{\includegraphics[height=7cm,width=7cm,clip=false]{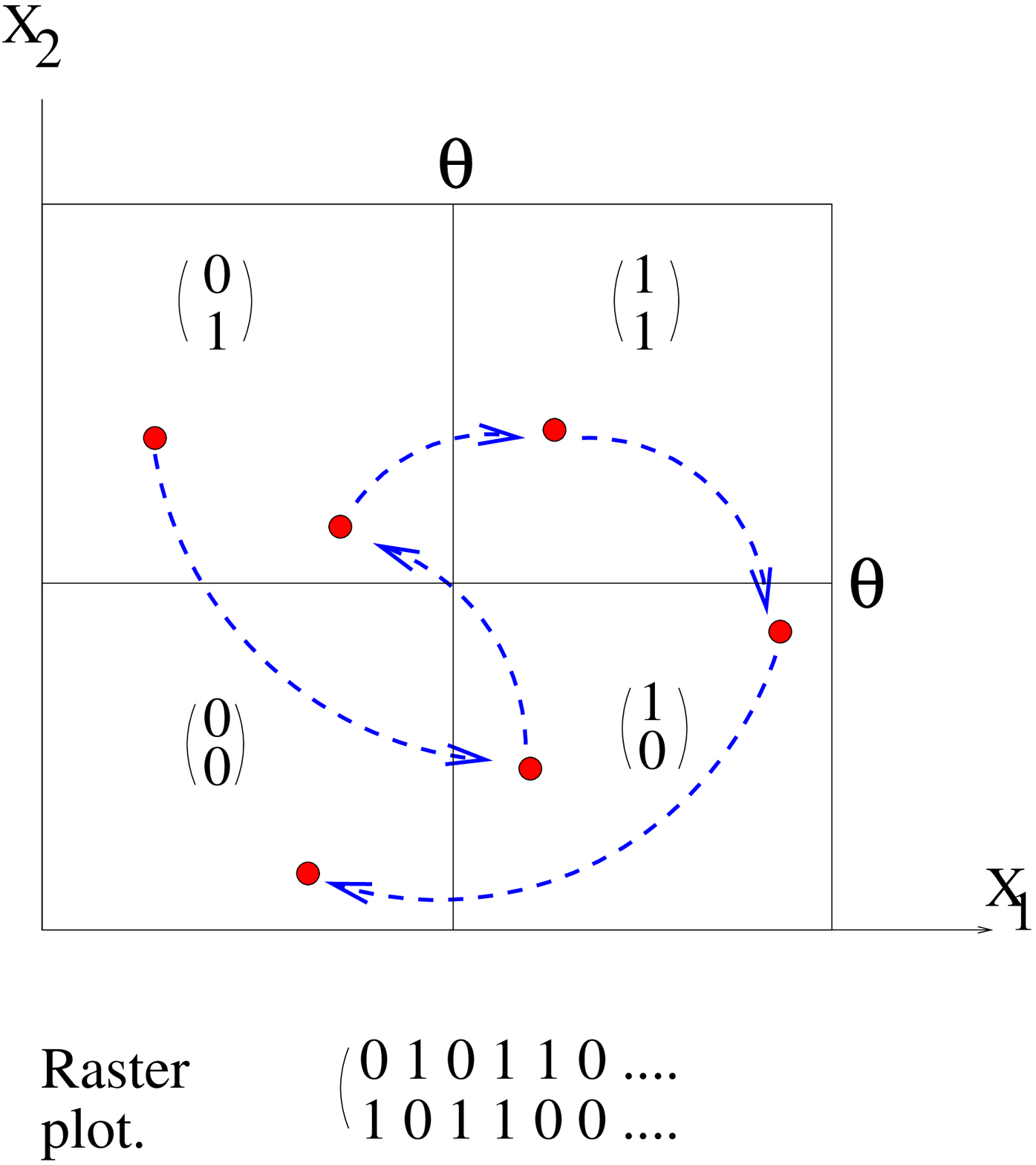}}
\caption{\footnotesize{%\label{Fnatpart}
  The phase space $\cM$ (here represented for $N=2$ neurons with constant threshold)
is partitioned in $2^N$ parts $\Po = \prod_{i=1}^N \cP_{\omega_i}$, $\bom$ being a spiking pattern.
In this way, one associates naturally to an orbit of (\ref{DNN}) a raster plot.}}
\end{figure}

The dynamics (\ref{DNN}) induce a dynamics on the set of raster plot in the following way:
 \\ \centerline{$\begin{array}{ccc}
    \X   & \begin{array}{cc} \Fg \\ \longrightarrow \\ \end{array} & \Fg(\X) \\
 \drep &                                                         & \drep \\
    \tom    & \begin{array}{cc} \sg \\ \longrightarrow \\ \end{array} & \sg(\tom)  \\
  \end{array}$} \\

In this construction $\sg$, called the ``left shift'',  shifts the raster plot left-wise at each time step of the dynamics.
Thus,  in some sense,
raster plots provide a code for the orbits of (\ref{DNN}). 
But, the correspondence may not be one-to-one.  That is why we use the notation 
$\rep$ instead of $\to$.

Let us introduce the following notation. If we are interested in a  prescribed sequence of  spiking patterns $\bom(s), \dots \bom(t)$,
 from time $s$ to time $t$, we denote by $\tost$, the set of raster plots whose spiking patterns from time $s$
to time $t$ match this sequence  (cylinder set).

\ssu{Asymptotic dynamics.} %\label{AsDyn}

\begin{figure}[htb]
 \centerline{\includegraphics[height=5cm,width=9cm,clip=false]{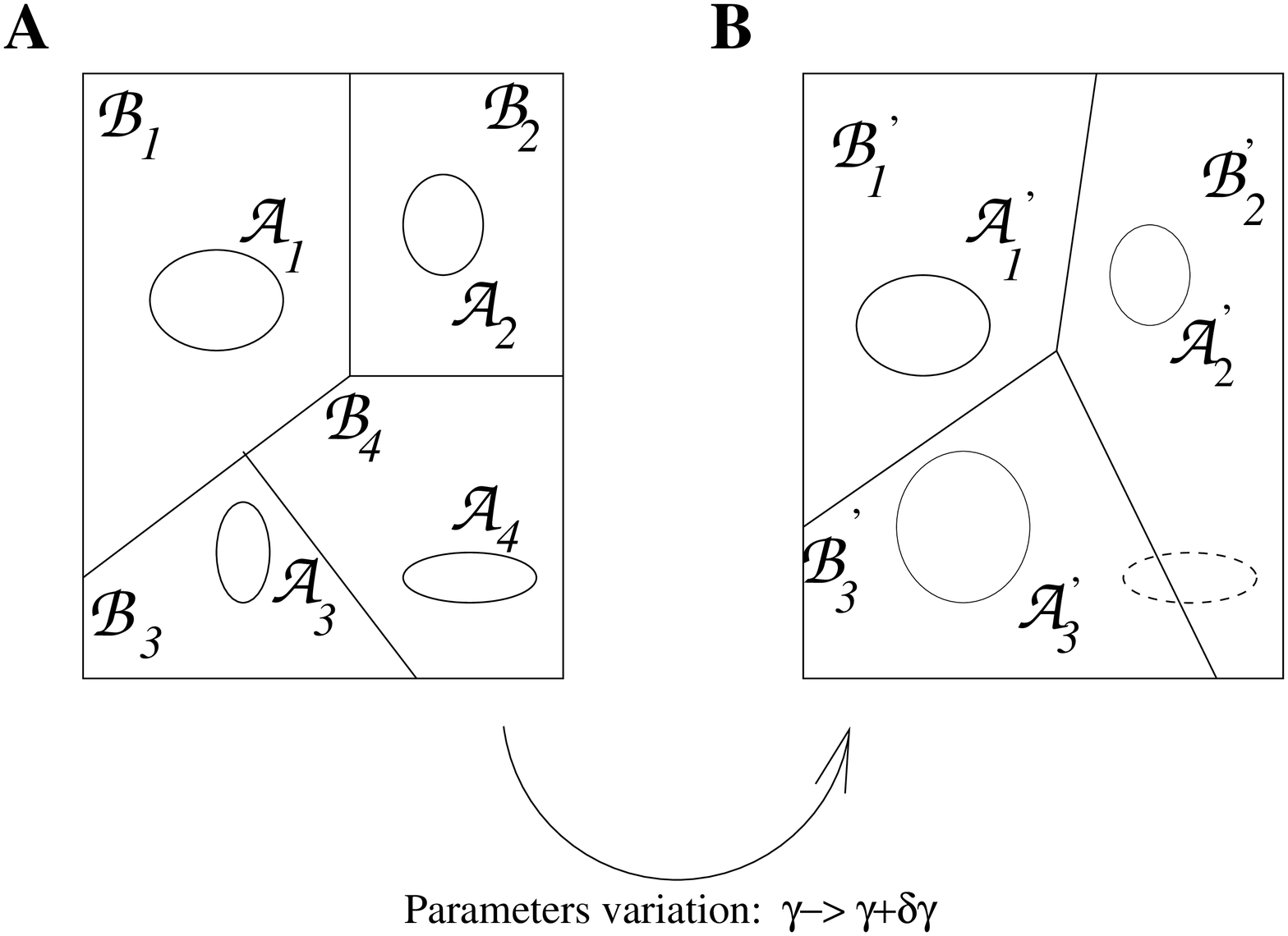}}
\caption{\footnotesize{\label{Fattractors} Schematic illustration of the attractor landscape for neural network models.
[{\bf A}] The phase space is partitioned into bounded domains ${\cal B}_l$ and for each initial condition in ${\cal B}_l$ the initial trajectory is attracted toward an
attractor (e.g. a fixed point, a periodic orbit or a more complex attractor) $\cA_l$.
[{\bf B}] If the parameters (external current, weights) change, the landscape is modified and several phenomena can occur: 
change in the basins shape, number of attractors, modification of the attractor as for $\cA_3$ in this example;
A point belonging to $\cA_4$ in Fig. \ref{Fattractors} A, can, after modification of the parameters, converge
 either to attractor $\cA'_2$ or $\cA'_3$
}}
\end{figure}

We are here mainly  interested in the asymptotic behavior of (\ref{DNN}) and set a few notations
and notions.
 The $\omega$-limit set, $\oM$, is  the set of accumulation points of $\Fgt(\cM)$.
Since $\cM$ is closed and invariant, we have
$\oM=\bigcap_{t=0}^\infty \Fgt(\cM)$. 
In dissipative systems (i.e. a volume element in the phase space  is dynamically contracted),
   the $\omega$-limit
set typically contains the attractors of the system.
A compact set $\cA \in \cM$ is called an \textit{attractor}
for $\Fg$ if there exists a neighborhood $\cU$ of $\cA$
and a time $n>0$ such that $\Fgn(\cU) \subset \cU$ and
$\cA=\D{\bigcap_{t=0}^\infty \Fgt(\cU)}$. In all examples considered in the present 
paper $\Fg$ is dissipative and the phase space is divided into finitely many attraction basins 
each of them containing an attractor (see Fig. \ref{Fattractors}). Simple examples
of attractors are stable fixed points, or stable periodic
orbits. More complex attractors such as chaotic attractors
can be encountered as well. The attraction basins and attractors change
when the parameters $\bg$ vary. These changes can be smooth (structural stability)
or sharp (typically this arises at bifurcations points).

%\paragraph{Remark} Note that fig. \ref{Fattractors} is a representation of the phase space where the asymptotic behaviour of \textit{all}
%possible initial conditions in $\cM$ are represented. When dealing with experiments, the situation is slightly
%different. One fixes \textit{one} initial condition and follows its evolution and the asymptotic behaviour that one observes
%depends on the way how the system is prepared. This is the point of view adopted in this paper. When dealing with empirical
%observations (and especially when dealing with empirical measures, section \ref{rastplotstat}) we select one trajectory and we observe its behaviour.
%This holds as well when dealing with statistical properties of this orbit (section \ref{SpikesStat}), and also when dealing with synaptic adaptation (section \ref{adapt})

\ssu{A generic example: generalized integrate and fire models}\label{Examples}

\sssu{Quiescent stage of the neuron.} Among the various models of neural dynamics, generalized integrate 
and fire models play a central role, due to their (relative) simplicity,
while the biological plausibility is well accepted \cite{rudolph-destexhe:06,gerstner-kistler:02}. A representative class of such  models is of form:
\beq\label{yvettenet}
\frac{dV_k}{dt}=-\frac{1}{\tau_L} \, (V_k - E_L) + i_k^{(ext)} - i_k^{(syn)}(V_k, t, \ltjn), 
\eeq
defining the evolution of the membrane potential $V_k$ of neuron $k$. Here $\tau_L = RC \simeq 10-20ms$ is the membrane time-constant related to the membrane resistance
and its electric capacity, while $E_L \simeq -80mV$ is the related reversal potential. The term $i_k^{(ext)}$ is an external ``current''\footnote{This is a slight abuse of
language since we have divided  eq. (\ref{yvettenet}) by the membrane capacity. } assumed
to be time constant in this paper. In the end of this section we shall however consider the case 
where some noise is superimposed upon the constant external current.
$\tjn$ is the $n$-th firing time\footnote{For continuous time systems, the firing times of neuron $k$, for the trajectory $\V$, is defined by:

%\beq\label{Tdech}
$$t_k^{(n)}(\V)=\inf\left\{t  \ | t > t_k^{(n-1)}(\V), \ V_k(t) \geq \theta  \right\},$$
%\eeq

\nid where $t_k^{(0)}=-\infty$.}
 of
neuron $j$ and $\ltjn$ is the list of firing times of all neurons up
to time $t$.

The synaptic currents reads:

$$i^{(syn)}_k(V_k,t,\ltjn)=(V_k-E^+)\, \sum_{j \in \cE} g_{kj}(t,\ltjn) + (V_k-E^-)\, \sum_{j \in \cI} g_{kj}(t,\ltjn),$$

\nid where $E^\pm$ are reversal potential (typically $E^+ \simeq 0 mV$ and $E^- \simeq -75 mV$).
$\cE$ and $\cI$ refers respectively to excitatory and inhibitory neurons and
the $+$ ($-$) sign is relative to excitatory (inhibitory) synapses. 
Note that conductances are always positive thus
the sign of the post-synaptic potential (PSP) is determined by the reversal potentials $E^\pm$.
At rest ($V_k \sim -70mV$) the $+$ term leads to a positive PSP while $-$ leads to a negative
PSP.

Conductances depend on  past spikes via the relation:

$$g_{kj}(t,\ltjn)=  G_{kj} \sum_{n=1}^{M_j(t,\V)} \alpha_j(t-\tjn).$$ 

\nid In this equation, $M_j(t,\V)$ is the number
of times neuron $j$ has fired at time $t$ and  $\ltjn$ is the list of firing times of all neurons up
to time $t$. $G_{kj}$ are positive constants, proportional to
 the \textit{synaptic efficacy}:

\beq\label{Wij}
\left\{
\baR{ccc}
 W_{kj}=E^+G_{kj} \quad &\mbox{if}&  \quad j \in \cE, \\
 W_{kj}=E^-G_{kj} \quad &\mbox{if}&  \quad j \in \cI. 
\eaR
\right. 
\eeq
 
We use the convention $W_{kj}=0$
if there is no synapse from $j$ to $k$.
Finally, $\alpha$ represents the unweighted shape (called a $\alpha$-shape) of the post-synaptic potentials.
Known examples  of $\alpha$-shapes are $\alpha(t)=K e^{-t/{\tau}} H(t)$
or $\alpha(t)=Kte^{-t/{\tau}} H(t)$, where $H$ is the Heaviside function.

Then, we may write (\ref{yvettenet}) in the form (see \cite{cessac-vieville:08} for more details):

$$\frac{dV_k}{dt}+g_k V_k=i_k,$$

\nid with:

%\beq\label{gklt}
$$g_k(t,\ltjn)=\frac{1}{\tau_L}+\sum_{j=1}^N g_{kj}(t,\ltjn),$$
%\eeq

\nid and:

%\beq\label{ik}
$$i_k(t,\ltjn)=\frac{E_L}{\tau_L} 
+  \, \sum_{j \in \cE} W_{kj}\sum_{n=1}^{M_j(t,\V)} \alpha_j(t-\tjn)
+  \, \sum_{j \in \cI} W_{kj}\sum_{n=1}^{M_j(t,\V)} \alpha_j(t-\tjn)
+i^{(ext)}_k.
$$
%\eeq 

\sssu{Fire regime.}
The previous equations hold whenever the neuron is quiescent, i.e. whenever membrane potential is smaller than a threshold $\theta>0$, usually depending on time
(to account for characteristics such as  refractory period of the neuron) and on the neuronal state. Here, however, we assume that $\theta$ is a constant
(in fact without loss of generality \cite{gerstner-kistler:02}).
When the membrane potential exceeds the threshold value, the neuron ``fires'' (emission of an action potential or ``spike''). The spike shape depends on the model.
In the present case, the membrane potential is reset instantaneously to a fixed value $\Vr \simeq E_L$, corresponding to the value of the membrane
potential when the neuron is at rest.
For simplicity we set $\Vr=0$ without loss of generality.
This modeling of the spike introduces a natural notion of spike time, but the price to pay is to introduce a discontinuity in dynamics.
Moreover, since spike is instantaneous the set of possible spike times occurring within time interval is uncountable in the continuous case. 
Introducing a minimal time discretisation at scale $\delta$ removes this pathology.

\sssu{Time discretisation.}

Assuming that spike times are only known within a precision $\delta>0$  \cite{cessac-vieville:08}
one shows that the dynamics of membrane potential, discretized at the time scale $\delta$ writes:

\beq \label{yvette}
V_k(t+\delta)= \rho_k(t,t+\delta,\tot) \, V_k(t)+\Jkto
\eeq

\nid where:

%\beq \label{yvette-2} 
$$
\begin{array}{rcl}
\Jkto &=& \int_{t}^{t+\delta} i_k(s,\tot) \, \rho_k(s,t+\delta,\tot) \, ds, \\
\rho_k(s,t+\delta,\tot) &=& e^{-\int_{s}^{t+\delta} \, g_k(s',\tot) \, ds'}, \\
g_k(t,\tot) &=& \frac{1}{\tau_L}+\sum_{j=1}^N G_{kj} \sum_{n=1}^{\mjt} \alpha^\pm(s-\tjn), \\
\end{array} 
$$
%\eeq

\nid where  $\mjt$ is the number of spikes emitted by neuron $j$ up to time $t$, in the raster plot $\tom$.
In the sequel we  assume that $\alpha^\pm(u)=0$ whenever $|u| \geq \tau_M$, i.e. $g_k(t,\tot)$ depends on past firing times over a finite
time horizon $\tau_M$. We also set $\delta=1$.

\sssu{Model I and model II.} A step further, two additional simplifications can be introduced:

 \ben

\item Assuming that $\rho_k(t,t+\delta,\tot) = \rho$ is almost constant, which is equivalent to 
considering ``current'' synapses instead of ``conductance'' synapses,
 i.e. neglect the dependency of $i_k^{(syn)}$ with respect to $V_k$;

 \item Considering a much simpler form for $\Jkto$, where each connected neuron simply increments the membrane potential during its firing state. This is equivalent to
 considering the post-synaptic profiles in the previous equations as instantaneous step-wise profiles.

\een

Considering these assumptions leads the following dynamics of membrane potential:
\beq\label{FiBMS}
\F_{\bg,i}(\V)=\rho \, V_i \left(1 - Z[V_i] \right)+ \sum_{j=1}^N W_{ij}Z[V_j]+ \Iei; \qquad i=1 \dots N,
\eeq
where $\F_{\bg,i}$ is the $i$-th component of $\Fg$, $Z(x)=1$ if $x\geq \theta$ and $0$ otherwise, both the integrate and firing regime being integrated in this equation.
It turns out that this corresponds exactly to the time discretisation of the standard integrate and fire neuron model, which as discussed 
in e.g.,\cite{izhikevich:03}
provides a rough but realistic approximation of biological neurons behaviors.
This model is called \textit{model I} in the sequel whereas the model based on~(\ref{yvette}) is called \textit{model II}.

\sssu{Genericity results for models I and II.}\label{SgenmodI-II}

The map $\Fg$ in models I and II is  locally contracting \cite{cessac:08,cessac-vieville:08},
but it is not smooth due to the sharp threshold  in neurons firing definition.
The \textit{singularity set} where $\Fg$ is not continuous is:
$$\cS=\left\{\V \in \cM | \exists i=1 \dots N, \mbox{\ such \ that} \
V_i=\theta \right\}.$$
This is the set of membrane potential vectors such that at least
one of the neurons has a membrane potential exactly equal to the threshold.
Because of this, dynamics exhibit sensitivity to perturbations
for orbits which approach too close the singularity set. 

Now, call

\beq \label{dOS}
\dOS=\inf_{\V \in \Omega} \inf_{t \geq 0} \min_{i=1 \dots N} |V_i(t)-\theta|.
\eeq

Then, the following theorem holds \cite{cessac:08,cessac-vieville:08}.

\bth\label{ThdAS}

For a generic (in a metric and topological sense) set of parameters $\bg$, $\dOS >0$. Consequently,

\ben 

\item $\Omega$ is composed of finitely many periodic orbits with a finite period,
Thus,  attractors are generically stable period orbits.

\item There is  a one-to-one correspondence between membrane potential trajectories
and raster plots, except  for a negligible set of $\cM$.

\item There is a finite Markov partition.
\een

\enth

\textbf{Remarks} 

\ben
\item Note however that, depending on parameters (synaptic weights, external current),  
 some
 periods can be quite large (\underline{well beyond  any accessible computational time})
Also, the number of stable periodic orbits can grow exponentially with the number of neurons.

\item  Result 2 means that spike trains provide a symbolic coding 
for the dynamics.

\item See section \ref{SGrammar} for the implications of result 3.
\een

These models constitute therefore nice examples where one has a good control on dynamics
and where one can apply the machinery of thermodynamic formalism
 (see Appendix). Note that
since dynamics is eventually periodic, one may figure out that there is nothing non trivial to
say from the respect of spike train statistics. The key point is that period are practically so
large that there is any hope to see the periodicity in real experiments. Thus, one may
do ``as if'' the system were ``chaotic''. This is a fortiori the case if one superimposes upon
the external current a small amount of noise to the external current $\Ie$.

Also, though dynamics of model I and II may look rather trivial compared to what is expected from
biological neural networks, it might be that such models are able to approximate  trajectories
of a real neural network, for suitable values of $\bg$ (e.g. after a suitable adaptation  
of the synaptic weights) and provided $N$, the number of neurons, is sufficiently large.
This type of properties are currently sought by computational neuroscientist with 
interesting indications that `IF models are good enough'' to approximate biological
neurons spike trains \cite{jolivet-et-al:06}. See \cite{rostro-cessac-etal:09b} for a recent illustration of this.

\ssu{Spikes dynamics and statistics.}%\label{SpikesStat}

\sssu{Grammar.} \label{SGrammar}

Though dynamics (\ref{DNN}) produces raster plots, 
it is important to remark that it is not able to produce \textit{any} possible sequence
of spiking patterns. This fundamental fact is most often neglected in computational neuroscience
literature and leads, e.g. to severe overestimation of the system's entropy. Also,
 it plays a central role in determining spike train statistics. 

There are therefore \textit{allowed} and \textit{forbidden} sequences 
depending on conditions involving the detailed form of $\Fg$ and on the parameters
$\bg$ (synaptic weights and external current).
Let $\bom,\bom'$ be two spiking patterns. The transition $\bom \to \bom'$ is \textit{legal}
or \textit{admissible}
if there exists a neural state $\X$ such that neurons
fire according to the firing pattern $\bom$ and, at the next time,
according to the firing pattern $\bom'$. Equivalently,
 $\Fg(\Po)  \cap \Pop \neq \emptyset$. An admissible transition  must satisfy
 \textit{compatibility conditions}
depending on synaptic weights and currents. 

As an example, for model I,
compatibility conditions write, for all $i =1 \dots N$: 
%
%\beq\label{Transitions}
$$
\baR{ccccc}
&(a) \quad & i \, \mbox{is such that} \, \omega_i=1 \, \mbox{and} \, \omega'_i=1 & \Leftrightarrow& \sum_{j =1}^N W_{ij}\omega_j +  \Iei \geq \theta\\
&(b) \quad & i \, \mbox{is such that} \, \omega_i=1 \, \mbox{and} \, \omega'_i=0 & \Leftrightarrow& \sum_{j =1}^N W_{ij}\omega_j +  \Iei < \theta\\
&(c) \quad & i \, \mbox{is such that} \, \omega_i=0 \, \mbox{and} \, \omega'_i=1 & \Leftrightarrow& \exists \X \in \Po \, \mbox{such that}, \,
\rho X_i +\sum_{j =1}^N W_{ij}\omega_j +  \Iei \geq \theta\\
&(d) \quad & i \, \mbox{is such that} \, \omega_i=0 \, \mbox{and} \, \omega'_i=0 & \Leftrightarrow& \exists \X  \in \Po \, \mbox{such that}, \,
\rho X_i +\sum_{j =1}^N W_{ij}\omega_j +  \Iei < \theta\\
\eaR
$$
%\eeq
%
\nid Note that, while the two first conditions only depend on the partition element
$\Po$, the two last ones depend on the point $\X \in \Po$.
Basically, this means that the natural partition is not a Markov
partition. The idea is then to find a refinement of the natural partition such that
allowed transitions depend only on the partition-elements.

 A  nice situation occurs when the refined partition is \textit{finite}.
 Equivalently, compatibility conditions can be
obtained by a \textit{finite} set of rules or a finite \textit{grammar}. This is possible if there exists some 
finite integer $r$ with the following property: Construct 
 blocks of spiking patterns $\bom(0) \dots \bom(r-1)$ of width $r$.
There are at most $2^{Nr}$ such possible blocks (and in general quite a bit less \textit{admissible} blocks). 
Label each of these blocks with a symbol $\alpha \in A$, where $A$ is called an \textit{alphabet}.
Each block corresponds to a connected domain $\cap_{s=0}^{r-1} \F^{-s}_{\bg}[\cP_{\bom(s)}] \subset \cM$
and to a cylinder set $\torm \subset \Spg$.
 Define a \textit{transition matrix} $\Gg: A \times A \to \left\{0,1\right\}$, depending on $\bg$, with entries $g_{\alpha\beta}$, such that
$g_{\alpha\beta}=1$ if the transition $\beta \to \alpha$ is admissible, and $0$ otherwise.
To alleviate the notations in the next equation write  $\alpha(t)=\totr$, $\forall t \geq 0$.
If 

%\beq\label{GramDef}
$$\Spg=\left\{\tom \ | \ g_{\alpha(t+1)\alpha(t)}=1, \forall t \geq 0  \right\},$$
%\eeq

\nid then all admissible raster plots are obtained via the finite transition matrix $\Gg$ which provides
the grammar of spiking patterns sequences. 

In models I, II  such a finite grammar exists for the (generic) set
of parameters $\bg$ such that $\dAS>0$. 
Moreover, there are open domains in the space of parameters
where the grammar is fixed and is not affected by small variations of synaptic weights
or current.
%\footnote{Call
%$\tin=\max \left\{1\leq k < n,\omega_i(k)=1\right\}$.
% On suppose que ce tin existe i.e. que le neurone i a decharge.
%In model I the raster plot $\tom$ is admissible iff
%$\forall n>0$,
%
%\beq
%\left(\sum_{j=1}^N W_{ij}\sum_{l=\tin+1}^n \rho^{n-l}\omega_j(l-1)+\Iei \frac{1-\gamma^{\tin}}{1-\rho}-\theta\right)
%\left(2\omega_i(n)-1 \right) > 0.
% On a enleve la condition initiale
%\eeq
%}
 
For more general models, it might be that 
there is no finite grammar to describe all admissible raster plots.
When dealing with experiments,
one has anyway only access to finite raster plots and
the grammar extracted from empirical observations has a finite number of rules (see \cite{collet-et-al:95,chazottes-etal:98,chazottes:99} 
for nice applications of this idea in the field of turbulence).
 This empirical
grammar is compatible with the dynamics at least for the time of observation.\\

The grammar depends on $\bg$. Assume now that we are continuously varying $\bg$. This corresponds to
following some path in the space of control parameters. Then, this path
generically crosses domains where grammar is fixed, while at the
boundary of these domains, there is a  grammar modification, and thus
a change in the set of admissible raster plots that dynamics is able to produce.
Synaptic adaptation, as described in section \ref{adapt}, corresponds precisely
to this situation.   
Thus, we expect that the neural network exhibits
``regular'' periods of adaptation where spike trains before and after synaptic changes,
correspond to the same grammar. Between these regular periods, sharp changes in
raster plots structure are expected, corresponding somehow
to displaying new transitions and new rules in the spike code.
Actually, the ability of neural networks to display, after adaptation, a \textit{specific} subset
of admissible sequences, provided by a specific grammar,
 is an important issue  of this paper.

\sssu{Spike responses of neurons.}

Neurons respond to excitations or stimuli by finite sequences of spikes.
In model I and II a stimulus is typically the external current but more general
forms (spike trains coming from external neurons) could be considered as well. 
Thus, the dynamical response $R$ of a neuronal network to a stimuli $S$ (which can be applied
to several neurons in the network), is a sequence $\bom(t) \dots \bom(t+n)$ of spiking
patterns. ``Reading the neural code'' means that one seeks a correspondence between
responses and stimuli. However, the spike response does not only depend on the stimulus, but also
on the network dynamics and therefore fluctuates randomly (note that this apparent randomness is provided by the 
dynamical evolution and does not require the invocation of an exogenous noise).
Thus, the spike response  is  sought as a conditional probability $P(R|S)$ \cite{rieke-etal:96}. Reading the code consists in
inferring $P(S|R)$ e.g. via Bayesian approaches, providing a loose dictionary where
the observation of a fixed spikes sequences $R$ does not provide a unique possible
stimulus, but a set of stimuli, with different probabilities. 
Having models
for conditional probabilities $P(R|S)$ is therefore of central importance.
For this, we need a good notion of statistics.

\sssu{Performing statistics.} 
These statistics
 can be obtained in two different ways. Either one repeats
a large number of experiments, submitting the system to the same stimulus $S$,
and computes $P(R|S)$ by an average over these experiments. This approach relies on the assumption
that the system has the same statistical properties during the whole set of experiments 
(i.e. the system has not evolved, adapted or undergone bifurcations meanwhile). 
In our setting this amounts to assuming that the parameters $\bg$ have not changed.

Or, one performs a time average. For example, to compute $P(R|S)$, one counts the number of times $n(R,T,\tom)$ when the finite sequence of
spiking patterns $R$,
appears in a spike train $\tom$ of length $T$, when the network is submitted to a stimulus $S$.
 Then, the probability $P(R|S)$ is estimated
by:
%
%\beq\label{PRStime}
$$P(R|S) = \lim_{T \to \infty} \frac{n(R,T,\tom)}{T}.$$
%\eeq
%
This approach implicitly assumes that the system is in a stationary
state. 
  
The empirical approach is often ``in-between''. One fixes a time window of length
$T$ to compute the time average and then performs an average over a finite number $\cN$
of experiments corresponding to selecting different initial conditions. 
In any case the implicit assumptions are  essentially impossible to control in real (biological)
experiments, and difficult to prove in models. 
So,  they are basically used as ``working'' assumptions.\\

To summarize, one observes, from $\cN$ repetitions of the same experiment,
 $\cN$ raster plots $\tom_m, m=1 \dots \cN$ on a finite time horizon 
of length $T$.  From this, one computes experimental averages allowing to estimate
$P(R|S)$ or, more generally, to estimate the average value, $\langle \phi \rangle$, of some prescribed
observable $\phi(\tom)$.  These averages are estimated by : 
\beq\label{empav}
\bar{\phi}^{(\cN,T)}=\frac{1}{\cN T}\sum_{m=1}^{\cN} \sum_{t=1}^T \phi(\stg \tom_m).
\eeq
Typical examples of such observables are $\phi(\tom)=\omega_i(0)$ in which case 
$\langle \phi \rangle$ is the firing rate of neuron $i$;  
$\phi(\tom)=\omega_i(0)\omega_j(0)$ then $\langle \phi \rangle$ measures the probability of spike coincidence
for neuron $j$ and $i$; $\phi(\tom)=\omega_i(\tau)\omega_j(0)$ then $\langle \phi \rangle$ measures the probability of 
the event ``neuron $j$ fires and neuron $i$ fires $\tau$ time step later'' (or sooner according to the sign
of $\tau$). In the same way  $P(R|S)$ is the average 
of the indicatrix function $\chi_R(\tom)=1$ if $\omega \in R$ and $0$ otherwise.
Note that in (\ref{empav}) we have used the shift $\stg$ for the time evolution of the raster plot. This notation is
more compact and more adapted to the next developments than the
classical formula, reading, e.g., for firing rates 
$\frac{1}{\cN T}\sum_{m=1}^{\cN} \sum_{t=1}^T \phi(\bom_m(t))$.

This estimation depends on $T$ and $\cN$. However, one expects that,
as $\cN,T \to \infty$, the empirical average $\bar{\phi}^{(\cN,T)} $ 
converges to the theoretical average $ \langle \phi \rangle$, as stated e.g. from the law of large numbers.
Unfortunately, one usually does not have access to these limits, and one is lead to extrapolate
theoretical averages from empirical estimations. The main difficulty is that 
these observed raster plots are produced by an underlying dynamics
which is usually not explicitly known (as it is the case in experiments) or impossible to fully characterize 
(as it is the case in most large dimensional neural networks models). Even for models I and II, where one has
a full characterization of generic orbits, a numerical generation of raster plots
can produce periodic orbits whose period is out of reach. 
Thus, one is constrained to propose ad hoc statistical models.
This amounts to assuming that the underlying dynamics satisfies specific constraints. 

\sssu{Inferring statistics from an hidden dynamics.}\label{InfStat}

We follow this track, proposing a generic method to construct statistical
models from a prescribed set of observables. For this, we shall assume that
the observed raster plots are generated by a \textit{uniformly hyperbolic} dynamical system,
where spike trains constitute a symbolic coding for dynamics
(see the appendix for more details). This a technical assumption
allowing us to develop the thermodynamic formalism on a safe mathematical ground.
Basically, we assume that the observed raster plots can be generated by 
a dynamical system which is chaotic with exponential correlation decay, and that a finite
Markov partition exists, obtained by a refinement of the natural partition.

This is obviously a questionable assumption, but let us give a few arguments in its favor.
First, it has been argued and numerically checked in \cite{rostro-cessac-etal:09b} that  finite
sequence of spiking patterns, produced by an hidden neural network, including data coming from biological experiments, can be
exactly reproduced by a gIF model of type II, by adding hidden neurons and noise to the dynamics.
This suggests that gIF models are somehow ``dense'' in the space of spiking neural networks dynamical systems,
meaning that any finite piece of  trajectory from a spiking neural network can be approached by a gIF model
trajectory. Second, adding noise to model I, II makes them uniformly hyperbolic\footnote{Contraction
is uniform in model I. In model II it can be bounded by a constant strictly lower that $1$.
Introducing noise amounts to adding, for each neuron, an unstable direction
where randomness is generated by a chaotic one-dimensional system. One forms in this way a skew product
where the unstable fiber is the random generator and the stable one corresponds to
the contracting dynamics of membrane potentials (eq. (\ref{yvette}), or (\ref{FiBMS}))
Because of the discontinuity of the map, the main difficulty
is to show that, in this extended system, generic points have a local stable manifold of sufficiently large
diameter (Cessac \& Fernandez, in preparation).
See \cite{blanchard-cessac-etal:00,cessac-blanchard-etal:04} for an application of this strategy
in the framework of Self-Organized Criticality.
 \label{fnuh}}, though not continuous.
 So, we expect the theory developed below to apply 
to model I, II. Now, dealing with spike train coming from a neural network whose mathematical properties
are unknown (this is especially the case with biological neural networks), and whose statistical properties
are sought, the idea is to do as if these data were generated by a dynamical system like model I or II. 

As a matter of fact, the choice of a statistical model always relies on assumptions. Here we make an attempt to formulate
these assumptions in a compact way with the widest range of application.
These assumptions are compatible
with the statistical models commonly used in the literature like Poisson models or  Ising like models \`a la Schneidman and collaborators\cite{schneidman-etal:06}, but 
lead also us to propose more general forms of statistics. Moreover, our approach
incorporates additional elements such as the consideration of neurons dynamics,
and the grammar. This last issue is, according to us, fundamental, and, to the best
of our knowledge, has never been considered before in this field. Finally,
this postulate allows us to propose algorithms that can be applied to real data
with an posteriori check of the initial hypotheses \cite{cessac-vasquez-etal:09}.\\ 
 
On this basis we propose the following definition.
 Fix a set $\phi_l$, $l=1 \dots K$,
of observables, i.e. functions $\Spg \to \bbbr$ which associate real numbers to sequences of spiking
patterns. Assume that the empirical average (\ref{empav}) of these functions has been computed, 
for a finite $T$ and $\cN$, and
that $\bar{\phi_l}^{(T,\cN)}=C_l$.

A \textit{statistical model} is a probability distribution $\nu$ on the set of raster plots such
that:

\ben
\item $\nu(\Spg)=1$, i.e. the set of non admissible raster plots has a  zero $\nu$-probability.
\item $\nu$ is ergodic for the left-shift $\sg$ (see section \ref{Serg} in the appendix for a definition).
\item For all $l=1 \dots K$, $\nu(\phi_l)=C_l$, i.e., $\nu$ is compatible with the empirical averages.
\een

Note that item 2 amounts to assuming that statistics are invariant under time translation.
On practical grounds, this hypothesis can be relaxed  using sliding time windows.
This issue is discussed in more details in  \cite{cessac-vasquez-etal:09}.
Note also that $\nu$ depends on the parameters $\bg$.

Assuming that $\nu$ is ergodic has the advantage that one does not have to average \textit{both}
over experiments \textit{and} time. It is sufficient to focus on time average for
a single raster plot, via the time-empirical average:

$$\pi_{\omega}^{(T)}(\phi) = \frac{1}{T} \sum_{t=1}^T \phi(\stg \omega)$$

\nid defined within more details in the appendix,
section \ref{rastplotstat}.\\

Item 3 can be generalized as follows. 

\ben
\item[3'.] $d(\pTo,\nu) \to 0 $, as $T \to \infty$,
\een

\nid where $d(\mu,\nu)$
is the relative entropy or Kullback-Leibler divergence between two
measures $\mu,\nu$ (see eq. (\ref{HKL}) in the appendix.).

Dealing with real or numerical data, it is obviously not possible to extrapolate to $T \to \infty$,
but the main idea here is to minimize the ``distance''  $d(\pTo,\nu)$ between the empirical
measure, coming from the data, and the statistical model. Especially, if several models can be proposed
then, the ``best'' one minimizes the Kullback-Leibler divergence (see th. \ref{ThKL} in the Appendix). The main advantage is that
 $d(\pTo,\nu)$ can be numerically estimated using the thermodynamic formalism.
Note however that this  approach may fail at phase transition points where $d(\mu,\nu)=0$ does
not necessarily imply $\mu=\nu$ \cite{chazottes:99}. Phase
transition can be numerically detected from empirical data \cite{comets:97}.
 
\sssu{Gibbs measures as statistical models.} \label{Statmod}

\st{Variational principle.} Statistical physics naturally proposes a canonical way to construct a statistical model:
``Maximizing the entropy under the constraints $\nu(\phi_l)=C_l$, $l=1 \dots K$'', (see \cite{jaynes:57} for
a beautiful and deep presentation of what became, since then, a ``folklore'' result).
In the context of  thermodynamic formalism  this amounts
to solving the following variational principle (see eq. (\ref{supmu}) in the Appendix).
$$\pres=\sup_{\nu \in m^{(inv)}} (h\left[\nu\right]+\nu\left[\bpsi\right]),$$
%\eeq
%
where $m^{(inv)}$ is the set of invariant measures for $\sigma$ and $h$ the Kolomogorov-Sinai entropy or entropy rate
(see Appendix for more details). The ``potential'' $\bpsi$ is given by $\bpsi=\sum_{l=1}^K \lambda_l \phi_l$ where
the $\lambda_l$'s are adjustable Lagrange multipliers.
 A measure $\mpg$ which realizes the supremum, i.e.
$$\pres=h\left[\mpg\right]+\mpg\left[\bpsi\right],$$
is called, in this context, an ``equilibrium state''. The function $\pres$ is called the ``topological pressure''.\\

\st{Gibbs states.} Uniformly hyperbolic dynamical systems have equilibrium states. Moreover, in this case, equilibrium states are \textit{Gibbs states}
(and vice-versa). A Gibbs state, or Gibbs measure, is a probability measure such that,
one can find some constants $c_1,c_2$ with $0 < c_1 \leq 1 \leq c_2$ such
that for all $n \geq 1$ and for all $\tom \in \Sigma$:
$$c_1 \leq \frac{\mpg\left(\tom  \in \Con\right)}{ \exp(-n\pres+\Snpo)} \leq c_2,$$
(Cf eq. (\ref{dGibbs}) in the Appendix), 
where $\Snpo = \sum_{t=0}^{n-1} \bpsi(\stg \tom)$. Basically, this means that the probability
that a raster plot  starts with the bloc $\Con$ behaves like $\frac{\exp(\Snpo)}{Z_n}$.
One recognizes the classical Gibbs form where 
space translation in lattice system is replaced by  time translation
(shift $\stg$) and where the normalization factor $Z_n$ is the partition function.
Note that $\pres=\limsup_{n \to \infty}\frac{1}{n}\log Z_n$, so that $\pres$ is the formal analog
of a thermodynamic potential (like free energy).\\

\st{Topological pressure.} The topological pressure is a convex function. Moreover, this is
the generating function for the cumulants of the Gibbs distribution.
Especially,  the Lagrange multipliers $\lambda_l$ can be tuned to a value $\lambda_l^\ast$
such that  $\mpg\left[\phi_l\right]=C_l$,
(item 3) using:

\beq\label{Gener}
\left. \frac{\partial \pres}{\partial  \lambda_l}\right|_{\lambda_l=\lambda_l^\ast}=C_l.
\eeq
It expresses that the Gibbs state is given the tangent of the pressure 
at $\lambda_l^\ast$ \cite{keller:98}. Note that we have here assumed that the topological pressure
is differentiable, namely that we are away from a  phase transition. Higher order moments 
are obtained in the same way, especially second order moments related to the central limit
theorem obeyed by Gibbs distributions \cite{bowen:75,bowen:98,keller:98}. 
It is also possible to obtain averages of more complex functions than moments \cite{ji:89}.
The topological pressure is obtained via the spectrum of the Ruelle-Perron-Frobenius operator
and can be calculated numerically when $\bpsi$ has a finite (and small enough) range 
(see appendix for more details).\\

\st{Validating a Gibbs statistical model.}
The Kullback-Leibler divergence provides some notion of ``distance'' between two measures.
For $\mu$ an invariant measure and $\mpg$ a Gibbs measure with a potential
$\bpsi$, both defined on the same set of sequences  $\Sigma$, one has 
(see eq. (\ref{dKLpres}) in the appendix):  
$$d\left(\mu,\mpg \right) = \pres - \int \bpsi d\mu - h(\mu).$$
Though $\pTo$ is  not invariant, this
 suggests to use this relation to compare 
different statistical models (i.e. when several choices of observables
$\phi_l$ are possible) by choosing the one which minimizes the
quantity:

\beq\label{KLemp}
d(\pTo,\mpg)=\pres-\pTo(\bpsi) -h(\pTo).
\eeq
(see th. \ref{ThKL} in the appendix).
The advantage is that this quantity can be numerically estimated \cite{cessac-vasquez-etal:09}.\\

\st{Probability of spiking patterns blocs.} In this context, the probability of a spiking pattern block 
$R=\left[\bom \right]_{0,n-1}$ of length $n$ corresponding
to the response $R$ to a stimuli $S$ ``behaves like'' (in the sense of eq. (\ref{dGibbs})):

\beq\label{P(R)}
P\left[R|S \right]=\nu\left[\omega \in R | S \right] \sim \frac{1}{Z_n\left[\bl^\ast(S)\right]}
\exp\left[
\sum_{l=1}^K \lambda^\ast_l(S) \sum_{t=0}^{n-1} \phi_l(\stg \tom)
\right],
\eeq
where the conditioning is  made explicit in
the dependence of $\bl^\ast$ in the stimulus $S$. (Typically, referring to model I and II, $S$ is an external current and is incorporated in $\bg$.)
This equation holds as well when $S=0$ (no stimulus). Obviously, for two different stimuli the probability $P(R|S)$ may drastically change.
 Indeed, different stimuli imply different parameters $\bg$ thus a different dynamics and different
raster plots, with different statistical weights on a specific bloc $R$. The grammar itself can also change, in which case $R$ may be forbidden
for a stimulus and allowed for another one.

\ssu{Examples.}

\sssu{Canonical examples.}

\st{Firing rates.} If $\phi_l(\tom)=\omega_l(0)$, then $\pTo(\phi_l)=r_l$ is the average firing rate of neuron $l$ within
the time period $T$. Then, the corresponding statistical model is a Bernoulli distribution where neuron $l$ has a probability
$r_l$ to fire at a given time. The probability that neuron $l$ fires $k$ times within a time delay $n$ is a binomial distribution 
and the inter-spike interval is Poisson distributed \cite{gerstner-kistler:02}.\\

\st{Spikes coincidence.} If $\phi_l(\tom)\equiv \phi_{(i,j)}(\tom)=\omega_i(0) \, \omega_j(0)$ where, here,
 the index $l$ is an enumeration for all (non-ordered) pairs
$(i,j)$, then the corresponding statistical models has the form of an Ising model, as discussed by Schneidman and collaborators in 
\cite{schneidman-etal:06,tkacik-etal:06}. As shown by these authors in experiments on the salamander retina, the probability of spike
blocs estimated from the ``Ising'' statistical model fits quite better to empirical date than the classical Poisson model.  \\

\st{Enlarged spikes coincidence.} As a generalization one may consider the probability of co-occurrence of spikes from neuron $i$ and $j$ within
some time interval $\tau$. The corresponding functions are $\phi_l(\tom)=\omega_i(0)\omega_j(\tau)$ and the probability
of a spike bloc $R$ writes:

$$
P\left[R|S \right] = \frac{1}{Z_n\left[\bl^\ast(S)\right]} 
\exp\left[
\sum_{i,j} \lambda^\ast_{i,j}(S) \sum_{t=0}^{n-1} \omega_i(t) \, \omega_j(t+\tau)
\right].
$$

\nid Further generalizations are considered below.

\ssu{Conclusion.}
The main conclusion of this section is that Gibbs measures constitute
\textit{optimal} (in the sense of entropy maximization) statistical
models whenever a set of observable has been prescribed. It also allows
to select a model between several choices by minimizing the Kullback-Leibler
divergence between the empirical measure and the statistical models.

In all the examples presented above the statistical model is determined by
the choice of an \textit{a priori} form for the potential.
In the next section, we  explicitly compute the potential
in neural networks with synaptic adaptation.

\su{Synaptic plasticity.} \label{adapt}

\ssu{Synapses update as an integration over spikes trains.} 

Synaptic plasticity corresponds to the 
evolution of  synaptic
efficacy (synaptic weights). More precisely, in our notations (eq. (\ref{Wij})), $W_{ij}$ 
essentially provides the maximal
amplitude of the post-synaptic potential induced, at the synapse
linking $j$ to $i$, when neuron $j$ fires a spike. Synaptic
weights evolve in time according to the spikes emitted 
by the pre- and post- synaptic neuron. 
In other words,
the variation of $W_{ij}$ at time $t$ is a function
of the spiking sequences of neurons $i$ and $j$ from time $t-T_s$ to
time $t$, where $T_s$ is time scale characterizing the width of the spike
trains influencing the synaptic change. 
In most examples the  synapse update writes:
\beq\label{DSyn}
\dWij(t+1)=g\left(\Wij(t),\omeit,\omejt \right), \ t > T_s,
\eeq
with $\omeit=[\omega_i(t-T_s) \dots \omega_i(t)]$. Thus, typically,
synaptic adaptation results from an integration of spikes over the time scale $T_s$. 

\ssu{Spikes responses of synapses.} \label{spikerep}

The synaptic variation $\dWij$ is the integrated response of the synapse from neuron $j$ to neuron $i$
when neuron $j$ sends a spike sequence $\omejt$ and neuron $i$ fires according to $\omeit$. This
response is not a deterministic function, while $g$ is. Thus (\ref{DSyn}) is an approximation.
As a matter of fact, the explicit form of $g$ 
is usually derived from phenomenological considerations
as well as experimental results where synaptic changes can be induced by \textit{specific} simulations conditions,
defined through  the firing frequency of pre- and post-synaptic
neurons \cite{bliss-gardner:73,dudek-bear:93}, the membrane potential of the post-synaptic
neuron \cite{artola-etal:90}, or spike timing \cite{levy-stewart:83,markram-etal:97,bi-poo:01}
(see \cite{malenka-nicoll:99} for a review). 
 Thus, these results are usually based
on a repetition of experiments involving the excitation
of pre- and post-synaptic neurons by specific spike trains. The phenomenological
plasticity rules derived from these experiments are therefore 
of \textit{statistical} nature. Namely, they do not tell us what will be the exact changes
induced on synapses when this or this spike train is applied
to pre- and post-synaptic neuron. Instead, they provide us the \textit{average} synaptic change. 
 Thus, the function
$g(W_{ij},\omeit,\omejt)$ in (\ref{DSyn}) is typically a statistical average of the synaptic response
when the spike train of neuron $j$ (resp. $i$)
is $\omejt$ (resp. $\omeit$), and the actual synaptic weight value is $W_{ij}$.

In this paper we investigate the effects of those synaptic plasticity rules when the characteristic
time scale $T_s$ is quite a bit larger than the time scale of evolution of the neurons. Namely,
we consider \emp{slow} adaptation rules. 
In this situation  the synaptic weights update can be written as a function of the 
empirical average  (\ref{pTo}).
We therefore consider adaptation rules of form:

\beq\label{gcondensed}
g\left(W_{ij},\omeit,\omejt \right)=\epsilon\pTo\left[ \phi_{ij}(\Wij,.) \right].
\eeq

\nid where $\epsilon$ is a  parameter that will be typically small.
$\phi_{ij}(\Wij,\tom)$ is a function that we now make explicit.\\

Following   \cite{gerstner-kistler:02} canonical form of adaptation rules
use an expansion in singlet, pairs, triplets etc of spikes. Formally, one may write
a generic form for these rules.  Fix $T_s<\infty$ a positive integer. Let $\cL$ be the finite set of ordered lists $L$ in $\left\{-T_s,-T_s+1, \dots, 0\right\}$
with the form $L= (t_1,t_2, \dots, t_L)$ with $-T_s \leq t_1 < t_2 < \dots < t_L \leq 0$. For $L \in \cL$, $1 \leq i \leq N$,
we call an \textit{$i$-monomial }a function $m_{i,L}(\tom)=\omega_i(t_1) \dots \omega_i(t_L)$. We call an \textit{$(i,j)$
polynomial} a function:

\beq\label{phiij}
\phi_{ij}(\Wij,\tom)=\sum_{L_1,L_2 \in \cL} h_{ijL_1L_2}(\Wij)  \,  m_{i,L_1}(\tom) \, m_{j,L_2}(\tom),
\eeq

\nid where $h_{ijL_1L_2}(\Wij)$ are smooth functions of $\Wij$.  They 
can be  constant, linear functions of $\Wij$ (``multiplicative'' rules)
or  nonlinear functions  allowing for example to constrain $\Wij$ within bounded values (e.g. 
``hard bound'' or ``soft bounds'' rules). The form (\ref{phiij}) is the most general form of synaptic adaptation rules considered
in this paper, while $g$ has the explicit form:

\begin{multline} \label{gexpl}
g\left(W_{ij},\omeit,\omejt \right)=\\
\epsilon\sum_{k,l=0}^K
\sum_{
\tiny{
\begin{array}{ccc}
-T_s \leq t_1 < \dots < t_k \leq 0& \\
-T_s \leq s_1 < \dots < s_l \leq 0&
\end{array}
}
} 
h_{ij;t_1,\dots, t_k, s_1, \dots, s_l}(W_{ij})  \, 
\pTo\left[\omega_i(t+t_1) \dots \omega_i(t+t_k) \, 
\omega_j(t+s_1) \dots \omega_j(t+s_l) \right].
\end{multline}

Though explicit, this formulation is heavy, and we use instead the form (\ref{gcondensed}), (\ref{phiij}).

\ssu{Examples of adaptation rule instantiation.}

\st{Hebbian learning}
uses firing rates. A typical example corresponds to 
$g_{ij}(W_{ij},\left[\omega_i\right]_{t-T_s,t},\left[\omega_j\right]_{t-T_s,t})=\epsilon \frac{1}{T_s}\sum_{s_1,s_2=t-T_s}^{t}(\omei(s_1)-r_i(s_1))(\omej(s_2)-r_j(s_2)$ 
(correlation rule) where $r_i(t)=\frac{1}{T_s}\sum_{s=t-T_s}^{t} \omega_i(s)$ is  the frequency
rate of neuron $i$ in the raster plot $\tom$, computed in the time windows $[t-T_s,t]$. \\

\st{Spike-Time Dependent Plasticity} as derived from Bi and Poo \cite{bi-poo:01}
provides the average amount of synaptic variation given 
the delay between the pre- and post-synaptic spike.
Thus, ``classical'' STDP writes \cite{gerstner-kistler:02,izhikevich-desai:03}:
\beq\label{STDPFC}
g\left(W_{ij},\omeit,\omejt \right)=
\frac{\epsilon}{T_s}\sum_{s_1,s_2=t-T_s}^{t} f(s_1-s_2) \, \omega_i(s_1) \, \omega_j(s_2)
\eeq
\nid with:
\beq\label{fSTDP}
f(x)=
\left\{
\baR{llll}
A_- e^{\frac{x}{\tau_-}}, \ &x <0, \quad A_- < 0;\\
A_+ e^{-\frac{x}{\tau_+}},\ &x >0, \quad A_+ > 0;\\
0, \ &x=0;
\eaR
\right.
\eeq 
\nid where the shape of $f$  has been obtained from statistical extrapolations of experimental data.
 Hence STDP is based on a second order statistics (spikes correlations).
There is, in this case, an evident time scale $T_s=\max(\tau_-,\tau_+)$, beyond which
$f$ is essentially zero. \\

\st{``Nearest neighbors'' STDP} (according to the terminology of\cite{izhikevich-desai:03}) writes:
\beq
g\left(W_{ij},\omeit,\omejt \right)=\frac{\epsilon}{T_s}\sum_{s=t-T_s}^t f(\tau_j(s)-s) \, \omega_i(\tau_j(s)) \, \omega_j(s),
\eeq
\nid with $\tau_j(s)=\min_{t, \omega_j(t)=1} |t-s|$. \\

\st{Generalized STDP} As a last example, \cite{gerstner-kistler:02}
propose a rule which corresponds to: 
\beq
g(W_{ij},\omeit,\omejt)= \epsilon\left[
a_1^{pre} \sum_{s=t-T_s}^t \omega_j(s)+  
a_1^{post} \sum_{s=t-T_s}^t \omega_i(s) +  
\sum_{s_1,s_2=t-T_s}^t f(s_1-s_2)\omega_i(s_1)\omega_j(s_2)\right].
\eeq

We capture thus the main standard synaptic adaptation rules with (\ref{phiij}).

\ssu{Coupled dynamics.}

We consider now the following coupled dynamics.
Neurons are evolving according to (\ref{DNN}). 
We focus here on \textit{slow} synapses dynamics.  
Namely, synaptic weights are  constant for $T \geq T_s$  consecutive dynamics steps, where $T$ is large. This defines an
``adaptation epoch''. At the end of the adaptation epoch,  synaptic weights are updated according to (\ref{DSyn}). 
 This  has the consequence
of modifying neurons dynamics and possibly spike trains. The weights are then updated and a new adaptation
epoch begins. We denote by $t$ the update index of neuron states (neuron
dynamics) inside an adaptation epoch, while $\tau$ indicates the update index
of synaptic weights (synaptic plasticity). Call $\X^{(\tau)}(t)$
the state of the neurons  at time $t$ within the adaptation
epoch $\tau$.  Let $W_{ij}^{(\tau)}$ be the synaptic weights 
from neuron $j$ to neuron at $i$ in the $\tau$-th adaptation epoch. 
At the end of each adaptation epoch, the neuron dynamics time indexes are reset, i.e.
$x_i^{(\tau+1)}(0)=x_i^{(\tau)}(T), i=1 \dots N$.
The coupled dynamics writes:

\beq\label{Dcoupled}
\left\{
\baR{ccc}
\XTtp&=&\FgT(\XTt) \\
\dWijT &\deq& \WijTp-\WijT=g\left(\WijT,\omeit,\omejt)\right)
\eaR
\right.
\eeq

Recall that $\bg=(\cW,\Ie)$ (see section \ref{NeurStat}) and $\gT$ is the set of parameters at adaptation epoch
$\tau$. In the present setting the external current
$\Ie$ is kept fixed and only synaptic weights are evolving. 
Basically, $\Ie$ is used as an \textit{external stimulus}.

\ssu{Statistical effects of synaptic plasticity.}

Synaptic plasticity has several prominent effects.
First, modifying the synaptic weights has an action on the dynamics resulting, either 
in smooth changes, or in sharp changes. In the first case, corresponding
to parameters variations in a domain where the dynamical system is structurally stable,
small variations of the synaptic weights induce smooth variations on the $\omega$-limit
set structure (e.g. points are slightly moved) and in the statistics of orbits. The grammar
is not changed. On the opposite, when crossing bifurcations manifolds, the slightest change
in one synaptic weight value results typically in a drastic reorganization of the $\omega$-limit set
structure where
 attractors and their attraction basin are modified (see Fig. \ref{Fattractors} 
and  \cite{dauce-etal:98,siri-etal:07} for an illustration
of this in the case of frequency rates neural networks with Hebbian learning). 
But this can also change  the grammar and the set of admissible raster plots.
Some forbidden transitions become allowed, some allowed transitions
become forbidden.  Finally, the spikes train statistics are also modified. Typically, the
time-empirical average of the raster plot changes ($\pToT \to \pTpoT$) and
 the corresponding statistical model also evolves.

%Though the structure of bifurcations manifolds in general neural networks models
%such as Hodkin-Huxley's are out of reach at the current state of the art, gIF models,
%and especially model I,II, provide a tractable example where the bifurcations
%manifold can be  characterized. This is the set of parameters such that $\dOS$ (eq. (\ref{dOS}, section
%\ref{SgenmodI-II}) vanishes, and where theorem \ref{ThdAS} does not apply.
%This non generic set is the boundary of open domains
%inside  which  grammar is fixed and, where, on the basis
%of the discussion in section \ref{InfStat}, we expect the statistics of spike trains, provided by a Gibbs distribution,
%to vary smoothly (i.e. the potential is varying smoothly) as well as the pressure.
%On the opposite, sharp changes in the statistics occur when crossing the boundaries of
%these domains. \\

But synaptic adaptation is not a mere variation of  synaptic weights.
Indeed, synaptic weights are associated with the coupled dynamics (\ref{Dcoupled}),
meaning that synaptic changes depends on neurons dynamics, itself depending
on parameters. Thus, a synaptic adaptation process corresponds to following
a path in the space of parameters $\bg$; this path is not determined
a priori but evolves according to neurons dynamics. 
 Though this path can belong to a unique domain
of structural stability, this is not the case in general. At some point
during the adaptation, this path crosses a bifurcation manifold
inducing sharp changes in the dynamics and in the grammar\footnote{There is here an analogy with phase transitions in statistical physics  \cite{beck-schloegl:95}.
Detecting and characterizing these phase transitions  from empirical data is possible provided that there are only a few control
parameters
 \cite{comets:97}.}. Hence, after several changes of this type one can end
up with a system displaying raster plots with a structure rather different from 
the initial situation.  
 These changes depend obviously upon the detailed form 
of neuron dynamics (\ref{DNN}) and upon the synaptic update mechanism (\ref{DSyn}); they
are also conditioned by parameters such as stimuli.
We now Analise these effects in the light of thermodynamic formalism and Gibbs distributions.

\sssu{Static synaptic weights.}\label{static}

Let us first consider the situation where $\dW=0$, corresponding to synaptic weights that do not evolve.
Typically, this is the case if synaptic weights matrix converges to an asymptotic value $\cW^{\ast}$.
From eq. (\ref{gcondensed}) this corresponds to :
\beq\label{condstat}
\pTo\left[\phi_{ij} \right]=0,  \ \forall i,j \in \left\{1, \dots N \right\}.
\eeq
This imposes  a condition on the average value of $\phi_{ij}$. Therefore, from section
\ref{Statmod}
 \textit{this imposes  that the statistical model is a Gibbs measure $\nu$ with a  potential of form}:

\beq\label{psis}
\bpsis=\Phi+\bl^\ast.\bphi,
\eeq 

\nid where $\bpsis=\left(\psi^\ast_{ij} \right)_{i,j=1}^N$, $\bphi=\left(\phi_{ij} \right)_{i,j=1}^N$, $\bl^\ast=\left(\lambda^\ast_{ij} \right)_{i,j=1}^N$ 
and $\bl^\ast.\bphi=\sum_{i,j=1}^N \lambda^\ast_{ij}\phi_{ij} $. The potential $\Phi$ in (\ref{psis}) is such that  $\Phi(\tom)=0$ if $\tom$ is admissible and
$\Phi(\tom)=-\infty$ if it is forbidden, so that forbidden raster plots have zero probability.
This is a way to include the grammar in the potential (see appendix). 
 The statistical parameters
$\lambda_{ij}^\ast$, are given by eq. (\ref{Gener}) in section \ref{Statmod}, and, making $\phi_{ij}$ explicit (eq. (\ref{phiij})):

\beq\label{Generstat}
\left. 
\frac{\partial P\left[\psi\right]}{\partial  \lambda_{ij}}\right|_{\bl= \bls}
=\mpgs(\phi_{ij})=\sum_{L_1,L_2 \in \cL} h_{ijL_1L_2}(\Wij^\ast)\mpgs\left[ m_{i,L_1}m_{j,L_2}\right]=0.
\eeq

Since 
$ m_{i,L_1}m_{j,L_2}$ are monomials, this equation thus imposes \textit{constraints} on the probability 

$$\mpgs\left[\omega_i(t_1) \dots \omega_i(t_{L_1})\omega_j(s_1) \dots \omega_j(s_{L_2})\right],$$
of spikes $n$-uplets $\omega_i(t_1) \dots \omega_i(t_L)\omega_j(s_1) \dots \omega_j(s_L)$.
Note that condition (\ref{Generstat}) corresponds to an \textit{extremum} for the topological pressure as a function
of $\bl$. 
Note also that the variational formulation of Gibbs (equilibrium) state writes, in this case 
$P\left[\bpsis \right]=
\sup_{\nu \in m^{(inv)}} h\left[\nu\right]$ since $\nu(\psi)=0$. Hence, the corresponding Gibbs measure has \textit{maximal entropy}. 
\\

Let us emphasize what we have obtained. The statistical model that fits with the condition (\ref{condstat})
 in section \ref{Statmod} is a Gibbs distribution such that the probability
of a spin block $R$ of length $n$ is given by :

%\beq
$$P\left[R|S \right] = \frac{1}{Z_n\left[\bl^\ast(S)\right]}
\exp\left[
  \sum_{t=1}^n \bpsi^\ast(\stg \tom)
\right].
$$
%\eeq

\nid where $\bpsi^\ast$ depends on $S$ via the statistical parameters $\bl^\ast$ and via $\Phi$ (grammar).

When the situation $\dW=0$ corresponds to the asymptotic state for a synaptic adaptation process,
this potential provides us the form of the statistical model \textit{after adaptation}, and \textit{integrates all past 
changes in the synaptic weights}. We now discuss this process within details.

\ssu{Variational formulation of synaptic plasticity.}%\label{Varplast}

\sssu{Synaptic weights update and related potentials.} 
Let us now formalize the coupled evolution (\ref{Dcoupled}) in the  context
of thermodynamic formalism. 
The main idea is to  make the assumption that at each adaptation step,
 $\pToT$ can be approximated by a Gibbs
measure $\npT$ with potential $\psiT$ 
and topological pressure $\PT$. In this case, when $T$ is large, synaptic adaptation writes :
\beq \label{dWij}
\dWijT  = \epsilon\npT\left[\phi_{ij}(\WijT,.) \right].
\eeq

The synaptic update results in a change of parameters $\bg$, $\gTp=\gT+\delta \gT$.
 This induces a variation of the potential
$\psiTp=\psiT+\delta\psiT$ and of the pressure $\PTp=\PT+\dPT$.
We now distinguish two situations both arising when synaptic weights
evolve.

\sssu{Smooth variations.} \label{smoothvar}
 Let us first assume that these variations are smooth,
%.  For models I, II this means that 
i.e. one stays 
inside a domain where dynamics is structurally stable and  the grammar $\GT$ is not modified.
We  assume moreover that the topological pressure is differentiable with respect
to the variation $\delta \gT$.

Let us define: 

\beq\label{FpT}
\FpT(\cW)=P\left[ \psiT+(\cW-\cWT).\bphi(\cWT) \right]-P\left[\psiT \right],
 \eeq

\nid so that $\FpT(\cWT)=0$.  Note that $\FpT(\cW)$ is
 \textit{convex}, due to the convexity of the topological pressure (eq. (\ref{Convex}) in appendix).

Then,  using eq. (\ref{dP}) in appendix, 
the adaptation rule (\ref{dWij}) can be written in the form:

\beq \label{dWgrad}
\dWT= \epsilon \nabla_{\tiny{\cW=\cWT}} \FpT(\cW).
\eeq

 Since, in this section,  pressure $P$ is assumed to be smooth,  one has,  using (\ref{dP}),(\ref{FDT}):
\beq\label{dPT}
\epsilon \left(P\left[\psiT+\dWT.\bphi(\cWT) \right]-P\left[\psiT\right]\right)=\dWT.\left[I+\frac{\epsilon}{2} \KT \right].\dWT + O({\dWT}^3),
\eeq
\nid where $\KT$ is the tensor with entries:
\beq\label{KabT}
\KabT= C^{(\tau)}_{\phia\phib}(0)+2\sum_{t=1}^{+\infty} C^{(\tau)}_{\phia\phib}(t); \qquad i_1,j_1,i_2,j_2=1 \dots N,
\eeq
\nid and $ C^{(\tau)}_{\phia\phib}(t)=\npT\left[\phia\circ{\stgT}\phib\right]-\npT\left[\phia\right]\npT\left[\phib\right]$
is the correlation function of $\phia,\phib$ for the measure $\npT$. Using the 
explicit form (\ref{phiij}) of $\phi_{ij}$ one can see that  $\KT$ is a sum of 
time correlations between uplets of spikes. This is a version of the fluctuation-dissipation theorem
where the response to a smooth variation of the potential $\psiT$ is given in terms of a series
involving the time correlations of the perturbation \cite{ruelle:69,ruelle:99}.
This series converges provided that dynamics is uniformly hyperbolic.

For sufficiently small $\epsilon$, the matrix $I+\frac{\epsilon}{2} \KT$ is positive and:

%\beq
$$P\left[\psiT+\dWT.\bphi(\cWT) \right] \geq P\left[\psiT\right].$$
%\eeq

It follows that the variation  $\delta^{(\tau)}\cF_\phi=\FpTp(\cWTp) - \FpT(\cWTp)$
is given by:

$$\delta^{(\tau)}\cF_{\small{\bphi}}= P\left[\psiT\right]-P\left[\psiT+\dWT.\bphi(\cWT) \right] =-\frac{1}{\epsilon}\dWT.\left[I+\frac{\epsilon}{2} \KT \right].\dWT - O({\dWT}^3)
$$
\beq\label{deltaFpT}
\delta^{(\tau)}\cF_{\small{\bphi}}=
-\epsilon \npT\left[\bphi \right] 
\left[I+\frac{\epsilon}{2} \KT \right].\npT\left[\bphi\right] - O({\dWT}^3)
\eeq

\nid This variation is therefore \textit{negative} when $\epsilon$ is sufficiently small.\\

We come therefore to the following important conclusion. The adaptation rule (\ref{gcondensed}) is a \textit{gradient} system where the function $\FpT$ \textit{decreases}
when iterating synaptic adaptation rules.
Were the transition $\tau \to \tau+1$ to be smooth for all $\tau$, would $\FpT$ reach a minimum\footnote{Additional constraints are required ensuring
that $\FpT$ does not tend to $-\infty$. Typically, such constraints amount to bound the synaptic 
weights variation (soft or hard-bounds rules) by a suitable choice of functions $h_{ijL_1L_2}$ in eq.
(\ref{phiij}). } at some $\cW^\ast$ as $\tau \to \infty$.
Such a minimum corresponds to $\nabla_{\tiny{\cW^\ast}} \FpT=0$, thus to $\dW=0$ according to eq. (\ref{dWgrad}). Hence, this minimum
corresponds to a \textit{static distribution} for the synaptic weights. Therefore, according to section \ref{static}
the potential $\bpsi^{(\tau)}$ would converge to the potential (\ref{psis}) as $\tau \to +\infty$.
However, we cannot expect these transitions to be smooth for all $\tau$. 
 
\sssu{Singular variations.}%\label{singvar}

As we saw, during the synaptic adaptation process,
the corresponding path in  the space of parameters usually crosses  bifurcations manifold
inducing sharp changes in the dynamics. At those points pressure may not be smooth corresponding
to a phase transition. These changes are not necessarily easy to detect
numerically, though algorithms allowing to detect phase
transitions from finite samples exist, based on rigorous results \cite{comets:97}.
Indeed, searching ``at blind'' for all possible kind of phase
transitions in so large dimensional dynamical systems, without any idea of what can
happen, seems to be desperate, even for model I and II,
 especially when thinking of what can already happen in one dimensional systems
\cite{beck-schloegl:95}. So we shall not discuss within more details this aspect,
focusing on grammar changes\footnote{We actually conjecture that the only possible phase transitions
in model I and II are grammar changes occurring when crossing the set where $\dOS=0$ (eq. (\ref{dOS})).}.

Indeed, from the point of view
of  neural network analysis, grammar changes implies  changes in the type of raster plots that the network
is able to produce.
An interesting situation occurs when the set of admissible
raster plots obtained after adaptation belongs to $\SgT \cap \SgTp$.
In this case, adaptation plays the role of a \textit{selective mechanism}
where the set of admissible raster plots, viewed as a neural
code, is gradually reducing, producing
after $n$ steps of adaptation a 
set $\cap_{m=1}^n \Sgm$ which can be 
rather small. This has been observed by Soula et al. in \cite{soula-etal:06}
for model I with Spike Time Dependent Plasticity, though not analyzed in those terms.
If we consider the situation where (\ref{DNN}) is
a neural network submitted to some stimulus, where
a raster plot $\tom$ encodes the spike response to the stimulus,
 then $\Spg$ is the set of all
possible raster plots encoding this stimulus.
Adaptation results in a reduction of the possible
coding, thus reducing the variability in the possible
responses.

\sssu{Evolution of the potential.}

Iterating the adaptation process, one expects to have periods of smooth variations,
punctuated by sharp transitions where potential and grammar change. We qualify
them under the generic name of ``phase transitions'' without further specifications.
  We write $\tau \equiv (l,n)$  where $l$ indexes the phase transition and $n$ the number of epoch since
the last phase transition. 
The adaptation process corresponds now to a sequence $\psiT \equiv \psinl$
of Gibbs potentials. Thus, $\psiOl$ is the potential arising just after the $l$-th transition,
 with the convention that $\psiOl(\tom)=-\infty$ if $\tom$ is forbidden,
such that $\psiOl$ characterizes the  grammar  after the $l$-th phase transition.
 We call a \textit{regular period} the succession of adaptation epochs between two phase transitions.\\

By definition, during a regular period the synaptic update  $\cWTp=\cWT+\dWT$ induces a smooth variation of the potential
$\psiTp=\psiT+\delta\psiT$, and  the variational principle (\ref{supmu}) selects a new Gibbs measure $\npTp$ 
with a potential: 

\beq\label{psiTpvspsiTreg}
\psiTp=\psiT+\delta\psiT,
\eeq

\nid It follows that:

\beq\label{psinl}
 \psinl=\psiOl+ \sum_{m=0}^n \delta\psi^{(\tau_l+m)}, 
\eeq

\nid where $\tau_l$ is the epoch where the $l$-th phase transition aroused.
The explicit form of $\delta\psiT$ (as well as $\psiT$) depends on the detailed form of the dynamics,
and there is little hope to determine it in general, except when the adaptation process converges
to a solution with $\dW=0$ (see section \ref{static}). 
According to section \ref{smoothvar}, the function $\FpT$ (eq. (\ref{FpT})) decreases
during regular periods. \\

When a phase transition occurs, the relation (\ref{psiTpvspsiTreg})
does not hold anymore. We now write:

\beq\label{psitrans}
\psilpO=\psilnl+\dpsirlpO+\dpsislpO
\eeq

\nid where $n_l$ is the last epoch before the phase transition $l+1$.
$\dpsirlpO$ contains the regular variations of the potential while $\dpsislpO$ contains the singular variations corresponding
to a change of grammar. 
%It can be defined as follows. Fix $L>0$ and very large.
%Set $\psi(\tom)=-L$ if the sequence $\tom$ is forbidden for the potential $\psi$.
%Then $\dpsislpO(\tom)=L$ if $\tom$ was forbidden and become allowed after adaptation
%and  $\dpsislpO(\tom)=-L$ if $\tom$ was allowed and become forbidden after adaptation.\\

We end up with the following  picture. During regular periods the grammar does not evolve,
the potential changes according to (\ref{psinl}), (\ref{psitrans}), and the function $\FpT$ decreases. Then,
there is a sharp change in the grammar and the potential. The function $\FpT$ may also have singularities
and may sharply increase. If the adaptation rule converges, the potential (\ref{psinl})
converges to $\bpsis=\Phi+\bl^\ast.\bphi$ (eq. (\ref{psis})). This potential  $\Phi$
characterizes the grammar and the statistical weight of allowed raster plots is given
by $\bl^\ast.\bphi$. The potential $\bpsis$ contains all changes in the grammar
and statistics arising during the evolution  (\ref{psinl},\ref{psitrans}).
Thus it contains the history of the system and the evolution of the grammar.

\su{A numerical example.}\label{numex}

\ssu{Model.}

\sssu{Adaptation rule.}

As an example we consider an adaptation rule  inspired from (\ref{STDPFC})
with an additional term $\ld \WijT$, $-1 <\ld <0$, corresponding to passive
LTD.

\beq\label{Rexample}
\dWijT=\epsilon
\left[\ld \WijT+  \frac{1}{T} \sum_{t=T_s}^{T+T_s} \omjt(t) \sum_{u=-T_s}^{T_s} f(u) \, \omit(t+u)\right],
\eeq

\nid where $f(x)$ is given by (\ref{fSTDP}) and with:

$$T_s \deq 2 \max(\tau_+,\tau_-).$$

Set :

\beq\label{Si}
S_i(\tom)= \sum_{u=-T_s}^{T_s} f(u) \, \omei(u),
\eeq

\nid 

%\beq\label{H}
$$H_{ij}(\tom)=\omega_j(0)S_i(\tom),$$
%\eeq

\nid $\bH=\left\{H_{ij}\right\}_{i,j=1}^N$, and:

%\beq\label{phiij_ex}
$$\phi_{ij}(\Wij,\tom)=\ld \Wij+H_{ij}(\tom),$$
%\eeq 
%
with $\bphi=\left\{\phi_{ij}\right\}_{i,j=1}^N$, where $\phi_{ij}$ is a finite
range potential with range $2T_s$. Then (\ref{Rexample}) has the form (\ref{gcondensed}),
$\dWijT=\epsilon \pToT\left[\phi_{ij}(\Wij,.) \right]$.

\sssu{Effects of the adaptation rule on the synaptic weights distribution.}\label{3reg}

The term  $S_i(\tom)$ (eq. \ref{Si}) can be either negative, inducing Long Term Depression,  or positive inducing Long Term Potentiation.
In particular, its average with respect to the empirical measure $\pToT$ reads:
\beq \label{PoTSi}
\pToT(S_i)=
\eta r_i(\tau)
\eeq
\nid
where:
\beq\label{eqbeta}
\eta=
\left[
A_- e^{-\frac{1}{\tau_-}}\frac{1-e^{-\frac{T_s}{\tau_-}}}{1-e^{-\frac{1}{\tau_-}}}
+
A_+ e^{-\frac{1}{\tau_+}}\frac{1-e^{-\frac{T_s}{\tau_+}}}{1-e^{-\frac{1}{\tau_+}}}.
\right]
\eeq
\nid and where $r_i(\tau)=\pToT(\omei)$ is the frequency rate of neuron $i$ in the $\tau$-th adaptation epoch.

The term $\eta$ neither depend on $\tom$ nor on $\tau$,
but only on the adaptation rule parameters $A_-,A_+,\tau_-,\tau_+,T_s$.
Equation (\ref{PoTSi}) makes explicit 3 regimes.

\bit
\item\textbf{Cooperative regime.} If $ \eta>0$ then $\pToT(S_i)>0$. Then synaptic weights have a tendency to become more positive.
 This corresponds to a cooperative system \cite{hirsch:89}.
When iterating adaptation, dynamics become trivial with neurons firing at each time step or
remaining quiescent forever.

\item\textbf{Competitive regime.} On the opposite if $ \eta<0$  synaptic weights  become  negative.
This corresponds to a competitive system \cite{hirsch:89}.

\item\textbf{Intermediate regime.} The intermediate regime corresponds to $ \eta \sim 0$.
Here no clear cut tendency can be distinguished from the average value of $S_i$
and spikes correlations  have to be considered as well.

\eit

\sssu{Static weights.}

Thanks to the soft bound term $\ld\Wij$ the synaptic adaptation rule admits a static solution given by:

\beq\label{Wstatex}
\Wij=-\frac{\pToT\left[\omega_j(0)S_i(\tom)\right]}{\ld}.
\eeq

\nid Note that this equation can have several solutions.

Using the same decomposition as the one leading to (\ref{PoTSi}) we obtain:

$$\pToT\left[\omega_j(0)S_i(\tom)\right]=
A_- \sum_{u=-T_s}^{-1} e^{\frac{u}{\tau_-}}\pToT\left[\omega_j(0) \, \omei(u)\right]+
A_+ \sum_{u=1}^{T_s} e^{-\frac{u}{\tau_+}}\pToT\left[\omega_j(0) \, \omei(u)\right].$$

Note that  $\omega_j(0) \, \omei(u) \geq 0$, thus the first 
term is negative and the second one is positive (see eq. (\ref{fSTDP})).
The sign of $\Wij$ depend on the parameters $A_-,A_+,T_s$, but also on the relative
strength of the terms $\pToT\left[\omega_j(0) \, \omei(u)\right]$. 

\sssu{Convergence to the static weights solution.}

The synaptic adaptation rule (\ref{Rexample}) defines
a mapping $\cQ$ on the set of synaptic weights:

%\beq\label{}
$$\WijTp=\WijT(1+\epsilon\ld) +  
\epsilon \pToT \left[H_{ij}\right]=\cQ_{ij}(\WT)$$
%\eeq

\nid with $|1+\epsilon\ld|<1$, where  $\pToT$
depends on $\WT$. Thus,

%\beq
$$\frac{\partial \cQ_{ij}}{\partial W_{kl}}= (1+\epsilon\ld)\delta_{ij,kl}
+ \epsilon\frac{\partial \pToT \left[H_{ij}\right]}{\partial W_{kl}},$$
%\eeq

\nid where $\delta_{ij,kl}=1$ if $i=k,j=l$ and $0$ otherwise.
The second term is a linear response  characterizing the variation
of $\pToT \left[H_{ij}\right]$ with respect to small variations
of $W_{kl}$. Approximating $\pToT$ by a Gibbs
distribution with a potential $\psiT$, it writes
$ C_{\psiT,H_{ij}}(0)+2\sum_{t=0}^{+\infty} C_{\psiT,H_{ij}}(t)$ where $ C_{\psiT,H_{ij}}(t)$
is the time-$t$ correlation function between $\psiT$ and $H_{ij}$. 
When $\npT$ is smooth with respect to $\cW$, the derivative
is dominated by the first term and $\cQ$ is contracting for sufficiently small $\epsilon$.
This is however not a sufficient condition for the convergence of (\ref{Rexample})
because $\cQ$ is not continuous everywhere. It has jumps whenever
$\bg$ crosses the boundary of a structural stability domain.

If the static solution is contained in a open ball where $\cQ$ is continuous
 the contraction property ensures the convergence to a static solution
for any initial condition in this ball (Brouwer theorem). 

In this paper we postulate the convergence and check it numerically.

\sssu{Spike train statistics in a static weights regime.}

 As emphasized  in section \ref{Statmod} and \ref{static}, when the synaptic
adaptation rule converges to a fixed point, the corresponding 
statistical model is a Gibbs
measure with a potential: 

%\beq
$$\bpsis=\Phi+\bl^\ast.\bphi,$$
%\eeq

\nid where $\Phi$ contains the grammar and $\bl$ are free statistical parameters.
The value ${\bl}^\ast$ of these parameters in the potential $\bpsis$ is determined by the relation:

\beq
\left.
\frac{\partial P\left[\bpsi\right]}{\partial \lambda_{ij}}\right|_{\bls}
=\ld \Wij^\ast+\nu_{\bpsis}[H_{ij}]=0, \
 \forall i,j \, ,
\eeq
\nid where the pressure is given by:

$$
P\left[\bpsi\right]=\ld \bl.\cW + \lim_{T \to \infty} \frac{1}{T}\log \sum_{\tom \in \SpT}  e^{\bl.S_T\bH(\tom)}.
$$

This procedure provides us  the explicit form of the raster plot probability distribution when
the adaptation rule converges. But the price to pay is that  we have to determine \textit{simultaneously} the $N^2$ parameters $\lambda_{ij}^\ast$
on which the Gibbs measure depends. Focusing on the joint probability of a small set of neurons (pairs, triplets)
this constraint can be relaxed in order to be numerically tractable.
\ssu{Numerical checks.}

The main goal of section \ref{numex} is to provide an 
example of the theoretical concepts developed in this paper. The emphasis
is however not put on numerical results which will be the main topic of a forthcoming paper.
Therefore, we  focus here on numerical simulations for model I only, since the simulations for model II
are quite more computer-time consuming, and we consider only one case of $\eta$ 
value corresponding to the intermediate regime defined above.

\sssu{Implementation.}

\paragraph{Neurons dynamics.}

 Previous numerical explorations have shown that a Model I- network of $N$ 
neurons, with synapses taken randomly from a distribution ${\cal N}(0,\frac{C^2}{N})$, 
where $C$ is a control parameter,
 exhibits a dynamics with very large periods in specific regions of values of the space $(\rho,C)$ 
\cite{cessac:08,cessac-vieville:08}. On this basis, we choose $N=100,\; \rho=0.95,\; C=4.0, N=100$. 
The external current $\Ie$ in eq. (\ref{FiBMS}) is given by
 $\Iei=0.06+0.01 \,{\cal N}(0,1)$. Note that fixing a sufficiently large average value for this current avoids a situation
where neurons stops firing after a certain time (``neural death'').

\paragraph{STDP implementation.}

An efficient implementation of a STDP rule does not only depend  on the analytic form of the rule
 but also on the respective time scales characterizing neurons and synapses evolutions. 
In ``on-line-protocols'', where the synaptic changes occur at the same time scale as neurons dynamics
 the so-called recursive approach  appears to be the more appropriated (see \cite{zou:06} and references therein). 
In our case, we use instead an offline protocol, where we register the dynamics of the system on a long time windows, then  compute the STDP modification, then let the system evolve again 
to its attractor (see section  \ref{adapt}). 
The offline protocols are very expensive in machine-time, especially 
when using long spike trains to get a reliable time average ($\pToT(\phi_{ij})$).
 For this reason we need to add some words about our implementation.\\

We register spike trains  in a binary code.
Indeed,  this is the cheapest way in memory requirements,
though it might be  expensive for accessing specific spike times.
Also, bit-wise 
operations are faster than their equivalents on other types of data. 
Finally, there exist very fast methods for computing the number of bits 
on any specific variable of type of integer 
(The faster for large number of iterations is a look-up table of precomputed values,
 but in-line methods - using parallel methods based on masks- are not so far in terms of performances.
 For details and speed comparison see the G. Manku website 
http://infolab.stanford.edu/~manku/bitcount/bitcount.html).
Numerical comparison of this method with the direct one that records the dynamics 
in Boolean arrays and compute STDP spike by spike shows enormous performance difference growing exponentially as the length of trains increases. 

\paragraph{Computation of $\delta^{(\tau)}\cF_\phi$.}

To check the result in section \ref{smoothvar}, stating that
$\FpT(\cW)$ is a decreasing function of $\tau$ (i.e. $\delta^{(\tau)}\cF_{\small{\bphi}}$ is negative)
during regular periods, we use the following method \cite{chazottes:99}.
 
Fix a potential $\bpsi$. 
If $T$ is the length of the experimental raster plot, one divides it into $k$ blocs of length $n$,
such that $T=k \times n$. Call:

\beq
P_{n,k}(\tom)=\frac{1}{n} \log\left(\frac{1}{k}\sum_{j=0}^{k-1} \exp(S_j^{(n)}(\tom)) \right),
\eeq

\nid with $S_j^{(n)}(\tom)=\sum_{t=jn}^{jn+n-1} \bpsi(\sigma^t \tom)$.
Then, the following result can be proved \cite{chazottes:99}:

\beq
P\left[\psi\right]=\lim_{n \to +\infty}\lim_{k \to +\infty}P_{n,k}(\tom),
\eeq

\nid for $\nu_{\bpsi}$ almost-every $\tom$. This computation requires the knowledge of $\bpsi$.

The result in  \cite{chazottes:99} can be extended straightforwardly 
to the following case. If $\bpsi_1,\bpsi_2$
are two potentials for the same grammar, and $\delta\bpsi=\bpsi_2 - \bpsi_1$ then:

\beq
P\left[\bpsi_2\right]-P\left[\bpsi_1\right]=
\lim_{n \to +\infty}\lim_{k \to +\infty}
\frac{1}{n} \log\left(\frac{1}{k}\sum_{j=0}^{k-1} \exp(\sum_{t=jn}^{jn+n-1} \delta \bpsi(\sigma^t \tom)) \right)
\eeq

Since $\delta^{(\tau)}\cF_\phi= P\left[\psiT\right]-P\left[\psiT+\dWT.\bphi(\cWT) \right]$,

\beq\label{numdF}
\delta^{(\tau)}\cF_\phi=-\lim_{n \to +\infty}\lim_{k \to +\infty}
\frac{1}{n} \log\left(\frac{1}{k}\sum_{j=0}^{k-1} \exp\Big(\sum_{t=jn}^{jn+n-1} \dWT.\bphi(\cWT,\sigma^t (\tom^{(\tau)}))\Big) \right)
 \eeq

Expanding this equation in power series of $\epsilon$ one recovers the expansion (\ref{dPT}).
In particular, subtracting to (\ref{numdF}) the leading term (of order $\dWT.\dWT$) one gets
an estimation of the ``linear response'' term $\KabT$ in eq. (\ref{KabT}).
However, on practical grounds, the computation of (\ref{numdF})
requires very long time series to reduce significantly the finite fluctuations of the
empirical average. Similar problems are encountered in the computation of linear response
from other methods  \cite{cessac:07}.

This result holds for any adaptation rule (any polynomial $\bphi$). 
In the case of the adaptation rule (\ref{Rexample}),
this reduces to computing:

%\beq
$$\delta^{(\tau)}\cF_\phi=-\ld \dWT.\WT -\lim_{n \to +\infty}\lim_{k \to +\infty}
\frac{1}{n} \log\left(\frac{1}{k}\sum_{j=0}^{k-1} \exp\Big(\sum_{t=jn}^{jn+n-1} \epsilon\dWT.\bH(\cWT,\sigma^t (\tom^{(\tau)}))\Big) \right).
$$
 %\eeq

\sssu{ Simulation results.} 

\begin{figure}[htbp]
 \begin{center}
\includegraphics[width=6cm,height=6cm]{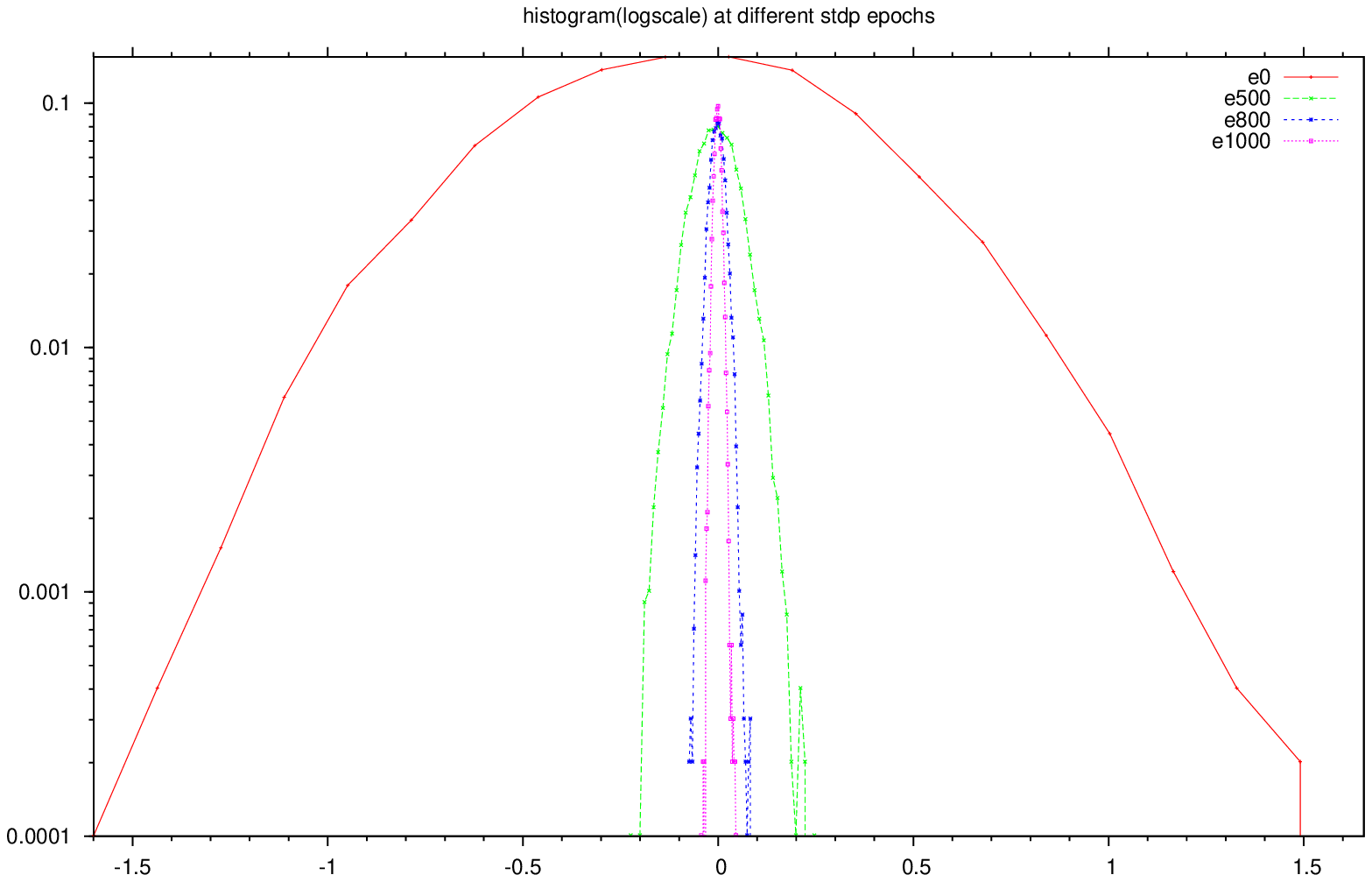}%Histogrammes
\includegraphics[width=6cm,height=6cm]{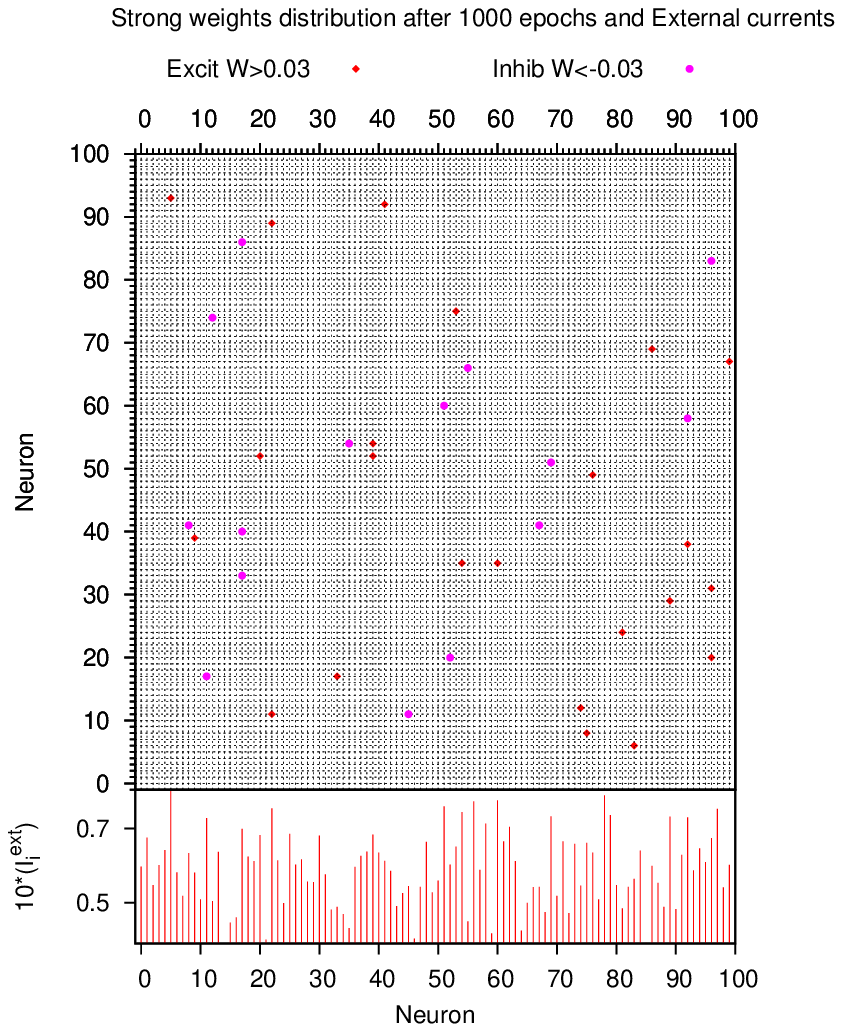}%Matrice 
\vspace{1cm}
\includegraphics[width=6cm,height=6cm]{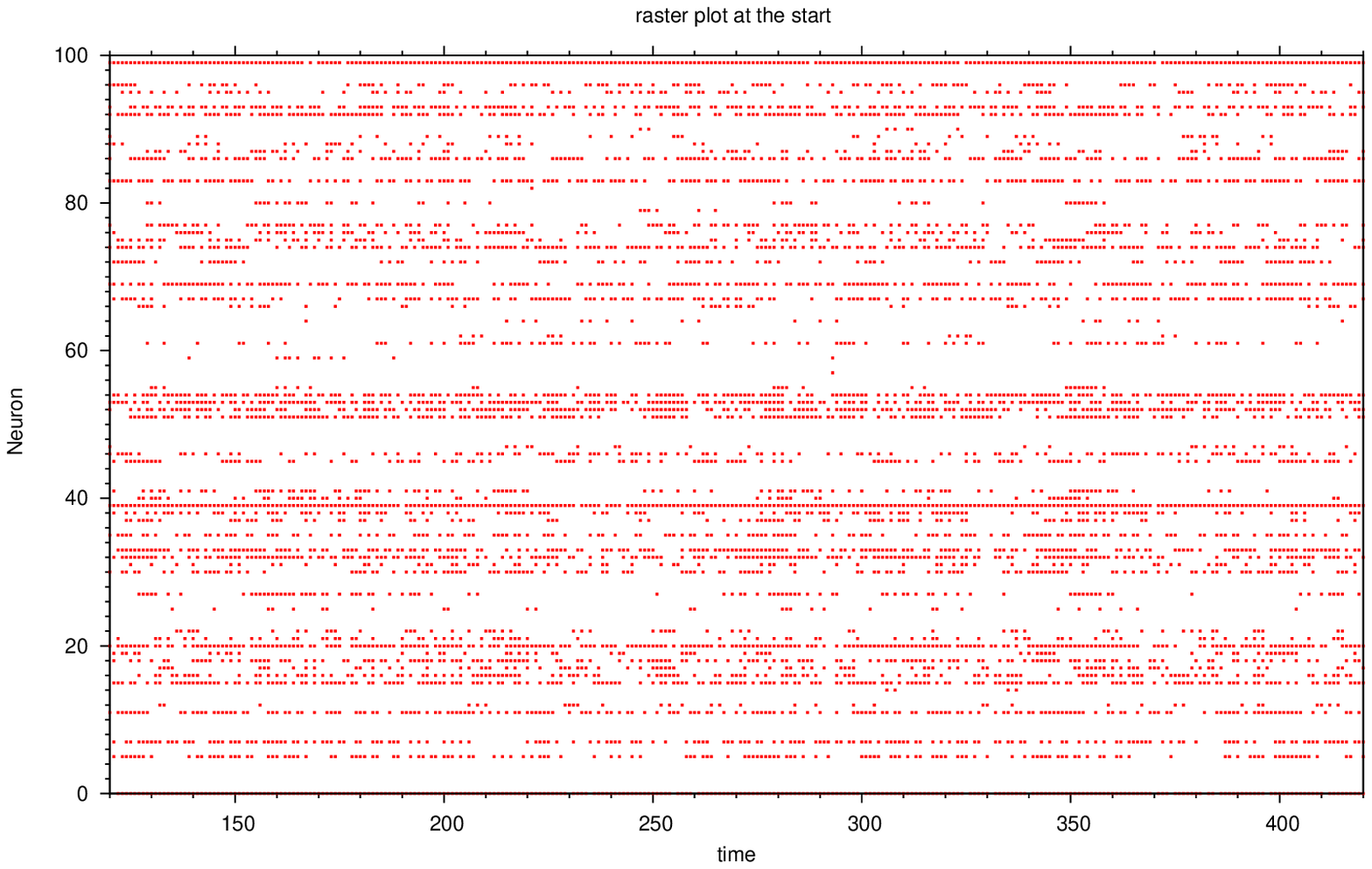}%Rasters
\includegraphics[width=6cm,height=6cm]{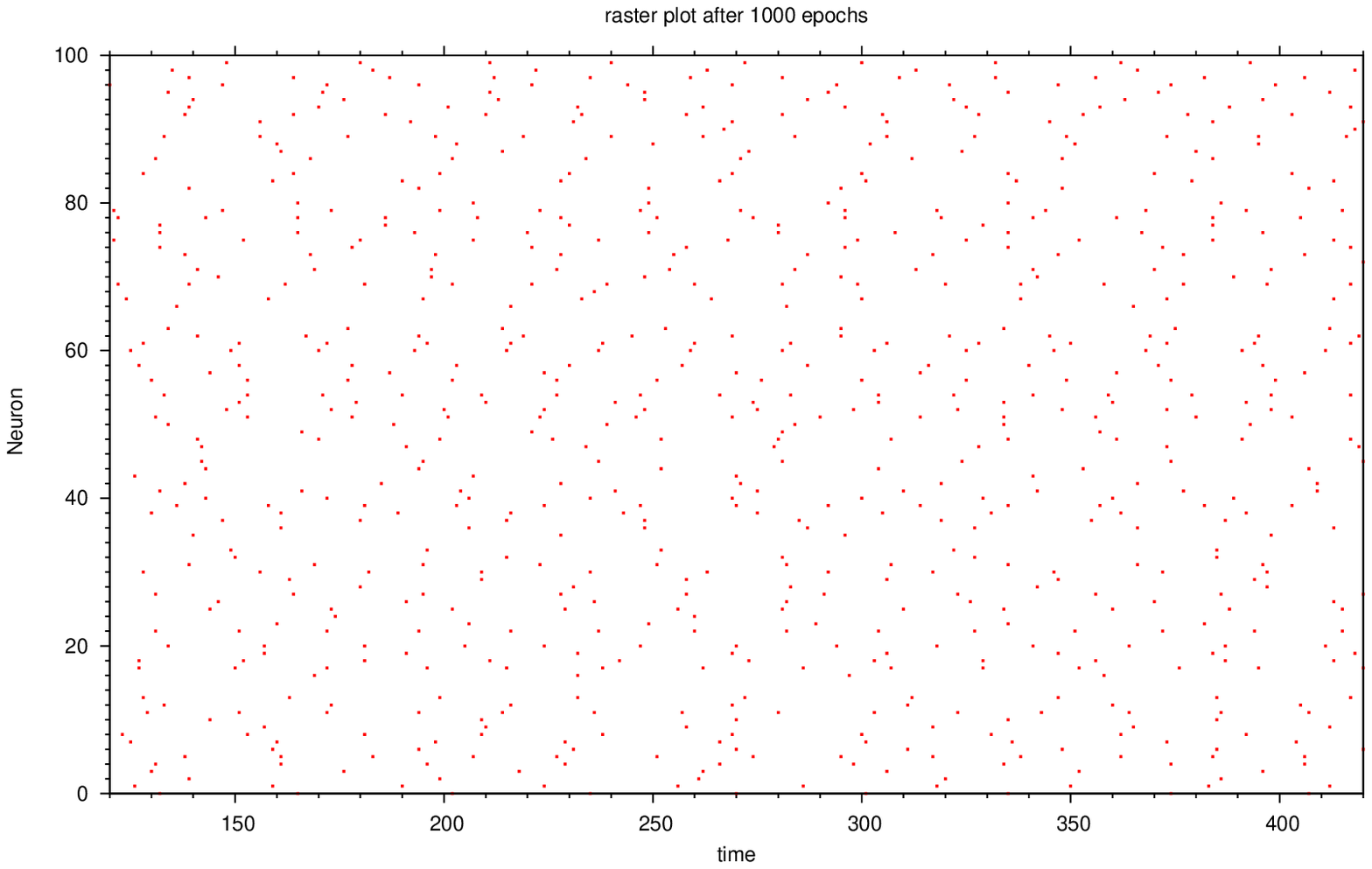}%Raster
\vspace{1cm}
\includegraphics[width=6cm,height=6cm]{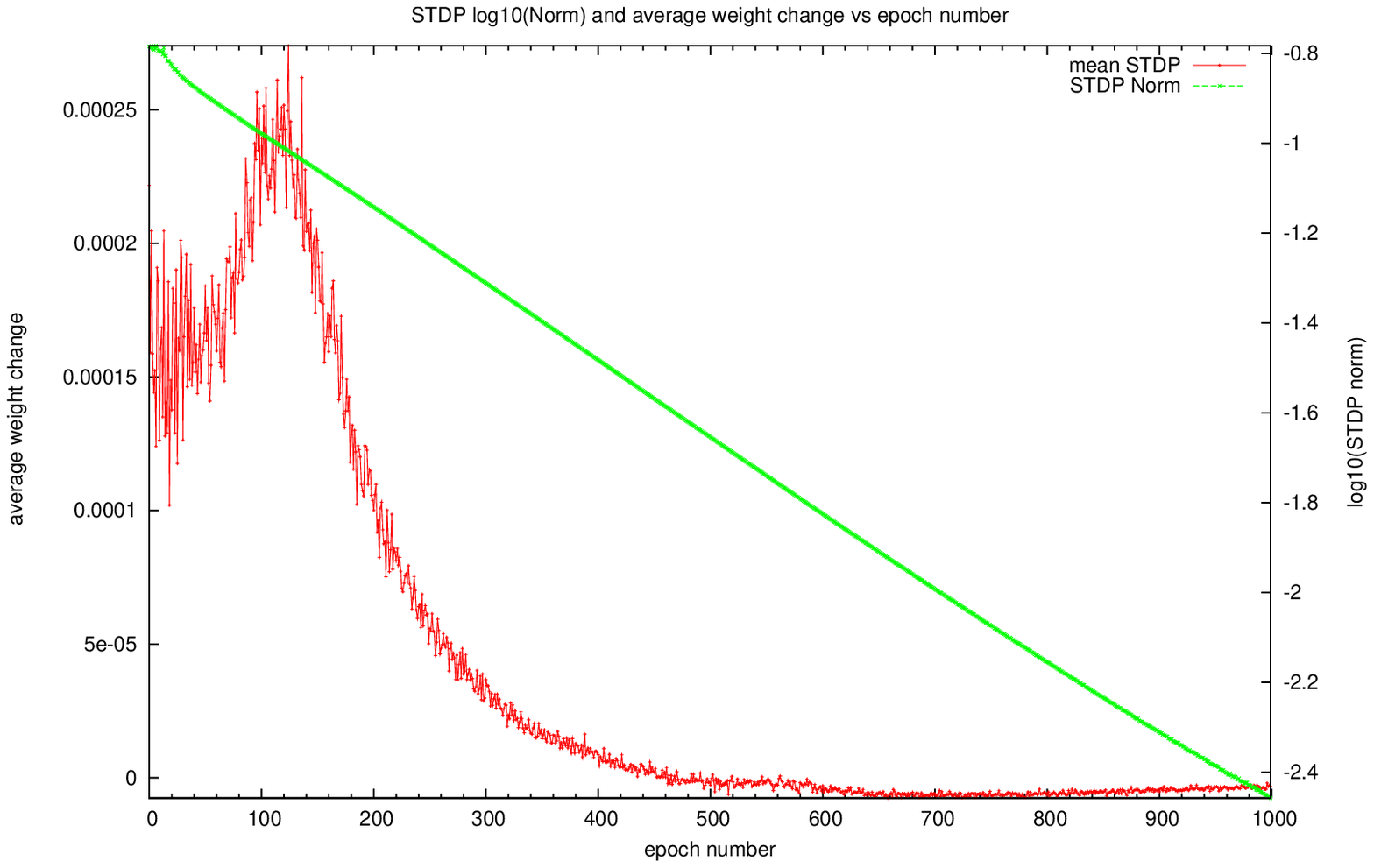}%Evolution de la norme des poids
\includegraphics[width=6cm,height=6cm]{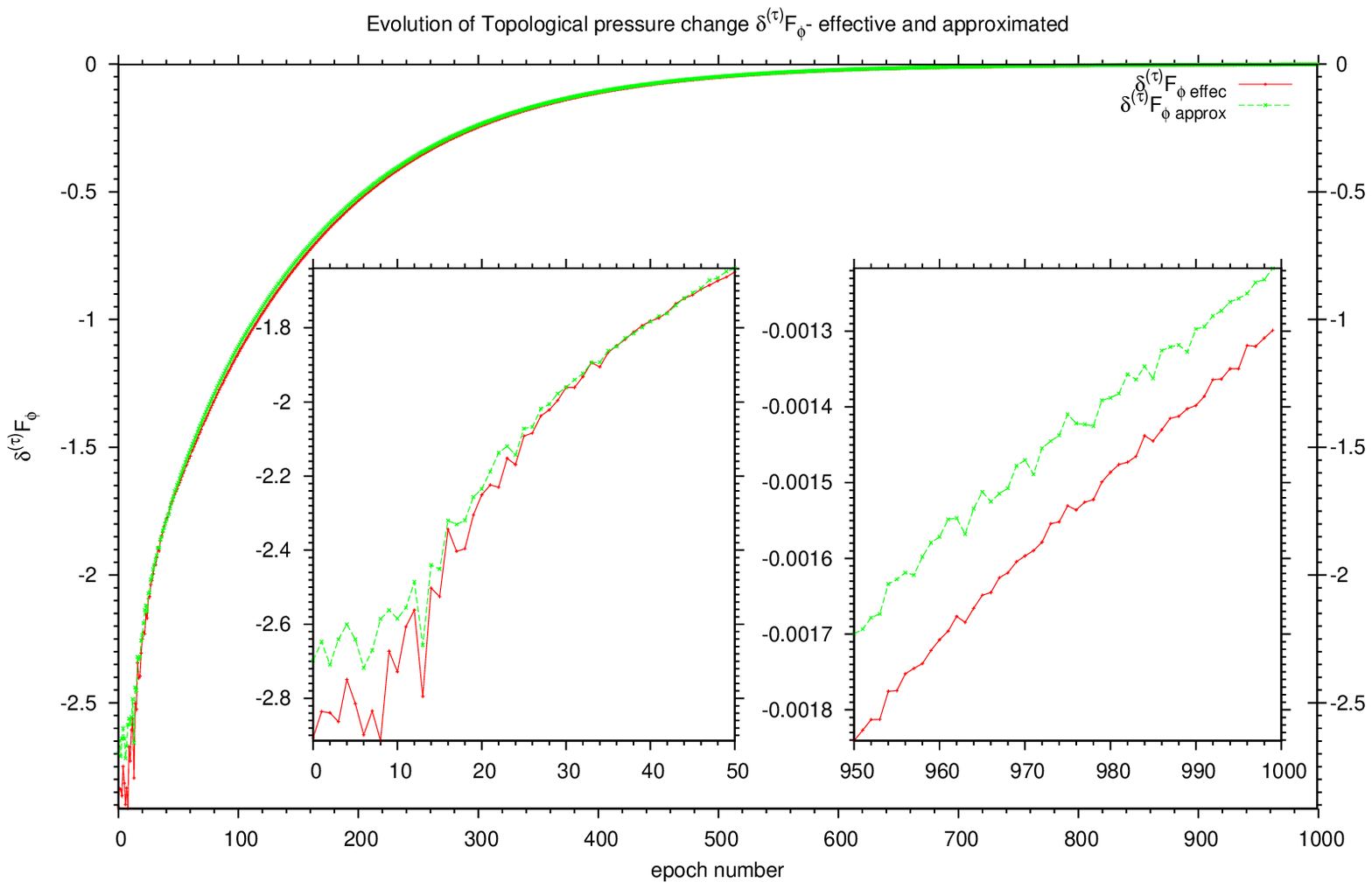}%delta Ft
\caption{\footnotesize{(color on line). The parameters are $N=100, \ld=-0.4, \eta=-0.01, \epsilon=0.01 $. 
 \textbf{Top}:(left) Histogram in log scale of synaptic connectivity in the network after 1000 
epochs of STDP;(right)  map of strongest connections, either inhibitory (pink dots) or excitatory (red diamonds) 
(i.e, $|W_{ij}|>0.03$). We also include at the bottom of this map the external current of each neuron (multiplied by 10).
 \textbf{Middle}:(left) A comparison of raster plots at the beginning  and (right) the end of the synapses evolution. 
\textbf{Bottom}:(left) Evolution of the mean synaptic change $\frac{1}{N^2}\sum_{i,j}^N \dWT_{ij}$(red) and the Frobenius norm for STDP matrix on log scale i.e, $\log_{10}\left(\sum_{i,j}^N |\dWT_{ij}|^2\right)$  (green); (right)  Variations of Topological Pressure (with 2 zoom graphics showing sharp changes)
 after each epoch as calculated explicitly in eq. \eqref{numdF} with 512 trains of size 512 (red),
and first order term approximated by $\frac{1}{\epsilon}\dWT.\dWT$(green). }}
\label{Fbeta-0.01}
\end{center}
\end{figure}

We have run simulations for STDP parameters 
values: $\ld=0.99,\;  \epsilon=0.001,\;\tau_+=16, \;\tau_-=32, \;A_+=1.0,$ and $T_s=64$,
corresponding to standard values \cite{izhikevich-desai:03}.
We fix $\eta=-0.01$ corresponding to the intermediate regime discussed in section \ref{3reg}.
This fixes the value of $A_-$ via  eq. (\ref{eqbeta}). 
The length of spike trains is $T=3941=4096-128=128\times 32 - 2 \times T_s$, where $32$ is the number of bits
in long integer, in our (machine-dependent) implementation. 
An extended description will be published elsewhere. The main results are summarized
in fig. \ref{Fbeta-0.01}.

In this figure  (Top) we have represented the evolution of the distribution of synaptic weights $\WT$
(see captions). For this value of $\eta$, the histogram evolves to a unimodal distribution with
some nodes having strong synaptic weights.  In fig.  \ref{Fbeta-0.01} (center) 
we depict the raster plots at the beginning and at the end of synaptic adaptation process.
In fig. \ref{Fbeta-0.01} (bottom-left) we have shown the evolution of the Frobenius norm for  
$\dWT$ (i.e, $\sum_{i,j}^N |\dWT_{ij}|^2$), which converges to $0$. This shows the convergence
of the rule in this case.
We have also plotted the average weights modification $\frac{1}{N^2}\sum_{i,j}\delta W_{ij}$.
 Finally in fig.  \ref{Fbeta-0.01} (bottom -right) is represented $\delta^{(\tau)}\cF_\phi$ computed with  $512$ block of size $512$. We see clearly that this quantity is negative
as expected from eq. (\ref{deltaFpT}) and tends to $0$. We have
also represented the term $-\frac{1}{\epsilon}\dWT.\dWT$ corresponding
to the first order expansion in eq. (\ref{deltaFpT}).

\su{Conclusion.}%\label{Conclusion}

In this paper, we have considered the questions of characterizing the 
statistics of spike trains generated by neuronal networks, and
how these statistics evolve during synaptic adaptation, from the angle
of dynamical systems theory and ergodic theory. We have introduced
a framework where notions such as raster plots, empirical averages
of observables, and synaptic plasticity rules, coming from neuroscience, can be formalized in the
language of dynamical systems. In this context, we propose a strategy to
produce optimal statistical models of spike trains, based on the maximization
of statistical entropy where expectation of observables must be compatible with
their empirical averages. These models are Gibbs measures.
Considering adaptation rules with slow dynamics, we also suggest
that such Gibbs measures may result from  related adaptation rules whenever
synaptic weights converge to a fix value. In this case,
synaptic weights resume the whole history of the neuronal network
contained in the structure of the generating function of cumulants
(the topological pressure) and in the structure of allowed/forbidden
raster plots (grammar). 

To our opinion this theoretical paper is a beginning. Our hope is now
to apply this strategy to data coming from biological neuronal networks,
while only formal neural networks were considered here. At the present
stage there is indeed a crucial issue to be able to characterize spike train
statistics and to discriminate statistical models. For example, it is still
an open problem to decide, for a given spike train in a given experiment, whether
firing rate are sufficient for the statistical characterization (leading to
uncorrelated Poisson models)  or if higher order orders statistics such as correlations are also
necessary. This is more than an academic question. Beyond the choice of such
or such statistical model, there are hypotheses on how neuronal network
convert stimuli from the external word into an action. Discriminating
statistical models is the way to select the best hypothesis.

Our approach opens up the possibility of producing numerical methods
for such a discrimination, e.g. based on Kullback-Leibler divergence minimization
obtained from eq. (\ref{KLemp}). Some of these methods are already available as open source code on
http://enas.gforge.inria.fr/. They are based on the theory developed in the present
paper, relying itself on ergodic theory and thermodynamic formalism. At the present
stage, the overwhelming richness of thermodynamic formalism has not really been considered
in neuroscience. We hope that this paper will attract the attention of the neuroscience
community on this point.

\bigskip

\textbf{Acknowledgments.}
This work has been supported by the INRIA ARC MACACC and the Doeblin CNRS federation.
We thank the Reviewer for helpful remarks, useful references and constructive criticism.
\pagebreak

%\appendix{Appendix}
\su{Appendix} 

 In this appendix we present the thermodynamic formalism material used in the paper. 
Though all  quantities defined below depend on $\bg$, we have dropped
this dependence, which is not central here, in order to alleviate notations.
We   follow  \cite{keller:98,parry-pollicott:90,chazottes-etal:98,chazottes:99,chazottes-keller:09}
in the  presentation of Gibbs measures in ergodic theory.
In accordance with the body of the text $\Sigma$ is the set of admissible
raster plots 
and we denote by $\cB$ the related Borel sigma-field.

\ssu{Ergodic measures.} \label{Serg}

 Fix $\phi$ a continuous function. The time average of $\phi$ along the orbit of $\X$ is given by
$$\pi_\X(\phi)=\lim_{T \to \infty} \frac{1}{T}\sum_{t=0}^T \phi(\X(t)).$$
If $\mu$ is an invariant measure\footnote{Namely $\mu(\Fg^{-1}A)=\Fg(A)$ for any measurable set $A \subset \cB$.}, then, according to Birkhoff theorem, 
this limit exists for $\mu$-almost every $\X$. Standard theorems in ergodic theory ensure that 
(\ref{DNN}) has at least one invariant measure \cite{katok-hasselblatt:98}
but there are typically many. 
Among them, ergodic measures play a distinguished role.
An invariant measure $\mu$ is ergodic if any real measurable invariant function
is $\mu$-almost surely constant. As a corollary of Birkhoff's
theorem $\pi_\X(\phi)=\mu(\phi)$, for $\mu$-almost every
$\X$, where $\mu(\phi)=\int \phi d\mu$
is the expectation of $\phi$ with respect to $\mu$. 
Namely, the empirical average $\pi_\X(\phi)$
is equal to the ``ensemble'' average $\mu(\phi)$
whatever the initial condition, provided it is chosen in the support
of $\mu$. Any invariant measure can be written as a convex
decomposition of ergodic measures.

\ssu{Raster plots statistics.}\label{rastplotstat}

 Fix $n>0$, a set of times $t_1, \dots t_n$,
and a set of prescribed spiking patterns  $\bom(t_1) \dots \bom(t_n)$. 
The set of raster plots  $\tom' \in \Sigma$ such
that $\bom'(t_k)=\bom(t_k), k=1 \dots n$ contains therefore 
raster plots where the spiking patterns at  prescribed
 times $t_1 \dots t_n$ are imposed (cylinder set). By (countable) intersection and 
union of cylinder sets one can generate all 
possible \textit{events} in $\Sigma$, such as ``neuron $i$ is firing at time
$t_1$'', or ``neuron  $i_1$ is firing at time
$t_1$, and neuron $i_2$ is firing at time $t_2$,
\dots and neuron $i_n$ is firing at time $t_n$'', etc \dots.

 The statistical properties of raster plots are 
inherited from the statistical properties of orbits
of (\ref{DNN}), via
the correspondence $\nu[C]=
\mu\left[\left\{\X \rep \tom, \tom \in C  \right\}\right]$, where $C$ is a cylinder set.
On practical grounds, these statistical properties are
obtained by  empirical average. The probability of a cylinder set $C$
is given by:
\beq\label{pTo}
 \pTo(C)=  \frac{1}{T}\sum_{t=1}^T \chi(\sit \in C)
\eeq
where $\chi$ is the indicatrix function, while the time average of some function $\phi : \Sigma \to \bbbr$ is given
by:
%
%\beq
$$\pi_{\tom}^{(T)}(\phi) = \frac{1}{T} \sum_{t=1}^T \phi(\sigma^t \tom).$$
%\eeq
%
%
When $\nu$ is ergodic the following holds.
$$\nu=\po \deq \lim_{T \to \infty} \pTo(.),$$
for $\nu$-almost every $\tom$,
where $\po(.)$ is called the \textit{empirical measure} for the raster plot $\tom$.
%This means that any raster plot, typical for the measure $\nu$, provides
%the same empirical average. On the other way round, 
%a probability measure $\nu$ which characterizes the typical statistical
%behaviour of raster plots obtained from experiments provides what we call a statistical
%model. 

\ssu{Thermodynamic formalism.}

\sssu{Working assumptions.} 
In this paper, we have adopted a pragmatic approach
based on the idea that statistical models obtained via the applications of
principles in thermodynamic formalism (which are basically the same as those
used in statistical physics) could constitute good prototypes for the analysis
of real spike trains statistics. In this way, our goal is not to prove theorems
but instead to use already proved theorems. However, theorems have hypotheses
that we now explicit and whose plausible validity is discussed.

Basically, to apply the machinery described below we need first
to assume that raster plots constitute a symbolic coding for the orbits
of (\ref{DNN}), i.e. there is a one-to-one correspondence between
membrane potential trajectories and raster plots, except for a negligible
subset of initial conditions. We also assume that symbolic trajectories
(raster plots) are generated by a finite grammar. This hypothesis is discussed
in section \ref{SGrammar}, and it can actually be rigorously established
for models I and II without noise. The case with noise is under current investigations.
 The corresponding transition matrix $G$ defines a 
sub-shift of finite type or topological Markov chain. Actually, part
of the results presented below are extensions of standard results on Markov chains.
%As discussed in section \ref{AsDyn}, dynamics (\ref{DNN}) admits typically
%several attractors. As a consequence, $G$ is block diagonal decomposable into
%bloc-submatrices, corresponding to a subset of the alphabet $A$ defined in section
%\ref{SGrammar}. The  symbols in this subset label the Markov partition elements that
%the corresponding attractor intersects.
We  also assumed that
$G$ is \textit{primitive} i.e. 
 $\exists n_0>0$ such that $\forall i,j$, $\forall  n >n_0$ $(G^n)_{ij}>0$.
This amounts to assuming that the dynamics on each attractor is topologically
mixing\footnote{A dynamical system $(\F,\cM)$ is topologically mixing
if for any pair of open sets $U,V \subset \cB$, there exists $n_0 \geq 0$ such that,
$\forall n >n_0$, $\F^nU \cap V \neq \emptyset$}. 

All these hypotheses hold if the dynamical system generating the raster plots is continuous and 
uniformly hyperbolic. Following the definition of \cite{parry-pollicott:90}, 
a map $\F: \cM \to \cM$ is uniformly hyperbolic on a set $\Lambda$ if:

\ben
\item There exists a splitting of the tangent space in unstable
($E^u$) and stable ($E^s$) subspaces such that there exists
$C,\mu>0$ with $\|D\F^t\|_{E^s},\|D\F^{-t}\|_{E^u}\leq C\mu^t$, $t>0$.
\item $\Lambda$ contains a dense orbit.
\item The periodic orbits in $\Lambda$ are dense (and $\Lambda$ consists of more than 
a single closed orbit).
\item There exists an open set $U \supset \Lambda$ such that $\Lambda=\bigcap_{t=-\infty}^\infty F^tU$.
\een

Adding noise to model I, II renders them uniformly hyperbolic (see footnote \ref{fnuh} in the text)
but they are not continuous due to the singularity of the firing threshold.
Hence, additional work is necessary to prove that there is a finite Markov partition (Cessac \& Fernandez, in preparation). 
Throughout this Appendix we assume that all these hypotheses hold.

\sssu{Potentials.} Fix $0 < \Theta < 1$. We define a metric on $\Sigma$,
the set of raster plots generated by the grammar $G$, by $\dT(\tom,\tom')=\Theta^p$,
where $p$ is the largest integer such that $\bom(t)=\bom(t'), 0 \leq t \leq p-1$.
(The related topology thus structures the fact that raster plots coincide until $t = p-1$).

For a function $\bpsi : \Sigma \to \bbbr$ and $n \geq 1$ define
$$
var_n\bpsi = \sup \left\{ |\bpsi(\tom) - \bpsi(\tom')| : \bom(t)=\bom'(t), 0 \leq i < n  \right\}.
$$
We denote by $C(\Sigma)$ the space of functions $\bpsi$ such that $var_n \bpsi \to 0$ as $n \to \infty$.
This is the set of  continuous real functions on $\Sigma$.
 
%i.e. it verifies a Lipschitz condition for the $L^\infty$ and cylindric metrics.
%
A  function $\bpsi \in C(\Sigma)$ is called a \textit{potential}.
A potential is \textit{regular}\footnote{This condition is analogous to the potential 
decay ensuring the existence of a thermodynamic limit in statistical mechanics \cite{ruelle:69,meyer:80}.
In the present context it ensures that the Gibbs measure related to the potential $\bpsi$
is unique and that it is also the unique equilibrium state.} if  $\sum_{n=0}^{+\infty}var_n(\bpsi) <  \infty$.
Equivalently, $\bpsi$ is H\"older.
For a positive integer $r$ a \textit{range-r} potential is a potential such that $\bpsi(\tom)=\bpsi(\tom')$, 
if $\bom(t)=\bom'(t), 0 \leq t < r$. That is, $\bpsi$ depends only
on the $r$ first spiking patterns $\bom(0), \dots, \bom(r-1)$. Thus, a finite range potential
is necessarily regular.
Examples are $\omega_{i_1}(0)$ (range-$1$ potential), 
$\omega_{i_1}(0)\omega_{i_2}(k)$ (range-$k$ potential),
$\omega_{i_1}(0)\omega_{i_2}(k)\omega_{i_3}(l)$ (range-$\max(k,l)$ potential),
etc, \dots. The potential $\phi_{ij}$ introduced in section \ref{spikerep} is a range-$T_s$ potential.
More generally, all potential introduced in the paper have finite range.

A specific example of potential, used in this paper, corresponds to integrating the grammar into a potential $\Phi$
such that $\Phi(\tom)=0$ is $\tom$ is allowed, and $\Phi(\tom)=-\infty$ otherwise. 
In this case, however, the potential is not continuous anymore but only upper semi-continuous.

\sssu{Equilibrium states}

Let $\nu$ be a $\sigma$-invariant measure and $\Spn$ the set of admissible cylinders of lenght $n$. Call:

%\beq
$$h^{(n)}\left[\nu\right]=-\sum_{\tom \in \Spn} \nu\left(\Con\right)\log\nu\left(\Con\right).$$
%\eeq

Then,
%
%\beq\label{hKS}
$$h\left[\nu\right]=\lim_{n \to \infty} \frac{h^{(n)}\left[\nu\right]}{n}$$
%\eeq
%
is the entropy of  $\nu$. Let
$m^{(inv)}$ be the set of invariant measures for $\sigma$, then the \textit{pressure} of a potential $\bpsi$ is defined by:
\beq \label{supmu}
\pres=\sup_{\nu \in m^{(inv)}} \left( h\left[\nu\right]+\nu\left[\bpsi\right] \right).
\eeq
This supremum is attained- not necessarily at a unique measure-  where:
$$\pres=h\left[\mpg\right]+\mpg\left[\bpsi\right].$$
$\mpg$ is called an \textit{equilibrium state}, as it maximises some version of the (neg) free energy.
Equivalently, a measure $\nu$ is an equilibrium state for a potential $\bpsi 
\in C(\Sigma)$ if and only if, for all potential $\Beta \in C(\Sigma)$:
\beq \label{Gibbstangent}
P\left[\bpsi+\Beta \right] -P\left[\bpsi  \right] \geq \nu\left[\bpsi \right]
\eeq 
This means that $\nu$ is a tangent functional for $P$ at $\bpsi$. In particular $\bpsi$
has a unique equilibrium state if and only if $P$ is differentiable at $\bpsi$, i.e.
$\lim_{t \to 0} \frac{1}{t}\left(P\left[\bpsi+t\bphi\right]-P\left[\bpsi\right]=\nu(\bpsi) \right)$,
for all $\bphi \in C(\Sigma)$. 

\sssu{Properties of the topological pressure.} We review some important properties of the topological pressure used in
this paper. Basically, the topological pressure has the same properties of thermodynamic potentials like free energy.
First,
\beq\label{PvsZn}
\pres=\limsup_{n \to \infty} \frac{1}{n} \log(Z_n(\bpsi)),
\eeq
where $Z_n(\psi)$ is a partition function:
\beq\label{Zn}
Z_n(\bpsi)=\sum_{\tom \in \Spn} e^{\Snpo},
\eeq 
and 
\beq\label{Snp}
\Snpo = \sum_{t=0}^{n-1} \bpsi(\sit \tom).
\eeq

Also, the topological pressure is a convex functional of
the potentials i.e.

\beq\label{Convex}
P\left[\alpha \bpsi_1 + (1-\alpha)\bpsi_2\right] \leq  \alpha P\left[\bpsi_1\right] + (1-\alpha) P\left[\bpsi_2\right] ; \quad \alpha \in [0,1].
\eeq

Like the free energy, the topological pressure is a generating functional (when it is differentiable).
For example,

\beq \label{dP}
\left.\frac{\partial  P\left[\bpsi +\alpha \bphi\right]}{\partial \alpha}\right|_{\alpha=0}=\mpg[\bphi],
\eeq

\nid
 and:

\beq \label{FDT}
\left.\frac{\partial^2  P\left[\bpsi +\alpha_1 \bphi_1 + \alpha_2 \bphi_2\right]}{\partial \alpha_1\partial \alpha_2}\right|_{\alpha_1=\alpha_2=0}= 
C_{\bphi_1\bphi_2}(0) + 2\sum_{t=0}^{+\infty} C_{\bphi_1\bphi_2}(t),
\eeq

\nid where:

%\beq\label{Cov}
$$C_{\bphi_1\bphi_2}(t)=\mpg\left(\bphi_1 \circ \sit \bphi_2 \right) - \mpg(\bphi_1)\mpg(\bphi_2),$$
%\eeq

\nid is the correlation function of $\phi_1,\phi_2$ at time $t$. 

\sssu{Gibbs measure.}
 Assume that the transition matrix of the shift $\sigma$ is primitive and that $\bpsi$
is regular.
Then, there is a unique ergodic measure $\mpg$, called a Gibbs measure, for which
one can find some constants $c_1,c_2$ with $0 < c_1 \leq 1 \leq c_2$ such
that for all $n \geq 1$ and for all $\tom \in \Sigma$:
\beq\label{dGibbs}
c_1 \leq \frac{\mpg\left(\tom \in \Con\right)}{\exp(-n\pres+\Snpo)} \leq c_2.
\eeq
\nid where $\Snpo$ is given by (\ref{Snp}).  $\mpg$ is also the unique equilibrium state.

\sssu{Ruelle-Perron-Frobenius operator.}

 The Ruelle-Perron-Frobenius (RPF) operator for the potential $\bpsi$, denoted by $L_{\bpsi}$, 
acts on functions $g \in C(\Sigma)$ as $L_{\bpsi} g(\tom)=\sum_{\tom',\, \tom=\sigma\tom'} e^{\bpsi(\tom')}g(\tom')$.
This operator has a unique maximal eigenvalue $s=e^{P\left[\bpsi \right]}$
 associated to a right eigenfunction $b_{\bpsi}$ and a left eigenfunction $\rho_{\bpsi}$ (probability measure)
such that $L_{\bpsi} b_{\bpsi}(\tom)=sb_{\bpsi}(\tom)$ and $\int L_{\bpsi} v d\rho_{\bpsi}=s\int v d\rho_{\bpsi}$, for all $v \in C(\Sigma)$.
The remaining part of the spectrum is located in a disk in the complex plane, of radius strictly lower than $s$.
Finally, for all   $v \in C(\Sigma)$ 

%\beq \label{ConvVect}
$$\frac{1}{s^n} L^n_{\bpsi} v \to b_{\bpsi}\int v d\rho_{\bpsi}.$$
%\eeq

\nid The Gibbs measure is $\nu_{\bpsi}=b_{\bpsi}\rho_{\bpsi}$.
 
Given $\bpsi$ and knowing its right eigenfunction $b_{\bpsi}$ and pressure $P\left[\bpsi \right]$ one can construct a new potential:

\beq
\bPsi(\tom)=\bpsi(\tom)-\log(b_{\bpsi}(\sigma \tom))+\log(b_{\bpsi}(\tom)) -P\left[\bpsi \right],
\eeq

\nid such that $L_{\bPsi} \textbf{1}=1$, where $\textbf{1}$ is the constant function $\textbf{1}(\tom)=1$.
Such a potential is called ``normalised''. Its pressure is zero.

If $\bpsi$ has a finite range-$r$ then the RPF operator reduces to the transition matrix of a $(r+1)$-step
Markov chain.
Thus, $b_{\bpsi}$ and $\rho_{\bpsi}$ can be easily determined.
The Gibbs measure is none other than the invariant mesure of this chain.
This 
can be used to generate a raster plot distributed according to a Gibbs distribution with a potential $\bpsi$.
Moreover, for $\nu_{\bpsi}$-almost-every raster plot $\tom$:

\beq
\lim_{T\to +\infty} \frac{\pTo\left[j \bom(1) \dots \bom(r-1) \right]}{\pTo\left[\bom(1) \dots \bom(r) \right]}=e^{\bpsi(j \bom(1) \dots \bom(r-1))} 
\eeq

This allows to obtain of $\bpsi$ numerically \cite{chazottes-etal:98}.

\sssu{ Kullack-Leibler divergence.}

There is a last important property. Let $\mu$ be an invariant measure and $\mpg$ a Gibbs measure with a potential
$\bpsi$, both defined on the same set of sequences  $\Sigma$.
Let 
\beq\label{HKL}
d(\mu,\mpg)=\limsup_{n \to \infty} \frac{1}{n}\sum_{\Con} 
\mu\left(\Con\right)
\log\left[\frac{\mu\left(\Con\right)}{\mpg\left(\Con\right)} \right].
\eeq
be the relative entropy (or Kullack-Leibler divergence) between $\mu$ and $\nu$.
 Then,
\beq\label{dKLpres}
d\left(\mu,\mpg \right) = \pres - \mu(\bpsi) - h(\mu).
\eeq
If $\mu=\nu$, $d(\mu,\nu)=0$. The converse may not be true if
 the potential is not regular.

If a raster $\tom$ is typical for the Gibbs measure $\mpg$ then one expects
that $\pTo$ becomes closer  to $\mpg$ than any other Gibbs measure (for another
potential $\bpsi'$) as $T$ growths. This provides a criterion to compare two
Gibbs measures (and to discrimate between several statistical models).
Indeed, the following theorem holds \cite{chazottes-etal:98}.

\bth\label{ThKL}
For any pair of distinct\footnote{Non cohomologous.} regular potentials $\bphi,\bpsi$,
there exists an integer $N \equiv N(\bphi,\bpsi,\tom)$ such that, for all $T \geq N$,
\beq
d\left(\pTo,\mpg \right) < d\left(\pTo,\nu_{\bphi} \right)
\eeq
for $\mpg$-almost every $\tom$.
\enth

This says that $d\left(\pTo,\mpg \right)$ becomes, for sufficiently large $T$, smaller
than the Kullback-Leibler divergence between $\pTo$ and any other Gibbs measure.

\pagebreak

\bibliographystyle{apacite}
%\bibliographystyle{abbrv}

% Ici 1 est a decommente selon ou on est
\bibliography{odyssee,biblio}

\begin{thebibliography}{}

\bibitem[\protect\citeauthoryear{%
Adrian%
\ \BBA{} Zotterman%
}{%
Adrian%
\ \BBA{} Zotterman%
}{%
{\protect\APACyear{1926}}%
}]{%
adrian-zotterman:1926}%
\APACinsertmetastar{%
adrian-zotterman:1926}%
Adrian, E.%
\BCBT{}\ \BBA{} Zotterman, Y.%
%
\unskip\
\newblock
\APACrefYearMonthDay{1926}{}{}.
\newblock
\BBOQ{}\APACrefatitle{The impulses produced by sensory nerve endings: Part II:
  The response of a single end organ}{The impulses produced by sensory nerve
  endings: Part ii: The response of a single end organ}.\BBCQ{}
\newblock
\APACjournalVolNumPages{J Physiol (Lond.)}{61}{}{151-71}.
\PrintBackRefs{\CurrentBib}

\bibitem[\protect\citeauthoryear{%
Amit%
}{%
Amit%
}{%
{\protect\APACyear{1989}}%
}]{%
amit:89}%
\APACinsertmetastar{%
amit:89}%
Amit, D\BPBI J.%
%
\unskip\
\newblock
\APACrefYear{1989}.
\newblock
\APACrefbtitle{Modeling brain function---the world of attractor neural
  networks}{Modeling brain function---the world of attractor neural networks}.
\newblock
\APACaddressPublisher{New York, NY, USA}{Cambridge University Press}.
\newblock
 \begin{APACrefURL} \url{http://portal.acm.org/citation.cfm?id=77051}
  \end{APACrefURL}
\PrintBackRefs{\CurrentBib}

\bibitem[\protect\citeauthoryear{%
Arabzadeh%
, Panzeri%
\BCBL{}\ \BBA{} Diamond%
}{%
Arabzadeh%
\ \protect\BOthers{.}}{%
{\protect\APACyear{2006}}%
}]{%
arabzadeh-et-al:06}%
\APACinsertmetastar{%
arabzadeh-et-al:06}%
Arabzadeh, E.%
, Panzeri, S.%
\BCBL{}\ \BBA{} Diamond, M.%
%
\unskip\
\newblock
\APACrefYearMonthDay{2006}{}{}.
\newblock
\BBOQ{}\APACrefatitle{Deciphering the Spike Train of a Sensory Neuron: Counts
  and Temporal Patterns in the Rat Whisker Pathway}{Deciphering the spike train
  of a sensory neuron: Counts and temporal patterns in the rat whisker
  pathway}.\BBCQ{}
\newblock
\APACjournalVolNumPages{The Journal of Neuroscience}{26}{36}{9216-9226}.
\PrintBackRefs{\CurrentBib}

\bibitem[\protect\citeauthoryear{%
Artola%
, Br\"ocher%
\BCBL{}\ \BBA{} Singer%
}{%
Artola%
\ \protect\BOthers{.}}{%
{\protect\APACyear{1990}}%
}]{%
artola-etal:90}%
\APACinsertmetastar{%
artola-etal:90}%
Artola, A.%
, Br\"ocher, S.%
\BCBL{}\ \BBA{} Singer, W.%
%
\unskip\
\newblock
\APACrefYearMonthDay{1990}{}{}.
\newblock
\BBOQ{}\APACrefatitle{Different voltage-dependent thresholds for inducing
  long-term depression and long-term potentiation in slices of rat visual
  cortex.}{Different voltage-dependent thresholds for inducing long-term
  depression and long-term potentiation in slices of rat visual cortex.}\BBCQ{}
\newblock
\APACjournalVolNumPages{Nature}{347}{6288}{69--72}.
\PrintBackRefs{\CurrentBib}

\bibitem[\protect\citeauthoryear{%
Barbieri%
\ \protect\BOthers{.}}{%
Barbieri%
\ \protect\BOthers{.}}{%
{\protect\APACyear{2004}}%
}]{%
barbieri-et-al:04}%
\APACinsertmetastar{%
barbieri-et-al:04}%
Barbieri, R.%
, Frank, L\BPBI M.%
, Nguyen, D\BPBI P.%
, Quirk, M\BPBI C.%
, Wilson, M\BPBI A.%
\BCBL{}\ \BBA{} Brown, E\BPBI N.%
%
\unskip\
\newblock
\APACrefYearMonthDay{2004}{}{}.
\newblock
\BBOQ{}\APACrefatitle{Dynamic Analyses of Information Encoding in Neural
  Ensembles}{Dynamic analyses of information encoding in neural
  ensembles}.\BBCQ{}
\newblock
\APACjournalVolNumPages{Neural Computation}{16}{}{277-307}.
\PrintBackRefs{\CurrentBib}

\bibitem[\protect\citeauthoryear{%
Beck%
\ \BBA{} Schloegl%
}{%
Beck%
\ \BBA{} Schloegl%
}{%
{\protect\APACyear{1995}}%
}]{%
beck-schloegl:95}%
\APACinsertmetastar{%
beck-schloegl:95}%
Beck, C.%
\BCBT{}\ \BBA{} Schloegl, F.%
%
\unskip\
\newblock
\APACrefYear{1995}.
\newblock
\APACrefbtitle{Thermodynamics of Chaotic Systems: An
  Introduction}{Thermodynamics of chaotic systems: An introduction}.
\newblock
\APACaddressPublisher{Cambridge}{Cambridge University Press}.
\PrintBackRefs{\CurrentBib}

\bibitem[\protect\citeauthoryear{%
Bi%
\ \BBA{} Poo%
}{%
Bi%
\ \BBA{} Poo%
}{%
{\protect\APACyear{2001}}%
}]{%
bi-poo:01}%
\APACinsertmetastar{%
bi-poo:01}%
Bi, G.%
\BCBT{}\ \BBA{} Poo, M.%
%
\unskip\
\newblock
\APACrefYearMonthDay{2001}{}{}.
\newblock
\BBOQ{}\APACrefatitle{Synaptic Modification by Correlated Activity: Hebb's
  Postulate Revisited}{Synaptic modification by correlated activity: Hebb's
  postulate revisited}.\BBCQ{}
\newblock
\APACjournalVolNumPages{Annual Review of Neuroscience}{24}{}{139--166}.
\PrintBackRefs{\CurrentBib}

\bibitem[\protect\citeauthoryear{%
Bienenstock%
, Cooper%
\BCBL{}\ \BBA{} Munroe%
}{%
Bienenstock%
\ \protect\BOthers{.}}{%
{\protect\APACyear{1982}}%
}]{%
bienenstock-etal:82}%
\APACinsertmetastar{%
bienenstock-etal:82}%
Bienenstock, E\BPBI L.%
, Cooper, L.%
\BCBL{}\ \BBA{} Munroe, P.%
%
\unskip\
\newblock
\APACrefYearMonthDay{1982}{}{}.
\newblock
\BBOQ{}\APACrefatitle{Theory for the development of neuron selectivity:
  orientation specificity and binocular interaction in visual cortex}{Theory
  for the development of neuron selectivity: orientation specificity and
  binocular interaction in visual cortex}.\BBCQ{}
\newblock
\APACjournalVolNumPages{The Journal of Neuroscience}{2}{1}{32–-48}.
\PrintBackRefs{\CurrentBib}

\bibitem[\protect\citeauthoryear{%
Blanchard%
, Cessac%
\BCBL{}\ \BBA{} Krueger%
}{%
Blanchard%
\ \protect\BOthers{.}}{%
{\protect\APACyear{2000}}%
}]{%
blanchard-cessac-etal:00}%
\APACinsertmetastar{%
blanchard-cessac-etal:00}%
Blanchard, P.%
, Cessac, B.%
\BCBL{}\ \BBA{} Krueger, T.%
%
\unskip\
\newblock
\APACrefYearMonthDay{2000}{}{}.
\newblock
\BBOQ{}\APACrefatitle{What can one learn about Self-Organized Criticality from
  Dynamical System theory ?}{What can one learn about self-organized
  criticality from dynamical system theory ?}\BBCQ{}
\newblock
\APACjournalVolNumPages{Journal of Statistical Physics}{98}{}{375--404}.
\PrintBackRefs{\CurrentBib}

\bibitem[\protect\citeauthoryear{%
Bliss%
\ \BBA{} Gardner-Medwin%
}{%
Bliss%
\ \BBA{} Gardner-Medwin%
}{%
{\protect\APACyear{1973}}%
}]{%
bliss-gardner:73}%
\APACinsertmetastar{%
bliss-gardner:73}%
Bliss, T.%
\BCBT{}\ \BBA{} Gardner-Medwin, A.%
%
\unskip\
\newblock
\APACrefYearMonthDay{1973}{}{}.
\newblock
\BBOQ{}\APACrefatitle{Long-lasting potentiation of synaptic transmission in the
  Dentate Area of the unanaesthetised rabbit following stimulation of the
  perforant path.}{Long-lasting potentiation of synaptic transmission in the
  dentate area of the unanaesthetised rabbit following stimulation of the
  perforant path.}\BBCQ{}
\newblock
\APACjournalVolNumPages{J Physiol}{232}{}{357-374}.
\PrintBackRefs{\CurrentBib}

\bibitem[\protect\citeauthoryear{%
Bohte%
\ \BBA{} Mozer%
}{%
Bohte%
\ \BBA{} Mozer%
}{%
{\protect\APACyear{2007}}%
}]{%
bohte-mozer:07}%
\APACinsertmetastar{%
bohte-mozer:07}%
Bohte, S\BPBI M.%
\BCBT{}\ \BBA{} Mozer, M\BPBI C.%
%
\unskip\
\newblock
\APACrefYearMonthDay{2007}{}{}.
\newblock
\BBOQ{}\APACrefatitle{Reducing the Variability of Neural Responses: A
  Computational Theory of Spike-Timing-Dependent Plasticity.}{Reducing the
  variability of neural responses: A computational theory of
  spike-timing-dependent plasticity.}\BBCQ{}
\newblock
\APACjournalVolNumPages{Neural Computation}{19}{2}{371--403}.
\PrintBackRefs{\CurrentBib}

\bibitem[\protect\citeauthoryear{%
Bowen%
}{%
Bowen%
}{%
{\protect\APACyear{1975}}%
}]{%
bowen:75}%
\APACinsertmetastar{%
bowen:75}%
Bowen, R.%
%
\unskip\
\newblock
\APACrefYear{1975}.
\newblock
\APACrefbtitle{Equilibrium states and the ergodic theory of Anosov
  diffeomorphisms}{Equilibrium states and the ergodic theory of anosov
  diffeomorphisms}\ (\BVOL~470).
\newblock
\APACaddressPublisher{New York}{Springer-Verlag}.
\PrintBackRefs{\CurrentBib}

\bibitem[\protect\citeauthoryear{%
Bowen%
}{%
Bowen%
}{%
{\protect\APACyear{2008}}%
}]{%
bowen:98}%
\APACinsertmetastar{%
bowen:98}%
Bowen, R.%
%
\unskip\
\newblock
\APACrefYear{2008}.
\newblock
\APACrefbtitle{Equilibrium states and the ergodic theory of Anosov
  diffeomorphisms. Second revised version.}{Equilibrium states and the ergodic
  theory of anosov diffeomorphisms. second revised version.}
\newblock
\APACaddressPublisher{}{Springer-Verlag}.
\PrintBackRefs{\CurrentBib}

\bibitem[\protect\citeauthoryear{%
Brette%
\ \BBA{} Gerstner%
}{%
Brette%
\ \BBA{} Gerstner%
}{%
{\protect\APACyear{2005}}%
}]{%
brette-gerstner:05}%
\APACinsertmetastar{%
brette-gerstner:05}%
Brette, R.%
\BCBT{}\ \BBA{} Gerstner, W.%
%
\unskip\
\newblock
\APACrefYearMonthDay{2005}{}{}.
\newblock
\BBOQ{}\APACrefatitle{Adaptive exponential integrate-and-fire model as an
  effective description of neuronal activity}{Adaptive exponential
  integrate-and-fire model as an effective description of neuronal
  activity}.\BBCQ{}
\newblock
\APACjournalVolNumPages{Journal of Neurophysiology}{94}{}{3637--3642}.
\PrintBackRefs{\CurrentBib}

\bibitem[\protect\citeauthoryear{%
Cessac%
}{%
Cessac%
}{%
{\protect\APACyear{2007}}%
}]{%
cessac:07}%
\APACinsertmetastar{%
cessac:07}%
Cessac, B.%
%
\unskip\
\newblock
\APACrefYearMonthDay{2007}{}{}.
\newblock
\BBOQ{}\APACrefatitle{Does the complex susceptibility of the H{\'e}non map have
  a pole in the upper-half plane ? A numerical investigation.}{Does the complex
  susceptibility of the h{\'e}non map have a pole in the upper-half plane ? a
  numerical investigation.}\BBCQ{}
\newblock
\APACjournalVolNumPages{Nonlinearity}{20}{}{2883--2895}.
\PrintBackRefs{\CurrentBib}

\bibitem[\protect\citeauthoryear{%
Cessac%
}{%
Cessac%
}{%
{\protect\APACyear{2008}}%
}]{%
cessac:08}%
\APACinsertmetastar{%
cessac:08}%
Cessac, B.%
%
\unskip\
\newblock
\APACrefYearMonthDay{2008}{}{}.
\newblock
\BBOQ{}\APACrefatitle{A discrete time neural network model with spiking
  neurons. Rigorous results on the spontaneous dynamics}{A discrete time neural
  network model with spiking neurons. rigorous results on the spontaneous
  dynamics}.\BBCQ{}
\newblock
\APACjournalVolNumPages{J. Math. Biol.}{56}{3}{311-345}.
\PrintBackRefs{\CurrentBib}

\bibitem[\protect\citeauthoryear{%
Cessac%
, Blanchard%
, Krueger%
\BCBL{}\ \BBA{} Meunier%
}{%
Cessac%
\ \protect\BOthers{.}}{%
{\protect\APACyear{2004}}%
}]{%
cessac-blanchard-etal:04}%
\APACinsertmetastar{%
cessac-blanchard-etal:04}%
Cessac, B.%
, Blanchard, P.%
, Krueger, T.%
\BCBL{}\ \BBA{} Meunier, J.%
%
\unskip\
\newblock
\APACrefYearMonthDay{2004}{}{}.
\newblock
\BBOQ{}\APACrefatitle{Self-Organized Criticality and Thermodynamic
  formalism}{Self-organized criticality and thermodynamic formalism}.\BBCQ{}
\newblock
\APACjournalVolNumPages{Journal of Statistical Physics}{115}{516}{1283--1326}.
\PrintBackRefs{\CurrentBib}

\bibitem[\protect\citeauthoryear{%
Cessac%
\ \BBA{} Samuelides%
}{%
Cessac%
\ \BBA{} Samuelides%
}{%
{\protect\APACyear{2007}}%
}]{%
cessac-samuelides:07}%
\APACinsertmetastar{%
cessac-samuelides:07}%
Cessac, B.%
\BCBT{}\ \BBA{} Samuelides, M.%
%
\unskip\
\newblock
\APACrefYearMonthDay{2007}{}{}.
\newblock
\BBOQ{}\APACrefatitle{From neuron to neural networks dynamics.}{From neuron to
  neural networks dynamics.}\BBCQ{}
\newblock
\APACjournalVolNumPages{EPJ Special topics: Topics in Dynamical Neural
  Networks}{142}{1}{7--88}.
\PrintBackRefs{\CurrentBib}

\bibitem[\protect\citeauthoryear{%
Cessac%
, Vasquez%
\BCBL{}\ \BBA{} Vi\'eville%
}{%
Cessac%
\ \protect\BOthers{.}}{%
{\protect\APACyear{2009}}%
}]{%
cessac-vasquez-etal:09}%
\APACinsertmetastar{%
cessac-vasquez-etal:09}%
Cessac, B.%
, Vasquez, J.%
\BCBL{}\ \BBA{} Vi\'eville, T.%
%
\unskip\
\newblock
\APACrefYearMonthDay{2009}{}{}.
\newblock
\BBOQ{}\APACrefatitle{Parametric estimation of spike train
  statistics}{Parametric estimation of spike train statistics}.\BBCQ{}
\newblock
\APACjournalVolNumPages{submitted}{}{}{}.
\PrintBackRefs{\CurrentBib}

\bibitem[\protect\citeauthoryear{%
Cessac%
\ \BBA{} Vi{\'e}ville%
}{%
Cessac%
\ \BBA{} Vi{\'e}ville%
}{%
{\protect\APACyear{2008}}%
}]{%
cessac-vieville:08}%
\APACinsertmetastar{%
cessac-vieville:08}%
Cessac, B.%
\BCBT{}\ \BBA{} Vi{\'e}ville, T.%
%
\unskip\
\newblock
\APACrefYearMonthDay{2008}{jul}{}.
\newblock
\BBOQ{}\APACrefatitle{On Dynamics of Integrate-and-Fire Neural Networks with
  Adaptive Conductances}{On dynamics of integrate-and-fire neural networks with
  adaptive conductances}.\BBCQ{}
\newblock
\APACjournalVolNumPages{Frontiers in neuroscience}{2}{2}{}.
\PrintBackRefs{\CurrentBib}

\bibitem[\protect\citeauthoryear{%
Chazottes%
}{%
Chazottes%
}{%
{\protect\APACyear{1999}}%
}]{%
chazottes:99}%
\APACinsertmetastar{%
chazottes:99}%
Chazottes, J.%
%
\unskip\
\newblock
\APACrefYear{1999}.
\newblock
\APACrefbtitle{Entropie Relative, Dynamique Symbolique et Turbulence}{Entropie
  relative, dynamique symbolique et turbulence}.
\newblock
\BUPhD, Universit\'e de Provence - Aix Marseille I.
\PrintBackRefs{\CurrentBib}

\bibitem[\protect\citeauthoryear{%
Chazottes%
, Floriani%
\BCBL{}\ \BBA{} Lima%
}{%
Chazottes%
\ \protect\BOthers{.}}{%
{\protect\APACyear{1998}}%
}]{%
chazottes-etal:98}%
\APACinsertmetastar{%
chazottes-etal:98}%
Chazottes, J.%
, Floriani, E.%
\BCBL{}\ \BBA{} Lima, R.%
%
\unskip\
\newblock
\APACrefYearMonthDay{1998}{}{}.
\newblock
\BBOQ{}\APACrefatitle{Relative entropy and identification of Gibbs measures in
  dynamical systems}{Relative entropy and identification of gibbs measures in
  dynamical systems}.\BBCQ{}
\newblock
\APACjournalVolNumPages{J. Statist. Phys.}{90}{3-4}{697-725}.
\PrintBackRefs{\CurrentBib}

\bibitem[\protect\citeauthoryear{%
Chazottes%
\ \BBA{} Keller%
}{%
Chazottes%
\ \BBA{} Keller%
}{%
{\protect\APACyear{2009}}%
}]{%
chazottes-keller:09}%
\APACinsertmetastar{%
chazottes-keller:09}%
Chazottes, J.%
\BCBT{}\ \BBA{} Keller, G.%
%
\unskip\
\newblock
\APACrefYearMonthDay{2009}{}{}.
\newblock
\BBOQ{}\APACrefatitle{Pressure and Equilibrium States in Ergodic
  Theory}{Pressure and equilibrium states in ergodic theory}.\BBCQ{}
\newblock
\BIn{} E.~of Complexity\ \BBA{} S.~Science\ (\BEDS), (\BCHAP\ Ergodic Theory).
\newblock
\APACaddressPublisher{}{Springer}.
\PrintBackRefs{\CurrentBib}

\bibitem[\protect\citeauthoryear{%
Chechik%
}{%
Chechik%
}{%
{\protect\APACyear{2003}}%
}]{%
chechik:03}%
\APACinsertmetastar{%
chechik:03}%
Chechik, G.%
%
\unskip\
\newblock
\APACrefYearMonthDay{2003}{}{}.
\newblock
\BBOQ{}\APACrefatitle{Spike-Timing-Dependent Plasticity and Relevant Mutual
  Information Maximization}{Spike-timing-dependent plasticity and relevant
  mutual information maximization}.\BBCQ{}
\newblock
\APACjournalVolNumPages{Neural Computation}{15}{7}{1481--1510}.
\PrintBackRefs{\CurrentBib}

\bibitem[\protect\citeauthoryear{%
Collet%
, Galves%
\BCBL{}\ \BBA{} Lopez%
}{%
Collet%
\ \protect\BOthers{.}}{%
{\protect\APACyear{1995}}%
}]{%
collet-et-al:95}%
\APACinsertmetastar{%
collet-et-al:95}%
Collet, P.%
, Galves, A.%
\BCBL{}\ \BBA{} Lopez, A.%
%
\unskip\
\newblock
\APACrefYearMonthDay{1995}{}{}.
\newblock
\BBOQ{}\APACrefatitle{Maximum likelihood and minimum entropy identification of
  grammars}{Maximum likelihood and minimum entropy identification of
  grammars}.\BBCQ{}
\newblock
\APACjournalVolNumPages{Random and Computational Dynamics}{3}{3/4}{241-250}.
\PrintBackRefs{\CurrentBib}

\bibitem[\protect\citeauthoryear{%
Comets%
}{%
Comets%
}{%
{\protect\APACyear{1997}}%
}]{%
comets:97}%
\APACinsertmetastar{%
comets:97}%
Comets, F.%
%
\unskip\
\newblock
\APACrefYearMonthDay{1997}{}{}.
\newblock
\BBOQ{}\APACrefatitle{Detecting phase transition for Gibbs measures}{Detecting
  phase transition for gibbs measures}.\BBCQ{}
\newblock
\APACjournalVolNumPages{Ann. Appl. Probab.}{7}{2}{545-563}.
\PrintBackRefs{\CurrentBib}

\bibitem[\protect\citeauthoryear{%
Cooper%
, Intrator%
, Blais%
\BCBL{}\ \BBA{} Shouval%
}{%
Cooper%
\ \protect\BOthers{.}}{%
{\protect\APACyear{2004}}%
}]{%
cooper-etal:04}%
\APACinsertmetastar{%
cooper-etal:04}%
Cooper, L.%
, Intrator, N.%
, Blais, B.%
\BCBL{}\ \BBA{} Shouval, H.%
%
\unskip\
\newblock
\APACrefYear{2004}.
\newblock
\APACrefbtitle{Theory of cortical plasticity}{Theory of cortical plasticity}.
\newblock
\APACaddressPublisher{}{World Scientific, Singapore}.
\PrintBackRefs{\CurrentBib}

\bibitem[\protect\citeauthoryear{%
Cronin%
}{%
Cronin%
}{%
{\protect\APACyear{1987}}%
}]{%
cronin:87}%
\APACinsertmetastar{%
cronin:87}%
Cronin, J.%
%
\unskip\
\newblock
\APACrefYear{1987}.
\newblock
\APACrefbtitle{Mathematical aspects of Hodgkin-Huxley theory}{Mathematical
  aspects of hodgkin-huxley theory}.
\newblock
\APACaddressPublisher{}{Cambridge University Press}.
\PrintBackRefs{\CurrentBib}

\bibitem[\protect\citeauthoryear{%
Dauc\'e%
, Quoy%
, Cessac%
, Doyon%
\BCBL{}\ \BBA{} Samuelides%
}{%
Dauc\'e%
\ \protect\BOthers{.}}{%
{\protect\APACyear{1998}}%
}]{%
dauce-etal:98}%
\APACinsertmetastar{%
dauce-etal:98}%
Dauc\'e, E.%
, Quoy, M.%
, Cessac, B.%
, Doyon, B.%
\BCBL{}\ \BBA{} Samuelides, M.%
%
\unskip\
\newblock
\APACrefYearMonthDay{1998}{}{}.
\newblock
\BBOQ{}\APACrefatitle{Self-Organization and dynamics reduction in recurrent
  networks: stimulus presentation and learning}{Self-organization and dynamics
  reduction in recurrent networks: stimulus presentation and learning}.\BBCQ{}
\newblock
\APACjournalVolNumPages{Neural Networks}{11}{}{521--33}.
\PrintBackRefs{\CurrentBib}

\bibitem[\protect\citeauthoryear{%
Dayan%
\ \BBA{} Abbott%
}{%
Dayan%
\ \BBA{} Abbott%
}{%
{\protect\APACyear{2001}}%
}]{%
dayan-abbott:01}%
\APACinsertmetastar{%
dayan-abbott:01}%
Dayan, P.%
\BCBT{}\ \BBA{} Abbott, L\BPBI F.%
%
\unskip\
\newblock
\APACrefYear{2001}.
\newblock
\APACrefbtitle{Theoretical Neuroscience : Computational and Mathematical
  Modeling of Neural Systems}{Theoretical neuroscience : Computational and
  mathematical modeling of neural systems}.
\newblock
\APACaddressPublisher{}{MIT Press}.
\PrintBackRefs{\CurrentBib}

\bibitem[\protect\citeauthoryear{%
Dayan%
\ \BBA{} Hausser%
}{%
Dayan%
\ \BBA{} Hausser%
}{%
{\protect\APACyear{2004}}%
}]{%
dayan-hausser:04}%
\APACinsertmetastar{%
dayan-hausser:04}%
Dayan, P.%
\BCBT{}\ \BBA{} Hausser, M.%
%
\unskip\
\newblock
\APACrefYear{2004}.
\newblock
\APACrefbtitle{Plasticity kernels and temporal statistics}{Plasticity kernels
  and temporal statistics}\ (\BVOL~16; S.~Thrun, L.~Saul\BCBL{}\ \BBA{}
  B.~Schoelkopf, \BEDS{}).
\newblock
\APACaddressPublisher{}{Cambridge MA: MIT Press}.
\PrintBackRefs{\CurrentBib}

\bibitem[\protect\citeauthoryear{%
Delorme%
, Perrinet%
\BCBL{}\ \BBA{} Thorpe%
}{%
Delorme%
\ \protect\BOthers{.}}{%
{\protect\APACyear{2001}}%
}]{%
delorme-et-al:01}%
\APACinsertmetastar{%
delorme-et-al:01}%
Delorme, A.%
, Perrinet, L.%
\BCBL{}\ \BBA{} Thorpe, S.%
%
\unskip\
\newblock
\APACrefYearMonthDay{2001}{}{}.
\newblock
\BBOQ{}\APACrefatitle{Networks of integrate-and-fire neurons using Rank Order
  Coding B: Spike timing dependent plasticity and emergence of orientation
  selectivity}{Networks of integrate-and-fire neurons using rank order coding
  b: Spike timing dependent plasticity and emergence of orientation
  selectivity}.\BBCQ{}
\newblock
\APACjournalVolNumPages{Neurocomputing}{38-40}{}{539-45}.
\PrintBackRefs{\CurrentBib}

\bibitem[\protect\citeauthoryear{%
Dudek%
\ \BBA{} Bear%
}{%
Dudek%
\ \BBA{} Bear%
}{%
{\protect\APACyear{1993}}%
}]{%
dudek-bear:93}%
\APACinsertmetastar{%
dudek-bear:93}%
Dudek, S.%
\BCBT{}\ \BBA{} Bear, M\BPBI F.%
%
\unskip\
\newblock
\APACrefYearMonthDay{1993}{}{}.
\newblock
\BBOQ{}\APACrefatitle{Bidirectional long-term modification of synaptic
  effectiveness in the adult and immature hippocampus.}{Bidirectional long-term
  modification of synaptic effectiveness in the adult and immature
  hippocampus.}\BBCQ{}
\newblock
\APACjournalVolNumPages{J Neurosci.}{13}{7}{2910--2918}.
\PrintBackRefs{\CurrentBib}

\bibitem[\protect\citeauthoryear{%
FitzHugh%
}{%
FitzHugh%
}{%
{\protect\APACyear{1955}}%
}]{%
fitzhugh:55}%
\APACinsertmetastar{%
fitzhugh:55}%
FitzHugh, R.%
%
\unskip\
\newblock
\APACrefYearMonthDay{1955}{}{}.
\newblock
\BBOQ{}\APACrefatitle{Mathematical models of threshold phenomena in the nerve
  membrane}{Mathematical models of threshold phenomena in the nerve
  membrane}.\BBCQ{}
\newblock
\APACjournalVolNumPages{Bull. Math. Biophysics}{17}{}{257--278}.
\PrintBackRefs{\CurrentBib}

\bibitem[\protect\citeauthoryear{%
FitzHugh%
}{%
FitzHugh%
}{%
{\protect\APACyear{1961}}%
}]{%
fitzhugh:61}%
\APACinsertmetastar{%
fitzhugh:61}%
FitzHugh, R.%
%
\unskip\
\newblock
\APACrefYearMonthDay{1961}{}{}.
\newblock
\BBOQ{}\APACrefatitle{Impulses and physiological states in models of nerve
  membrane}{Impulses and physiological states in models of nerve
  membrane}.\BBCQ{}
\newblock
\APACjournalVolNumPages{Biophys. J.}{1}{}{445-466}.
\PrintBackRefs{\CurrentBib}

\bibitem[\protect\citeauthoryear{%
Gao%
, Kontoyiannis%
\BCBL{}\ \BBA{} Bienenstock%
}{%
Gao%
\ \protect\BOthers{.}}{%
{\protect\APACyear{2008}}%
}]{%
gao-et-al:08}%
\APACinsertmetastar{%
gao-et-al:08}%
Gao, Y.%
, Kontoyiannis, I.%
\BCBL{}\ \BBA{} Bienenstock, E.%
%
\unskip\
\newblock
\APACrefYearMonthDay{2008}{}{}.
\newblock
\BBOQ{}\APACrefatitle{Estimating the Entropy of Binary Time Series:
  Methodology, Some Theory and a Simulation Study}{Estimating the entropy of
  binary time series: Methodology, some theory and a simulation study}.\BBCQ{}
\newblock
\APACjournalVolNumPages{Entropy}{10}{2}{71-99}.
\PrintBackRefs{\CurrentBib}

\bibitem[\protect\citeauthoryear{%
Georgeopoulos%
, Merchant%
, Naselaris%
\BCBL{}\ \BBA{} Amirikian%
}{%
Georgeopoulos%
\ \protect\BOthers{.}}{%
{\protect\APACyear{2007}}%
}]{%
georgeopoulos:07}%
\APACinsertmetastar{%
georgeopoulos:07}%
Georgeopoulos, A\BPBI P.%
, Merchant, H.%
, Naselaris, T.%
\BCBL{}\ \BBA{} Amirikian, B.%
%
\unskip\
\newblock
\APACrefYearMonthDay{2007}{}{}.
\newblock
\BBOQ{}\APACrefatitle{Mapping of the preferred direction in the motor
  cortex}{Mapping of the preferred direction in the motor cortex}.\BBCQ{}
\newblock
\APACjournalVolNumPages{PNAS}{104}{26}{11068-11072}.
\PrintBackRefs{\CurrentBib}

\bibitem[\protect\citeauthoryear{%
Georgopoulos%
, Kalaska%
, Caminiti%
\BCBL{}\ \BBA{} Massey%
}{%
Georgopoulos%
\ \protect\BOthers{.}}{%
{\protect\APACyear{1982}}%
}]{%
georgeopoulos:82}%
\APACinsertmetastar{%
georgeopoulos:82}%
Georgopoulos, A.%
, Kalaska, J.%
, Caminiti, R.%
\BCBL{}\ \BBA{} Massey, J.%
%
\unskip\
\newblock
\APACrefYearMonthDay{1982}{}{}.
\newblock
\BBOQ{}\APACrefatitle{On the relations between the direction of two-dimensional
  arm movements and cell discharge in primary motor cortex}{On the relations
  between the direction of two-dimensional arm movements and cell discharge in
  primary motor cortex}.\BBCQ{}
\newblock
\APACjournalVolNumPages{J Neurosci}{2}{1527-1537}{}.
\PrintBackRefs{\CurrentBib}

\bibitem[\protect\citeauthoryear{%
Gerstner%
\ \BBA{} Kistler%
}{%
Gerstner%
\ \BBA{} Kistler%
}{%
{\protect\APACyear{2002}}%
{\protect\APACexlab{{\protect\BCnt{2}}}}}]{%
gerstner-kistler:02b}%
\APACinsertmetastar{%
gerstner-kistler:02b}%
Gerstner, W.%
\BCBT{}\ \BBA{} Kistler, W.%
%
\unskip\
\newblock
\APACrefYear{2002{\protect\BCnt{2}}}.
\newblock
\APACrefbtitle{Spiking Neuron Models}{Spiking neuron models}.
\newblock
\APACaddressPublisher{}{Cambridge University Press}.
\PrintBackRefs{\CurrentBib}

\bibitem[\protect\citeauthoryear{%
Gerstner%
\ \BBA{} Kistler%
}{%
Gerstner%
\ \BBA{} Kistler%
}{%
{\protect\APACyear{2002}}%
{\protect\APACexlab{{\protect\BCnt{1}}}}}]{%
gerstner-kistler:02}%
\APACinsertmetastar{%
gerstner-kistler:02}%
Gerstner, W.%
\BCBT{}\ \BBA{} Kistler, W\BPBI M.%
%
\unskip\
\newblock
\APACrefYearMonthDay{2002{\protect\BCnt{1}}}{}{}.
\newblock
\BBOQ{}\APACrefatitle{Mathematical formulations of Hebbian
  learning.}{Mathematical formulations of hebbian learning.}\BBCQ{}
\newblock
\APACjournalVolNumPages{Biological Cybernetics}{87}{}{404--415}.
\PrintBackRefs{\CurrentBib}

\bibitem[\protect\citeauthoryear{%
Grammont%
\ \BBA{} Riehle%
}{%
Grammont%
\ \BBA{} Riehle%
}{%
{\protect\APACyear{1999}}%
}]{%
grammont-riehle:99}%
\APACinsertmetastar{%
grammont-riehle:99}%
Grammont, F.%
\BCBT{}\ \BBA{} Riehle, A.%
%
\unskip\
\newblock
\APACrefYearMonthDay{1999}{}{}.
\newblock
\BBOQ{}\APACrefatitle{Precise spike synchronization in monkey motor cortex
  involved in preparation for movement}{Precise spike synchronization in monkey
  motor cortex involved in preparation for movement}.\BBCQ{}
\newblock
\APACjournalVolNumPages{Exp. Brain Res.}{128}{}{118--122}.
\PrintBackRefs{\CurrentBib}

\bibitem[\protect\citeauthoryear{%
Grammont%
\ \BBA{} Riehle%
}{%
Grammont%
\ \BBA{} Riehle%
}{%
{\protect\APACyear{2003}}%
}]{%
grammont-riehle:03}%
\APACinsertmetastar{%
grammont-riehle:03}%
Grammont, F.%
\BCBT{}\ \BBA{} Riehle, A.%
%
\unskip\
\newblock
\APACrefYearMonthDay{2003}{}{}.
\newblock
\BBOQ{}\APACrefatitle{Spike synchronization and firing rate in a population of
  motor cortical neurons in relation to movement direction and reaction
  time}{Spike synchronization and firing rate in a population of motor cortical
  neurons in relation to movement direction and reaction time}.\BBCQ{}
\newblock
\APACjournalVolNumPages{Biol Cybern}{88}{}{360-373}.
\PrintBackRefs{\CurrentBib}

\bibitem[\protect\citeauthoryear{%
Guckenheimer%
\ \BBA{} Labouriau%
}{%
Guckenheimer%
\ \BBA{} Labouriau%
}{%
{\protect\APACyear{1993}}%
}]{%
guckenheimer-labouriau:93}%
\APACinsertmetastar{%
guckenheimer-labouriau:93}%
Guckenheimer, J.%
\BCBT{}\ \BBA{} Labouriau, I\BPBI S.%
%
\unskip\
\newblock
\APACrefYearMonthDay{1993}{}{}.
\newblock
\BBOQ{}\APACrefatitle{Bifurcation of the Hodgkin-Huxley equations: A new
  twist}{Bifurcation of the hodgkin-huxley equations: A new twist}.\BBCQ{}
\newblock
\APACjournalVolNumPages{Bull. Math. Biol.}{55}{}{937-952}.
\PrintBackRefs{\CurrentBib}

\bibitem[\protect\citeauthoryear{%
Hebb%
}{%
Hebb%
}{%
{\protect\APACyear{1949}}%
}]{%
hebb:49}%
\APACinsertmetastar{%
hebb:49}%
Hebb, D.%
%
\unskip\
\newblock
\APACrefYear{1949}.
\newblock
\APACrefbtitle{The organization of behavior: a neuropsychological theory.}{The
  organization of behavior: a neuropsychological theory.}
\newblock
\APACaddressPublisher{}{Wiley, NY}.
\PrintBackRefs{\CurrentBib}

\bibitem[\protect\citeauthoryear{%
Hirsch%
}{%
Hirsch%
}{%
{\protect\APACyear{1989}}%
}]{%
hirsch:89}%
\APACinsertmetastar{%
hirsch:89}%
Hirsch, M.%
%
\unskip\
\newblock
\APACrefYearMonthDay{1989}{}{}.
\newblock
\BBOQ{}\APACrefatitle{Convergent activation dynamics in continuous time
  networks}{Convergent activation dynamics in continuous time networks}.\BBCQ{}
\newblock
\APACjournalVolNumPages{Neur. Networks}{2}{}{331--349}.
\PrintBackRefs{\CurrentBib}

\bibitem[\protect\citeauthoryear{%
Hodgkin%
\ \BBA{} Huxley%
}{%
Hodgkin%
\ \BBA{} Huxley%
}{%
{\protect\APACyear{1952}}%
}]{%
hodgkin-huxley:52}%
\APACinsertmetastar{%
hodgkin-huxley:52}%
Hodgkin, A.%
\BCBT{}\ \BBA{} Huxley, A.%
%
\unskip\
\newblock
\APACrefYearMonthDay{1952}{}{}.
\newblock
\BBOQ{}\APACrefatitle{A quantitative description of membrane current and its
  application to conduction and excitation in nerve.}{A quantitative
  description of membrane current and its application to conduction and
  excitation in nerve.}\BBCQ{}
\newblock
\APACjournalVolNumPages{Journal of Physiology}{117}{}{500--544}.
\PrintBackRefs{\CurrentBib}

\bibitem[\protect\citeauthoryear{%
Izhikevich%
}{%
Izhikevich%
}{%
{\protect\APACyear{2003}}%
}]{%
izhikevich:03}%
\APACinsertmetastar{%
izhikevich:03}%
Izhikevich, E.%
%
\unskip\
\newblock
\APACrefYearMonthDay{2003}{}{}.
\newblock
\BBOQ{}\APACrefatitle{Simple Model of Spiking Neurons}{Simple model of spiking
  neurons}.\BBCQ{}
\newblock
\APACjournalVolNumPages{IEEE Transactions on Neural
  Networks}{14}{6}{1569--1572}.
\PrintBackRefs{\CurrentBib}

\bibitem[\protect\citeauthoryear{%
Izhikevich%
}{%
Izhikevich%
}{%
{\protect\APACyear{2004}}%
}]{%
izhikevich:04}%
\APACinsertmetastar{%
izhikevich:04}%
Izhikevich, E.%
%
\unskip\
\newblock
\APACrefYearMonthDay{2004}{September}{}.
\newblock
\BBOQ{}\APACrefatitle{Which model to use for cortical spiking neurons?}{Which
  model to use for cortical spiking neurons?}\BBCQ{}
\newblock
\APACjournalVolNumPages{IEEE Trans Neural Netw}{15}{5}{1063--1070}.
\PrintBackRefs{\CurrentBib}

\bibitem[\protect\citeauthoryear{%
Izhikevich%
\ \BBA{} Desai%
}{%
Izhikevich%
\ \BBA{} Desai%
}{%
{\protect\APACyear{2003}}%
}]{%
izhikevich-desai:03}%
\APACinsertmetastar{%
izhikevich-desai:03}%
Izhikevich, E.%
\BCBT{}\ \BBA{} Desai, N.%
%
\unskip\
\newblock
\APACrefYearMonthDay{2003}{}{}.
\newblock
\BBOQ{}\APACrefatitle{Relating STDP to BCM}{Relating stdp to bcm}.\BBCQ{}
\newblock
\APACjournalVolNumPages{Neural Computation}{15}{}{1511--1523}.
\newblock
 \begin{APACrefURL}
  \url{http://vesicle.nsi.edu/users/izhikevich/publications/bcm.htm}
  \end{APACrefURL}
\PrintBackRefs{\CurrentBib}

\bibitem[\protect\citeauthoryear{%
Jaynes%
}{%
Jaynes%
}{%
{\protect\APACyear{1957}}%
}]{%
jaynes:57}%
\APACinsertmetastar{%
jaynes:57}%
Jaynes, E.%
%
\unskip\
\newblock
\APACrefYearMonthDay{1957}{}{}.
\newblock
\BBOQ{}\APACrefatitle{Information theory and statistical mechanics}{Information
  theory and statistical mechanics}.\BBCQ{}
\newblock
\APACjournalVolNumPages{Phys. Rev.}{106}{620}{}.
\PrintBackRefs{\CurrentBib}

\bibitem[\protect\citeauthoryear{%
Ji%
}{%
Ji%
}{%
{\protect\APACyear{1989}}%
}]{%
ji:89}%
\APACinsertmetastar{%
ji:89}%
Ji, C.%
%
\unskip\
\newblock
\APACrefYearMonthDay{1989}{}{}.
\newblock
\BBOQ{}\APACrefatitle{Estimating functionals of one-dimensional Gibbs
  states}{Estimating functionals of one-dimensional gibbs states}.\BBCQ{}
\newblock
\APACjournalVolNumPages{Probab. Theory Related Fields}{82}{2}{155-175}.
\PrintBackRefs{\CurrentBib}

\bibitem[\protect\citeauthoryear{%
Johnson%
}{%
Johnson%
}{%
{\protect\APACyear{1980}}%
}]{%
johnson:80}%
\APACinsertmetastar{%
johnson:80}%
Johnson.%
%
\unskip\
\newblock
\APACrefYearMonthDay{1980}{}{}.
\newblock
\BBOQ{}\APACrefatitle{Sensory discrimination: neural processes preceding
  discrimination decision.}{Sensory discrimination: neural processes preceding
  discrimination decision.}\BBCQ{}
\newblock
\APACjournalVolNumPages{J Neurophysiol}{43}{6}{1793-1815}.
\PrintBackRefs{\CurrentBib}

\bibitem[\protect\citeauthoryear{%
D.~Johnson%
}{%
D.~Johnson%
}{%
{\protect\APACyear{2004}}%
}]{%
johnson:04}%
\APACinsertmetastar{%
johnson:04}%
Johnson, D.%
%
\unskip\
\newblock
\APACrefYearMonthDay{2004}{}{}.
\newblock
\BBOQ{}\APACrefatitle{Neural Population Structure and Consequences for Neural
  Coding}{Neural population structure and consequences for neural
  coding}.\BBCQ{}
\newblock
\APACjournalVolNumPages{Journal of Computational Neuroscience}{16}{1}{69-80}.
\PrintBackRefs{\CurrentBib}

\bibitem[\protect\citeauthoryear{%
Jolivet%
, Rauch%
, Lescher%
\BCBL{}\ \BBA{} Gerstner%
}{%
Jolivet%
\ \protect\BOthers{.}}{%
{\protect\APACyear{2006}}%
}]{%
jolivet-et-al:06}%
\APACinsertmetastar{%
jolivet-et-al:06}%
Jolivet, R.%
, Rauch, A.%
, Lescher, H\BHBI R.%
\BCBL{}\ \BBA{} Gerstner, W.%
%
\unskip\
\newblock
\APACrefYear{2006}.
\newblock
\APACrefbtitle{"Integrate-and-Fire models with adaptation are good
  enough"}{"integrate-and-fire models with adaptation are good enough"}.
\newblock
\APACaddressPublisher{}{MIT Press, Cambridge}.
\PrintBackRefs{\CurrentBib}

\bibitem[\protect\citeauthoryear{%
Kang%
\ \BBA{} Amari%
}{%
Kang%
\ \BBA{} Amari%
}{%
{\protect\APACyear{2008}}%
}]{%
kang-amari:08}%
\APACinsertmetastar{%
kang-amari:08}%
Kang, K.%
\BCBT{}\ \BBA{} Amari, S. ichi.%
%
\unskip\
\newblock
\APACrefYearMonthDay{2008}{}{}.
\newblock
\BBOQ{}\APACrefatitle{Discrimination with Spike Times and ISI
  Distributions}{Discrimination with spike times and isi distributions}.\BBCQ{}
\newblock
\APACjournalVolNumPages{Neural Computation}{20}{}{1411-1426}.
\PrintBackRefs{\CurrentBib}

\bibitem[\protect\citeauthoryear{%
Katok%
\ \BBA{} Hasselblatt%
}{%
Katok%
\ \BBA{} Hasselblatt%
}{%
{\protect\APACyear{1998}}%
}]{%
katok-hasselblatt:98}%
\APACinsertmetastar{%
katok-hasselblatt:98}%
Katok, A.%
\BCBT{}\ \BBA{} Hasselblatt, B.%
%
\unskip\
\newblock
\APACrefYear{1998}.
\newblock
\APACrefbtitle{Introduction to the modern theory of dynamical
  systems}{Introduction to the modern theory of dynamical systems}.
\newblock
\APACaddressPublisher{}{Kluwer}.
\PrintBackRefs{\CurrentBib}

\bibitem[\protect\citeauthoryear{%
Keller%
}{%
Keller%
}{%
{\protect\APACyear{1998}}%
}]{%
keller:98}%
\APACinsertmetastar{%
keller:98}%
Keller, G.%
%
\unskip\
\newblock
\APACrefYear{1998}.
\newblock
\APACrefbtitle{Equilibrium States in Ergodic Theory}{Equilibrium states in
  ergodic theory}.
\newblock
\APACaddressPublisher{}{Cambridge University Press}.
\PrintBackRefs{\CurrentBib}

\bibitem[\protect\citeauthoryear{%
Levy%
\ \BBA{} Stewart%
}{%
Levy%
\ \BBA{} Stewart%
}{%
{\protect\APACyear{1983}}%
}]{%
levy-stewart:83}%
\APACinsertmetastar{%
levy-stewart:83}%
Levy, W.%
\BCBT{}\ \BBA{} Stewart, D.%
%
\unskip\
\newblock
\APACrefYearMonthDay{1983}{}{}.
\newblock
\BBOQ{}\APACrefatitle{Temporal contiguity requirements for long-term
  associative potentiation/depression in the hippocampus.}{Temporal contiguity
  requirements for long-term associative potentiation/depression in the
  hippocampus.}\BBCQ{}
\newblock
\APACjournalVolNumPages{Neuroscience}{8}{4}{791-–797}.
\PrintBackRefs{\CurrentBib}

\bibitem[\protect\citeauthoryear{%
Malenka%
\ \BBA{} Nicoll%
}{%
Malenka%
\ \BBA{} Nicoll%
}{%
{\protect\APACyear{1999}}%
}]{%
malenka-nicoll:99}%
\APACinsertmetastar{%
malenka-nicoll:99}%
Malenka, R\BPBI C.%
\BCBT{}\ \BBA{} Nicoll, R\BPBI A.%
%
\unskip\
\newblock
\APACrefYearMonthDay{1999}{}{}.
\newblock
\BBOQ{}\APACrefatitle{Long-Term Potentiation - A Decade of Progress
  ?}{Long-term potentiation - a decade of progress ?}\BBCQ{}
\newblock
\APACjournalVolNumPages{Science}{285}{5435}{1870 - 1874}.
\PrintBackRefs{\CurrentBib}

\bibitem[\protect\citeauthoryear{%
Malsburg%
}{%
Malsburg%
}{%
{\protect\APACyear{1973}}%
}]{%
vondermalsburg:73}%
\APACinsertmetastar{%
vondermalsburg:73}%
Malsburg, C. von-der.%
%
\unskip\
\newblock
\APACrefYearMonthDay{1973}{}{}.
\newblock
\BBOQ{}\APACrefatitle{Self-organisation of orientation sensitive cells in the
  striate cortex}{Self-organisation of orientation sensitive cells in the
  striate cortex}.\BBCQ{}
\newblock
\APACjournalVolNumPages{Kybernetik}{14}{}{85--100}.
\PrintBackRefs{\CurrentBib}

\bibitem[\protect\citeauthoryear{%
Markram%
, L\"{u}bke%
, Frotscher%
\BCBL{}\ \BBA{} Sakmann%
}{%
Markram%
\ \protect\BOthers{.}}{%
{\protect\APACyear{1997}}%
}]{%
markram-etal:97}%
\APACinsertmetastar{%
markram-etal:97}%
Markram, H.%
, L\"{u}bke, J.%
, Frotscher, M.%
\BCBL{}\ \BBA{} Sakmann, B.%
%
\unskip\
\newblock
\APACrefYearMonthDay{1997}{}{}.
\newblock
\BBOQ{}\APACrefatitle{Regulation of synaptic efficacy by coincidence of
  postsynaptic AP and EPSP}{Regulation of synaptic efficacy by coincidence of
  postsynaptic ap and epsp}.\BBCQ{}
\newblock
\APACjournalVolNumPages{Science}{275}{213}{}.
\PrintBackRefs{\CurrentBib}

\bibitem[\protect\citeauthoryear{%
Meyer%
}{%
Meyer%
}{%
{\protect\APACyear{1980}}%
}]{%
meyer:80}%
\APACinsertmetastar{%
meyer:80}%
Meyer, D.%
%
\unskip\
\newblock
\APACrefYear{1980}.
\newblock
\APACrefbtitle{The Ruelle-Araki transfer operator in classical statistical
  mechanics}{The ruelle-araki transfer operator in classical statistical
  mechanics}\ (\BVOL~123; L\BPBI N.~in Physics, \BED{}).
\newblock
\APACaddressPublisher{}{Springer-Verlag}.
\PrintBackRefs{\CurrentBib}

\bibitem[\protect\citeauthoryear{%
Miller%
, Keller%
\BCBL{}\ \BBA{} Stryker%
}{%
Miller%
\ \protect\BOthers{.}}{%
{\protect\APACyear{1989}}%
}]{%
miller-etal:89}%
\APACinsertmetastar{%
miller-etal:89}%
Miller, K.%
, Keller, J.%
\BCBL{}\ \BBA{} Stryker, M.%
%
\unskip\
\newblock
\APACrefYearMonthDay{1989}{}{}.
\newblock
\BBOQ{}\APACrefatitle{Ocular dominance column development: analysis and
  simulation}{Ocular dominance column development: analysis and
  simulation}.\BBCQ{}
\newblock
\APACjournalVolNumPages{Science}{245}{4918}{605--615}.
\PrintBackRefs{\CurrentBib}

\bibitem[\protect\citeauthoryear{%
Nagumo%
, Arimoto%
\BCBL{}\ \BBA{} Yoshizawa%
}{%
Nagumo%
\ \protect\BOthers{.}}{%
{\protect\APACyear{1962}}%
}]{%
nagumo-etal:62}%
\APACinsertmetastar{%
nagumo-etal:62}%
Nagumo, J.%
, Arimoto, S.%
\BCBL{}\ \BBA{} Yoshizawa, S.%
%
\unskip\
\newblock
\APACrefYearMonthDay{1962}{}{}.
\newblock
\BBOQ{}\APACrefatitle{An active pulse transmission line simulating nerve
  axon}{An active pulse transmission line simulating nerve axon}.\BBCQ{}
\newblock
\APACjournalVolNumPages{Proc.IRE}{50}{}{2061--2070}.
\PrintBackRefs{\CurrentBib}

\bibitem[\protect\citeauthoryear{%
Nemenman%
, Lewen%
, Bialek%
\BCBL{}\ \BBA{} Steveninck%
}{%
Nemenman%
\ \protect\BOthers{.}}{%
{\protect\APACyear{2006}}%
}]{%
nemenman-et-al:06}%
\APACinsertmetastar{%
nemenman-et-al:06}%
Nemenman, I.%
, Lewen, G.%
, Bialek, W.%
\BCBL{}\ \BBA{} Steveninck, R. de~Ruyter~van.%
%
\unskip\
\newblock
\APACrefYearMonthDay{2006}{}{}.
\newblock
\BBOQ{}\APACrefatitle{Neural coding of a natural stimulus ensemble: Information
  at sub-millisecond resolution.}{Neural coding of a natural stimulus ensemble:
  Information at sub-millisecond resolution.}\BBCQ{}
\newblock
\APACjournalVolNumPages{PLoS Comp Bio}{4}{}{e1000025}.
\PrintBackRefs{\CurrentBib}

\bibitem[\protect\citeauthoryear{%
Nirenberg%
\ \BBA{} Latham%
}{%
Nirenberg%
\ \BBA{} Latham%
}{%
{\protect\APACyear{2003}}%
}]{%
nirenberg-latham:03}%
\APACinsertmetastar{%
nirenberg-latham:03}%
Nirenberg, S.%
\BCBT{}\ \BBA{} Latham, P.%
%
\unskip\
\newblock
\APACrefYearMonthDay{2003}{}{}.
\newblock
\BBOQ{}\APACrefatitle{Decoding neuronal spike trains: how important are
  correlations}{Decoding neuronal spike trains: how important are
  correlations}.\BBCQ{}
\newblock
\APACjournalVolNumPages{Proceeding of the Natural Academy of
  Science}{100}{12}{7348--7353}.
\PrintBackRefs{\CurrentBib}

\bibitem[\protect\citeauthoryear{%
Osbone%
, Palmer%
, Lisberger%
\BCBL{}\ \BBA{} Bialek%
}{%
Osbone%
\ \protect\BOthers{.}}{%
{\protect\APACyear{2008}}%
}]{%
osbone-et-al:08}%
\APACinsertmetastar{%
osbone-et-al:08}%
Osbone, L.%
, Palmer, S.%
, Lisberger, S.%
\BCBL{}\ \BBA{} Bialek, W.%
%
\unskip\
\newblock
\APACrefYearMonthDay{2008}{}{}.
\newblock
\BBOQ{}\APACrefatitle{Combinatorial coding in neural populations}{Combinatorial
  coding in neural populations}.\BBCQ{}
\newblock
\APACjournalVolNumPages{arXiv.org:0803.3837}{}{}{}.
\PrintBackRefs{\CurrentBib}

\bibitem[\protect\citeauthoryear{%
Parry%
\ \BBA{} Pollicott%
}{%
Parry%
\ \BBA{} Pollicott%
}{%
{\protect\APACyear{1990}}%
}]{%
parry-pollicott:90}%
\APACinsertmetastar{%
parry-pollicott:90}%
Parry, W.%
\BCBT{}\ \BBA{} Pollicott, M.%
%
\unskip\
\newblock
\APACrefYear{1990}.
\newblock
\APACrefbtitle{Zeta functions and the periodic orbit structure of hyperbolic
  dynamics}{Zeta functions and the periodic orbit structure of hyperbolic
  dynamics}\ (\BVOLS\ 187--188).
\newblock
\APACaddressPublisher{}{Asterisque}.
\PrintBackRefs{\CurrentBib}

\bibitem[\protect\citeauthoryear{%
Perrinet%
, Delorme%
, Samuelides%
\BCBL{}\ \BBA{} Thorpe%
}{%
Perrinet%
\ \protect\BOthers{.}}{%
{\protect\APACyear{2001}}%
}]{%
perrinet-et-al:01}%
\APACinsertmetastar{%
perrinet-et-al:01}%
Perrinet, L.%
, Delorme, A.%
, Samuelides, M.%
\BCBL{}\ \BBA{} Thorpe, S.%
%
\unskip\
\newblock
\APACrefYearMonthDay{2001}{}{}.
\newblock
\BBOQ{}\APACrefatitle{Networks of integrate-and-fire neuron using rank order
  coding A: How to implement spike time dependent Hebbian plasticity.}{Networks
  of integrate-and-fire neuron using rank order coding a: How to implement
  spike time dependent hebbian plasticity.}\BBCQ{}
\newblock
\APACjournalVolNumPages{Neurocomputing}{38}{}{}.
\PrintBackRefs{\CurrentBib}

\bibitem[\protect\citeauthoryear{%
Rao%
\ \BBA{} Sejnowski%
}{%
Rao%
\ \BBA{} Sejnowski%
}{%
{\protect\APACyear{1991}}%
}]{%
rao-sejnowski:99}%
\APACinsertmetastar{%
rao-sejnowski:99}%
Rao, R.%
\BCBT{}\ \BBA{} Sejnowski, T\BPBI J.%
%
\unskip\
\newblock
\APACrefYear{1991}.
\newblock
\APACrefbtitle{Predictive sequence learning in recurrent neocortical
  circuits}{Predictive sequence learning in recurrent neocortical circuits}\
  (\BVOL~12; S.~Solla, T.~Leen\BCBL{}\ \BBA{} K.~Muller, \BEDS{}).
\newblock
\APACaddressPublisher{}{Cambridge MA, MIT Press}.
\PrintBackRefs{\CurrentBib}

\bibitem[\protect\citeauthoryear{%
Rao%
\ \BBA{} Sejnowski%
}{%
Rao%
\ \BBA{} Sejnowski%
}{%
{\protect\APACyear{2001}}%
}]{%
rao-sejnowski:01}%
\APACinsertmetastar{%
rao-sejnowski:01}%
Rao, R.%
\BCBT{}\ \BBA{} Sejnowski, T\BPBI J.%
%
\unskip\
\newblock
\APACrefYearMonthDay{2001}{}{}.
\newblock
\BBOQ{}\APACrefatitle{Spike-timing-dependent Hebbian plasticity as temporal
  difference learning.}{Spike-timing-dependent hebbian plasticity as temporal
  difference learning.}\BBCQ{}
\newblock
\APACjournalVolNumPages{Neural Comput.}{13}{10}{2221--2237}.
\PrintBackRefs{\CurrentBib}

\bibitem[\protect\citeauthoryear{%
Rieke%
, Warland%
, Steveninck%
\BCBL{}\ \BBA{} Bialek%
}{%
Rieke%
\ \protect\BOthers{.}}{%
{\protect\APACyear{1996}}%
}]{%
rieke-etal:96}%
\APACinsertmetastar{%
rieke-etal:96}%
Rieke, F.%
, Warland, D.%
, Steveninck, R. de~Ruyter~van%
\BCBL{}\ \BBA{} Bialek, W.%
%
\unskip\
\newblock
\APACrefYear{1996}.
\newblock
\APACrefbtitle{Spikes, Exploring the Neural Code}{Spikes, exploring the neural
  code}.
\newblock
\APACaddressPublisher{}{The M.I.T. Press}.
\PrintBackRefs{\CurrentBib}

\bibitem[\protect\citeauthoryear{%
Rostro-Gonzalez%
, Cessac%
, Vasquez%
\BCBL{}\ \BBA{} Vi\'eville%
}{%
Rostro-Gonzalez%
\ \protect\BOthers{.}}{%
{\protect\APACyear{2009}}%
}]{%
rostro-cessac-etal:09b}%
\APACinsertmetastar{%
rostro-cessac-etal:09b}%
Rostro-Gonzalez, H.%
, Cessac, B.%
, Vasquez, J.%
\BCBL{}\ \BBA{} Vi\'eville, T.%
%
\unskip\
\newblock
\APACrefYearMonthDay{2009}{}{}.
\newblock
\BBOQ{}\APACrefatitle{Back-engineering of spiking neural networks
  parameters}{Back-engineering of spiking neural networks parameters}.\BBCQ{}
\newblock
\APACjournalVolNumPages{Journal of Computational Neuroscience}{}{}{}.
\newblock
\APACrefnote{submitted}
\PrintBackRefs{\CurrentBib}

\bibitem[\protect\citeauthoryear{%
Rudolph%
\ \BBA{} Destexhe%
}{%
Rudolph%
\ \BBA{} Destexhe%
}{%
{\protect\APACyear{2006}}%
}]{%
rudolph-destexhe:06}%
\APACinsertmetastar{%
rudolph-destexhe:06}%
Rudolph, M.%
\BCBT{}\ \BBA{} Destexhe, A.%
%
\unskip\
\newblock
\APACrefYearMonthDay{2006}{}{}.
\newblock
\BBOQ{}\APACrefatitle{Analytical Integrate and Fire Neuron models with
  conductance-based dynamics for event driven simulation strategies}{Analytical
  integrate and fire neuron models with conductance-based dynamics for event
  driven simulation strategies}.\BBCQ{}
\newblock
\APACjournalVolNumPages{Neural Computation}{18}{}{2146--2210}.
\newblock
 \begin{APACrefURL}
  \url{http://www.mitpressjournals.org/doi/abs/10.1162/neco.2006.18.9.2146}
  \end{APACrefURL}
\PrintBackRefs{\CurrentBib}

\bibitem[\protect\citeauthoryear{%
Ruelle%
}{%
Ruelle%
}{%
{\protect\APACyear{1969}}%
}]{%
ruelle:69}%
\APACinsertmetastar{%
ruelle:69}%
Ruelle, D.%
%
\unskip\
\newblock
\APACrefYear{1969}.
\newblock
\APACrefbtitle{Statistical Mechanics: Rigorous results}{Statistical mechanics:
  Rigorous results}.
\newblock
\APACaddressPublisher{}{Benjamin, New York}.
\PrintBackRefs{\CurrentBib}

\bibitem[\protect\citeauthoryear{%
Ruelle%
}{%
Ruelle%
}{%
{\protect\APACyear{1999}}%
}]{%
ruelle:99}%
\APACinsertmetastar{%
ruelle:99}%
Ruelle, D.%
%
\unskip\
\newblock
\APACrefYearMonthDay{1999}{}{}.
\newblock
\BBOQ{}\APACrefatitle{Smooth dynamics and new theoretical ideas in
  nonequilibrium statistical mechanics.}{Smooth dynamics and new theoretical
  ideas in nonequilibrium statistical mechanics.}\BBCQ{}
\newblock
\APACjournalVolNumPages{J. Statist. Phys.}{95}{}{393-468}.
\PrintBackRefs{\CurrentBib}

\bibitem[\protect\citeauthoryear{%
Samuelides%
\ \BBA{} Cessac%
}{%
Samuelides%
\ \BBA{} Cessac%
}{%
{\protect\APACyear{2007}}%
}]{%
samuelides-cessac:07}%
\APACinsertmetastar{%
samuelides-cessac:07}%
Samuelides, M.%
\BCBT{}\ \BBA{} Cessac, B.%
%
\unskip\
\newblock
\APACrefYearMonthDay{2007}{}{}.
\newblock
\BBOQ{}\APACrefatitle{Random Recurrent Neural Networks}{Random recurrent neural
  networks}.\BBCQ{}
\newblock
\APACjournalVolNumPages{European Physical Journal - Special
  Topics}{142}{}{7--88}.
\PrintBackRefs{\CurrentBib}

\bibitem[\protect\citeauthoryear{%
Schneidman%
, Berry%
, Segev%
\BCBL{}\ \BBA{} Bialek%
}{%
Schneidman%
\ \protect\BOthers{.}}{%
{\protect\APACyear{2006}}%
}]{%
schneidman-etal:06}%
\APACinsertmetastar{%
schneidman-etal:06}%
Schneidman, E.%
, Berry, M.%
, Segev, R.%
\BCBL{}\ \BBA{} Bialek, W.%
%
\unskip\
\newblock
\APACrefYearMonthDay{2006}{}{}.
\newblock
\BBOQ{}\APACrefatitle{Weak pairwise correlations imply string correlated
  network states in a neural population}{Weak pairwise correlations imply
  string correlated network states in a neural population}.\BBCQ{}
\newblock
\APACjournalVolNumPages{Nature}{440}{}{1007-- 1012}.
\PrintBackRefs{\CurrentBib}

\bibitem[\protect\citeauthoryear{%
Sinanovi\'{c}%
\ \BBA{} Johnson%
}{%
Sinanovi\'{c}%
\ \BBA{} Johnson%
}{%
{\protect\APACyear{2006}}%
}]{%
sinanovic-johnson:06}%
\APACinsertmetastar{%
sinanovic-johnson:06}%
Sinanovi\'{c}, A.%
\BCBT{}\ \BBA{} Johnson, D.%
%
\unskip\
\newblock
\APACrefYearMonthDay{2006}{}{}.
\newblock
\BBOQ{}\APACrefatitle{Toward a theory of information processing}{Toward a
  theory of information processing}.\BBCQ{}
\newblock
\APACjournalVolNumPages{signal processing}{}{}{}.
\newblock
\APACrefnote{submitted}
\PrintBackRefs{\CurrentBib}

\bibitem[\protect\citeauthoryear{%
Siri%
, Berry%
, Cessac%
, Delord%
\BCBL{}\ \BBA{} Quoy%
}{%
Siri%
\ \protect\BOthers{.}}{%
{\protect\APACyear{2007}}%
}]{%
siri-etal:07}%
\APACinsertmetastar{%
siri-etal:07}%
Siri, B.%
, Berry, H.%
, Cessac, B.%
, Delord, B.%
\BCBL{}\ \BBA{} Quoy, M.%
%
\unskip\
\newblock
\APACrefYearMonthDay{2007}{}{}.
\newblock
\BBOQ{}\APACrefatitle{Effects of Hebbian learning on the dynamics and structure
  of random networks with inhibitory and excitatory neurons.}{Effects of
  hebbian learning on the dynamics and structure of random networks with
  inhibitory and excitatory neurons.}\BBCQ{}
\newblock
\APACjournalVolNumPages{Journal of Physiology, Paris}{101}{1-3}{138--150}.
\newblock
\APACrefnote{e-print: arXiv:0706.2602}
\PrintBackRefs{\CurrentBib}

\bibitem[\protect\citeauthoryear{%
Siri%
, Berry%
, Cessac%
, Delord%
\BCBL{}\ \BBA{} Quoy%
}{%
Siri%
\ \protect\BOthers{.}}{%
{\protect\APACyear{2008}}%
}]{%
siri-etal:08}%
\APACinsertmetastar{%
siri-etal:08}%
Siri, B.%
, Berry, H.%
, Cessac, B.%
, Delord, B.%
\BCBL{}\ \BBA{} Quoy, M.%
%
\unskip\
\newblock
\APACrefYearMonthDay{2008}{}{}.
\newblock
\BBOQ{}\APACrefatitle{A Mathematical Analysis of the Effects of Hebbian
  Learning Rules on the Dynamics and Structure of Discrete-Time Random
  Recurrent Neural Networks}{A mathematical analysis of the effects of hebbian
  learning rules on the dynamics and structure of discrete-time random
  recurrent neural networks}.\BBCQ{}
\newblock
\APACjournalVolNumPages{Neural Comp.}{20}{12}{12}.
\newblock
\APACrefnote{e-print: arXiv:0705.3690v1}
\PrintBackRefs{\CurrentBib}

\bibitem[\protect\citeauthoryear{%
Soula%
}{%
Soula%
}{%
{\protect\APACyear{2005}}%
}]{%
soula:05}%
\APACinsertmetastar{%
soula:05}%
Soula, H.%
%
\unskip\
\newblock
\APACrefYear{2005}.
\newblock
\APACrefbtitle{Dynamique et plasticit\'e dans les r\'eseaux de neurones \`a
  impulsions}{Dynamique et plasticit\'e dans les r\'eseaux de neurones \`a
  impulsions}.
\newblock
\BUPhD, INSA Lyon.
\PrintBackRefs{\CurrentBib}

\bibitem[\protect\citeauthoryear{%
Soula%
, Beslon%
\BCBL{}\ \BBA{} Mazet%
}{%
Soula%
\ \protect\BOthers{.}}{%
{\protect\APACyear{2006}}%
}]{%
soula-etal:06}%
\APACinsertmetastar{%
soula-etal:06}%
Soula, H.%
, Beslon, G.%
\BCBL{}\ \BBA{} Mazet, O.%
%
\unskip\
\newblock
\APACrefYearMonthDay{2006}{}{}.
\newblock
\BBOQ{}\APACrefatitle{Spontaneous dynamics of asymmetric random recurrent
  spiking neural networks}{Spontaneous dynamics of asymmetric random recurrent
  spiking neural networks}.\BBCQ{}
\newblock
\APACjournalVolNumPages{Neural Computation}{18}{1}{}.
\PrintBackRefs{\CurrentBib}

\bibitem[\protect\citeauthoryear{%
Soula%
\ \BBA{} Chow%
}{%
Soula%
\ \BBA{} Chow%
}{%
{\protect\APACyear{2007}}%
}]{%
soula-chow:07}%
\APACinsertmetastar{%
soula-chow:07}%
Soula, H.%
\BCBT{}\ \BBA{} Chow, C\BPBI C.%
%
\unskip\
\newblock
\APACrefYearMonthDay{2007}{}{}.
\newblock
\BBOQ{}\APACrefatitle{Stochastic Dynamics of a Finite-Size Spiking Neural
  Networks}{Stochastic dynamics of a finite-size spiking neural
  networks}.\BBCQ{}
\newblock
\APACjournalVolNumPages{Neural Computation}{19}{}{3262--3292}.
\PrintBackRefs{\CurrentBib}

\bibitem[\protect\citeauthoryear{%
Theunissen%
\ \BBA{} Miller%
}{%
Theunissen%
\ \BBA{} Miller%
}{%
{\protect\APACyear{1995}}%
}]{%
theunissen-miller:95}%
\APACinsertmetastar{%
theunissen-miller:95}%
Theunissen, F.%
\BCBT{}\ \BBA{} Miller, J.%
%
\unskip\
\newblock
\APACrefYearMonthDay{1995}{}{}.
\newblock
\BBOQ{}\APACrefatitle{Temporal Encoding in Nervous Systems: A Rigorous
  Definition.}{Temporal encoding in nervous systems: A rigorous
  definition.}\BBCQ{}
\newblock
\APACjournalVolNumPages{Journal of Computational Neuroscience}{2}{}{149—162}.
\PrintBackRefs{\CurrentBib}

\bibitem[\protect\citeauthoryear{%
Tkacik%
, Schneidman%
, Berry%
\BCBL{}\ \BBA{} Bialek%
}{%
Tkacik%
\ \protect\BOthers{.}}{%
{\protect\APACyear{2006}}%
}]{%
tkacik-etal:06}%
\APACinsertmetastar{%
tkacik-etal:06}%
Tkacik, G.%
, Schneidman, E.%
, Berry, M.%
\BCBL{}\ \BBA{} Bialek, W.%
%
\unskip\
\newblock
\APACrefYearMonthDay{2006}{}{}.
\newblock
\BBOQ{}\APACrefatitle{Ising models for networks of real neurons}{Ising models
  for networks of real neurons}.\BBCQ{}
\newblock
\APACjournalVolNumPages{arXiv}{q-bio/0611072}{}{}.
\PrintBackRefs{\CurrentBib}

\bibitem[\protect\citeauthoryear{%
Touboul%
}{%
Touboul%
}{%
{\protect\APACyear{2008}}%
}]{%
touboul:08}%
\APACinsertmetastar{%
touboul:08}%
Touboul, J.%
%
\unskip\
\newblock
\APACrefYearMonthDay{2008}{}{}.
\newblock
\BBOQ{}\APACrefatitle{Bifurcation Analysis of a General Class of Nonlinear
  Integrate-and-Fire Neurons}{Bifurcation analysis of a general class of
  nonlinear integrate-and-fire neurons}.\BBCQ{}
\newblock
\APACjournalVolNumPages{SIAM Journal on Applied Mathematics}{68}{4}{1045-1079}.
\newblock
 \begin{APACrefURL} \url{http://link.aip.org/link/?SMM/68/1045/1}
  \end{APACrefURL}
\PrintBackRefs{\CurrentBib}

\bibitem[\protect\citeauthoryear{%
Toyoizumi%
, Pfister%
, Aihara%
\BCBL{}\ \BBA{} Gerstner%
}{%
Toyoizumi%
\ \protect\BOthers{.}}{%
{\protect\APACyear{2005}}%
}]{%
toyoizumi-etal:05}%
\APACinsertmetastar{%
toyoizumi-etal:05}%
Toyoizumi, T.%
, Pfister, J\BHBI P.%
, Aihara, K.%
\BCBL{}\ \BBA{} Gerstner, W.%
%
\unskip\
\newblock
\APACrefYearMonthDay{2005}{}{}.
\newblock
\BBOQ{}\APACrefatitle{Generalized Bienenstock-Cooper-Munro rule for spiking
  neurons that maximizes information transmission}{Generalized
  bienenstock-cooper-munro rule for spiking neurons that maximizes information
  transmission}.\BBCQ{}
\newblock
\APACjournalVolNumPages{Proceedings of the National Academy of
  Science}{102}{}{5239--5244}.
\PrintBackRefs{\CurrentBib}

\bibitem[\protect\citeauthoryear{%
Toyoizumi%
, Pfister%
, Aihara%
\BCBL{}\ \BBA{} Gerstner%
}{%
Toyoizumi%
\ \protect\BOthers{.}}{%
{\protect\APACyear{2007}}%
}]{%
toyoizumi-etal:07}%
\APACinsertmetastar{%
toyoizumi-etal:07}%
Toyoizumi, T.%
, Pfister, J\BHBI P.%
, Aihara, K.%
\BCBL{}\ \BBA{} Gerstner, W.%
%
\unskip\
\newblock
\APACrefYearMonthDay{2007}{}{}.
\newblock
\BBOQ{}\APACrefatitle{Optimality Model of Unsupervised Spike-Timing Dependent
  Plasticity: Synaptic Memory and Weight Distribution}{Optimality model of
  unsupervised spike-timing dependent plasticity: Synaptic memory and weight
  distribution}.\BBCQ{}
\newblock
\APACjournalVolNumPages{Neural Computation}{19}{}{639--671}.
\PrintBackRefs{\CurrentBib}

\bibitem[\protect\citeauthoryear{%
Wood%
, Roth%
\BCBL{}\ \BBA{} Black%
}{%
Wood%
\ \protect\BOthers{.}}{%
{\protect\APACyear{2006}}%
}]{%
wood-et-al:06}%
\APACinsertmetastar{%
wood-et-al:06}%
Wood, F.%
, Roth, S.%
\BCBL{}\ \BBA{} Black, M.%
%
\unskip\
\newblock
\APACrefYearMonthDay{2006}{}{}.
\newblock
\BBOQ{}\APACrefatitle{Modeling Neural Population Spiking Activity with Gibbs
  Distributions}{Modeling neural population spiking activity with gibbs
  distributions}.\BBCQ{}
\newblock
\BIn{} Y.~Weiss, B.~Sch\"{o}lkopf\BCBL{}\ \BBA{} J.~Platt\ (\BEDS),
  \APACrefbtitle{Advances in Neural Information Processing Systems 18}{Advances
  in neural information processing systems 18}\ (\BPGS\ 1537--1544).
\newblock
\APACaddressPublisher{Cambridge, MA}{MIT Press}.
\PrintBackRefs{\CurrentBib}

\bibitem[\protect\citeauthoryear{%
Zou%
}{%
Zou%
}{%
{\protect\APACyear{2006}}%
}]{%
zou:06}%
\APACinsertmetastar{%
zou:06}%
Zou, Q.%
%
\unskip\
\newblock
\APACrefYear{2006}.
\newblock
\APACrefbtitle{Mod{\`e}les computationnels de la plasticit{\'e} impulsionnelle:
  synapses, neurones et circuits}{Mod{\`e}les computationnels de la
  plasticit{\'e} impulsionnelle: synapses, neurones et circuits}.
\newblock
\BUPhD, Universit{\'e} Paris VI.
\PrintBackRefs{\CurrentBib}

\end{thebibliography}
%\bibliography{/home/vthierry/Odyssee/Bibliography/string,/home/vthierry/Odyssee/Bibliography/odyssee,../../../Latex/biblio}

\end{document}